\documentclass[10pt,aps,pre,longbibliography,amsmath,amssymb,groupedaddress]{revtex4-2}

\usepackage{graphicx}  
\usepackage{bm}      
\usepackage{dcolumn}   
\usepackage{amsthm}
\usepackage{booktabs}
\usepackage{flafter}
\usepackage{float}
\usepackage{hyperref}

\usepackage{etoolbox}
\pretocmd{\paragraph}{\par\addvspace{0.3cm}}{}{}

\newcommand{\be}{\begin{equation}}
\newcommand{\ee}{\end{equation}}
\newcommand{\bea}{\begin{eqnarray}}
\newcommand{\eea}{\end{eqnarray}}
\def\bse{\begin{subequations}}
\def\ese{\end{subequations}}

\def\IZ{\relax\ifmmode\hbox{Z\kern-.4em Z}\else{Z\kern-.4em Z}\fi}

\newcommand{\non}{\nonumber \\}

\def\half{\frac{1}{2}}

\def\bi{\begin{itemize}} \def\ei{\end{itemize}}

\begin{document}

\title{Flip rate prediction in the double pendulum}

\author{Peleg Haham}
\email{peleg.haham@mail.huji.ac.il}

\author{Barak Kol}
\email{barak.kol@mail.huji.ac.il}

\affiliation{Racah Institute of Physics,
Hebrew University,
Jerusalem, 9190401, Israel
}

\date{\today}
\begin{abstract}
The intermediate-energy double pendulum is a prototypical chaotic system. Despite its irregular motion, it exhibits recurrent flips--events in which one of the arms passes over the top. We develop a statistical prediction for the mean flip rate in terms of phase-space flux. Applying this flux-based approach to the equal-mass, equal-arm (``egalitarian'') double pendulum, we find excellent agreement between the statistical prediction and numerical simulations: ensemble-averaged flip rates agree to within 1\%, while even the statistics of individual chaotic trajectories agree at the few-percent level, quantifying the validity of the ergodic approximation. This agreement holds for flips of either arm and over a broad range of energies. This flux-based approach is closely related to that used for the egalitarian three-body system.

A central role is played by the saddle orbits: periodic orbits that tend to the periodic saddle eigenmode as the energy approaches the saddle energy from above and approximately follow the ridge of the potential at higher energies. These orbits provide a natural dividing surface for defining flips while avoiding recrossings. Indeed, the resulting distribution of crossing times exhibits a distinct gap. It is argued that the corresponding flux-over-a-saddle prediction provides a preferred realization of the flux-based approach. We also present two simplified alternative dividing surfaces, including one based on a remarkably accurate analytic approximation to the saddle orbits, and show that both yield statistical predictions in agreement with the corresponding simulated rates.

\end{abstract}

\maketitle

\tableofcontents

\vspace{0.5cm}
\noindent
\begin{minipage}[H]{0.48\textwidth}
\begin{flushleft}
\begin{tabular}{@{}c@{}} 
PH dedicates this work\\ to the memory of his father,\\
Amnon Haham \\
4 December 1966 -- \\ 23 May 2021\\
with love and gratefulness.
\end{tabular}
\end{flushleft}
\end{minipage}
\begin{minipage}[H]{0.48\textwidth}  
\begin{flushright}
\begin{tabular}{@{}c@{}} 
BK dedicates this work\\ to the memory of his father,\\
Yoram Kol \\
11 June 1938, Gvat -- \\ 17 July 2026, Kfar Saba\\
with gratitude, love and pride.
\end{tabular}
\end{flushright}
\end{minipage}

\section{Introduction}
The double pendulum is a prototypical chaotic system with only two degrees of freedom; see, e.g. \cite{ShinbrotGrebogiWisdomYorke1992DoublePendulum,Richter_Scholz_1984,Richter_Scholz_1985} and references therein. As such, it is a good case study for fundamental aspects of chaos. 

\begin{figure}[H]
\centering\includegraphics[width=0.45\linewidth]{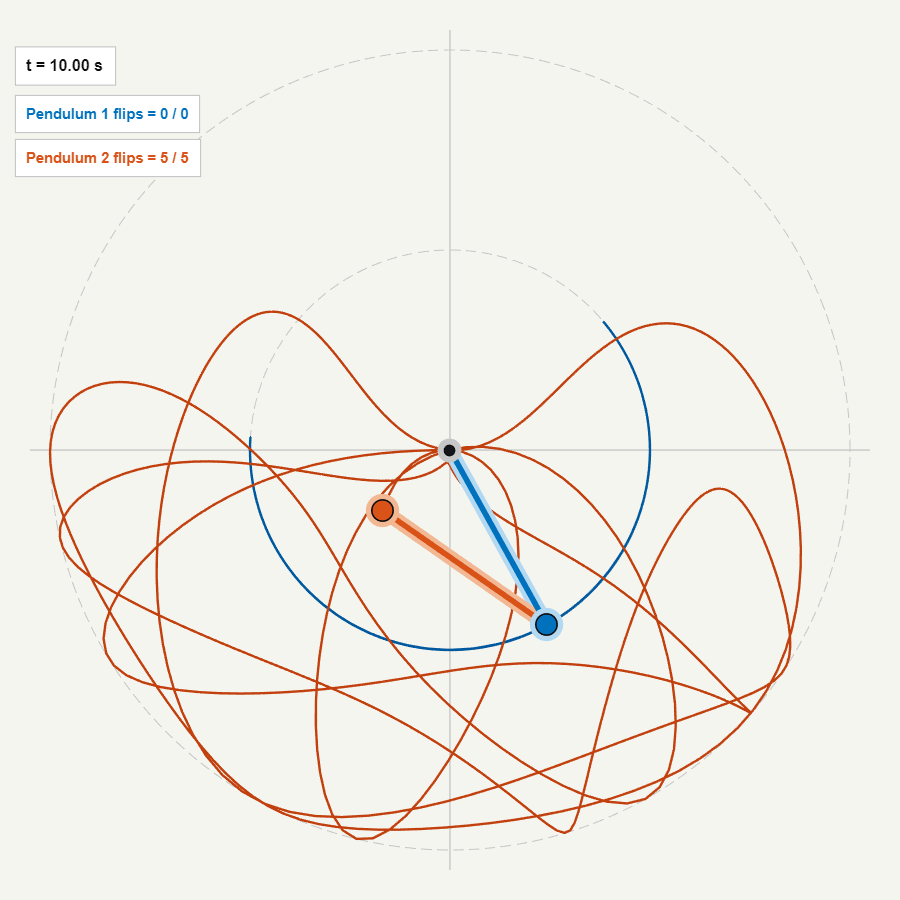}
    \caption{A typical chaotic trajectory of the double pendulum. The colored curves trace the paths of the two point masses; in this example, the second arm undergoes several flips while the first does not.
}
    \label{fig:typical_trajectory}
\end{figure}

Watching the chaotic motion of the double pendulum, one notices dramatic flips -- events in which one of the pendulum arms passes over the top, and possibly completes a rotation. Chaotic time-evolutions display erratic and random-like occurrences of flips, see Fig.~\ref{fig:typical_trajectory}. At first sight, it appears impossible to predict them.

Despite their chaotic random-like nature, flips are observed to persist -- they continue to occur as time goes on. This suggests that a \emph{mean flip rate} could be defined and measured in simulations. 

This observable raises a natural research question: can one formulate a theory that \emph{predicts} it? Since the mean flip rate is statistical in nature, such a theory must itself be statistical.

In this paper, we present such a statistical theory, based on the flux of phase-space volume (phase volume and phase flux, for short). We begin in Section~\ref{sec:system} by describing and formulating the double-pendulum system. Next, in Section~\ref{sec:criteria}, following \cite{MacKay1990,MacKay1994}, we define a natural criterion for a flip in terms of crossing the saddle orbit, a periodic and flux-minimizing orbit. An online, mobile-compatible interactive viewer of the saddle orbits for the double pendulum is available in \cite{interactive_viewer}. We also introduce two alternative flip criteria. In Section~\ref{sec:prediction}, we present the statistical prediction for the corresponding mean flip rates in terms of phase-space integrals. 

In Section~\ref{sec:implementation}, we describe our implementation of the computations required to compare the statistical prediction with numerical simulations. In Section~\ref{sec:results}, we present the observed flip rate statistics and compare them with the statistical prediction. Agreement is demonstrated not only at the level of ensemble averages, but also for individual chaotic trajectories, reflecting the ergodic nature of the system. Finally, in Section~\ref{sec:summary} we summarize the results and discuss their significance. Appendix~\ref{app:story} describes the unusual path that led to this work and its connection to the Newtonian three-body problem, while the remaining appendices provide details of several derivations.

Recent work on non-integrable dynamics 
in the double pendulum includes studies of tubes of stable and unstable manifolds associated with the saddle orbit \cite{Tubes_2023}, the use of the saddle orbit for swing-up control \cite{SwingUp_2024}, and the chaotic fraction of the system \cite{Chaotic_frac_2023,Chaotic_frac_2024}. 

\paragraph*{Notation.} We collect here several notational conventions:
\begin{center}    
\noindent \begin{tabular}{l@{\qquad}l}
     $\Theta(x)$    &  Heaviside step function \\
     $x_+$    & Ramp function, defined in \eqref{def:ramp} \\
     $R_a(E)$ & Flip rate according to criterion $a$ at energy $E$  \\
     $f_a(E)$ & Bidirectional phase-volume flux through dividing surface $a$ at energy $E$ \\
     $\Omega(E)$  & Total phase volume at energy $E$ \\
     $\varphi_a(E)$   &   Action of a closed curve $a$ with energy $E$ \\
     $T_a(E)$     &   Period of periodic orbit $a$ with energy $E$ 
    \end{tabular}
\end{center}

\section{The double pendulum}
\label{sec:system}

\begin{figure}[H]
    \centering
\includegraphics[width=0.42\linewidth]{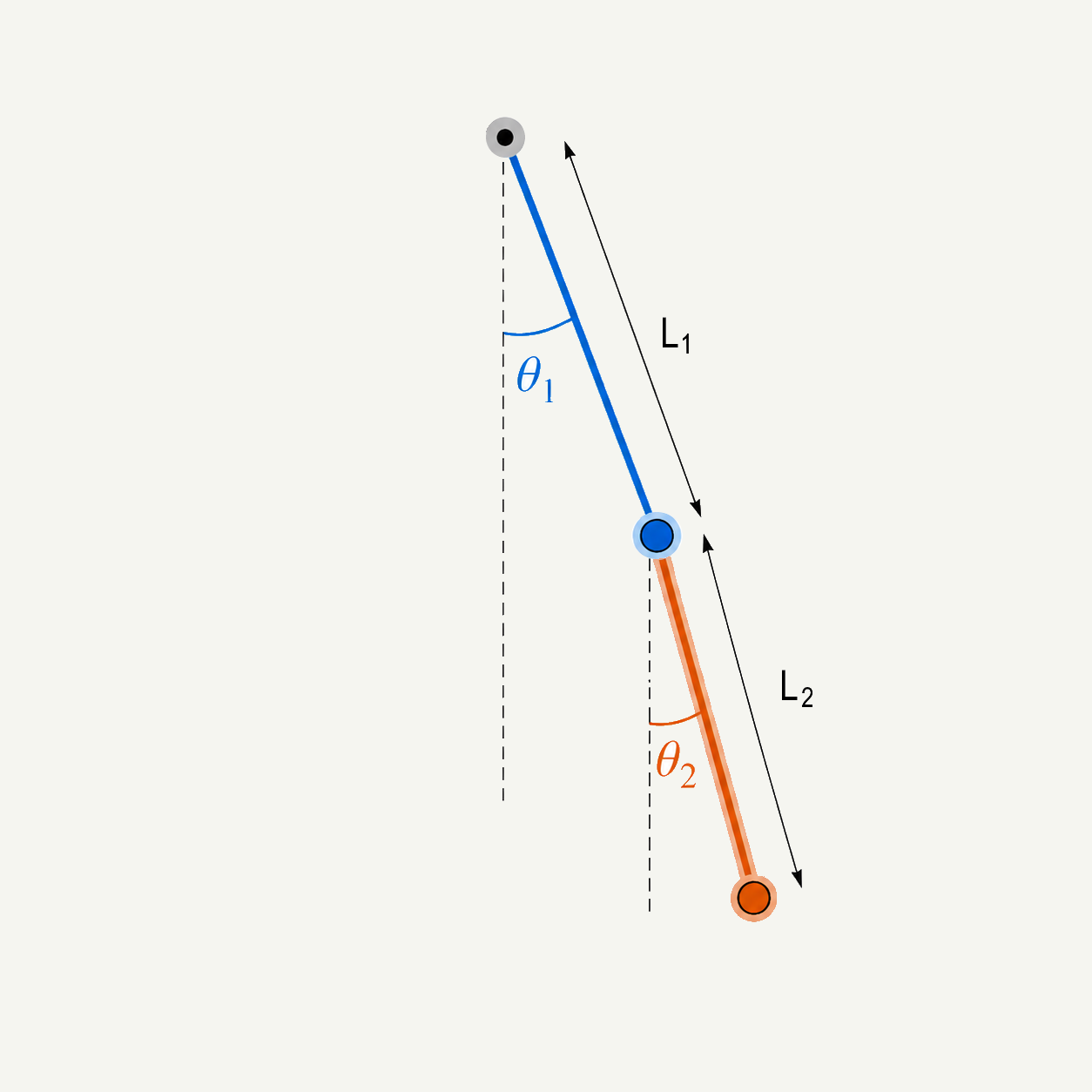}
    \caption{The double pendulum}
\label{fig:double_pendulum_system_example}
\end{figure}

The double pendulum is a mechanical system composed of two rigid pendulums connected in series: arm 1 hangs from a fixed pivot and arm 2 hangs from the end of arm 1. The arm lengths are $L_1$ and $L_2$, respectively, and the masses attached to their ends are $m_1$ and $m_2$, see Fig.~\ref{fig:double_pendulum_system_example}. The nonlinear coupling between the pendulums produces rich dynamical behavior, ranging from regular to chaotic motion, depending on the system's energy and initial conditions.

\subsection{Formulation}

The configuration of the double pendulum is described by two angles \begin{equation}
    \theta_1, ~ \theta_2 ~,
    \label{def:coordinates} 
\end{equation}
 where $\theta_1$ is the orientation of arm 1  measured from the downward vertical and \(\theta_2\) is defined analogously for arm 2; see Fig.~\ref{fig:double_pendulum_system_example}. We denote the relative angle by 
\begin{equation}
    \Delta \theta :=\theta_2 - \theta_1 ~.
    \label{def:Delta_theta}
\end{equation}

The kinetic energy is  \begin{equation}
     T := \half M_{ij} \, \dot{\theta}_i \dot{\theta}_j ~,
    \label{def:T}
\end{equation}
where the summation convention is assumed throughout (repeated indices are summed), $i,j=1,2$ label the two arms, and the configuration-dependent inertia matrix is \begin{equation}
     M :=\begin{bmatrix}
        (m_1 + m_2) L_1^2 &   m_2\, L_1\, L_2\, \cos \Delta \theta \\
        m_2\, L_1\, L_2\, \cos \Delta \theta       & m_2\, L_2^2
    \end{bmatrix} ~.
    \label{def:M}
\end{equation}
Equivalently, the kinetic energy can be written succinctly as $2 T=m_1\, L_1^2 \, \dot{\theta}_1^2 + m_2 \left| L_1\, \dot{\theta}_1 + L_2\, \dot{\theta}_2\, e^{i \Delta \theta} \right|^2$. 
For later use, we record \begin{equation}
    \det M = m_2 \left( m_1 + m_2 \, \sin^2 \Delta \theta \right) L_1^2\, L_2^2 ~,
    \label{detM}
\end{equation}
and  
\begin{equation}
     M^{-1} :=\frac{1}{\det M} \begin{bmatrix}
         m_2\, L_2^2    &   - m_2\, L_1\, L_2\, \cos \Delta \theta \\
        - m_2\, L_1\, L_2\, \cos \Delta \theta       & (m_1 + m_2) L_1^2 
    \end{bmatrix} ~.
    \label{def:M_inverse}
\end{equation} 

The potential energy is  \begin{equation}
    V := ( m_1+m_2 ) \, g\, L_1 \left( 1- \cos \theta_1 \right) + m_2  \, g\, L_2 \left( 1- \cos \theta_2 \right) ~.
    \label{def:V}
\end{equation}
The additive constant has been chosen such that the minimum of the potential energy is zero, so that $V(\theta_1,\theta_2)\geq 0$.

We can now define the Lagrangian to be \begin{equation}
    \mathcal{L} := T - V ~.
    \label{def:Lagrangian}
\end{equation} 
The equations of motion follow from the corresponding Euler-Lagrange equations \begin{equation}
    \frac{d}{dt}\frac{\partial \mathcal{L}}{\partial \dot{\theta}_i}
    =
    \frac{\partial \mathcal{L}}{\partial \theta_i} ~,
    \qquad i=1,2 ~.
    \label{Euler-Lagrange}
\end{equation}

The conjugate angular momenta are \begin{equation}
    l_i := \frac{\partial \mathcal{L}}{\partial \dot{\theta}_i}
    = M_{ij} \, \dot{\theta}_j ~.
\end{equation}
In terms of the angular momenta, the kinetic energy is \begin{equation}
    T(l) := \half M^{-1}_{ij} \, l_i \, l_j ~,
    \label{def:Tl}
\end{equation}
 where the inverse inertia matrix is given in \eqref{def:M_inverse}. The Hamiltonian is therefore $\mathcal{H} :=T(l) + V$.
 In Hamiltonian form, the equations of motion are
\begin{equation}
    \dot{\theta}_i = \frac{\partial \mathcal{H}}{\partial l_i},
    \qquad
    \dot{l}_i = -\frac{\partial \mathcal{H}}{\partial \theta_i} ~.
\end{equation}

For concreteness, we specialize from here on to equal arms and equal masses, a case that we call the egalitarian double pendulum. We also choose units such that \begin{equation}
    m_1=m_2=1 ~, \quad
    L_1=L_2=1 ~, \quad
    g=1 ~.
    \label{units}
\end{equation}
With this choice, the kinetic and potential energies in (\ref{def:T},\ref{def:Tl},\ref{def:V}) become \begin{align}
    2 T &= 2 \dot{\theta}_1^2 + 2 \cos \Delta \theta \, \dot{\theta}_1  \dot{\theta}_2 + \dot{\theta}_2^2 ~, \non 
    2 T(l) &= \frac{1}{1+\sin^2\Delta \theta} \left( l_1^2 - 2 \cos \Delta \theta \, l_1  l_2 + 2 l_2^2 \right) ~, \non
    V &= 3 - 2 \cos \theta_1 -\cos \theta_2 ~, \label{egalV} \\
    \mathcal{H} &= T(l) + V ~.
    \label{eq:egalitarian double pendulum hamiltonian}
\end{align}

\subsection{Basic dynamical properties}

We proceed by presenting several basic dynamical properties of the double pendulum: its equilibrium points, symmetries and conserved quantities, and the essential dependence of its dynamics on the total energy.

\begin{figure}[H]
    \centering
\includegraphics[width=0.6\linewidth]{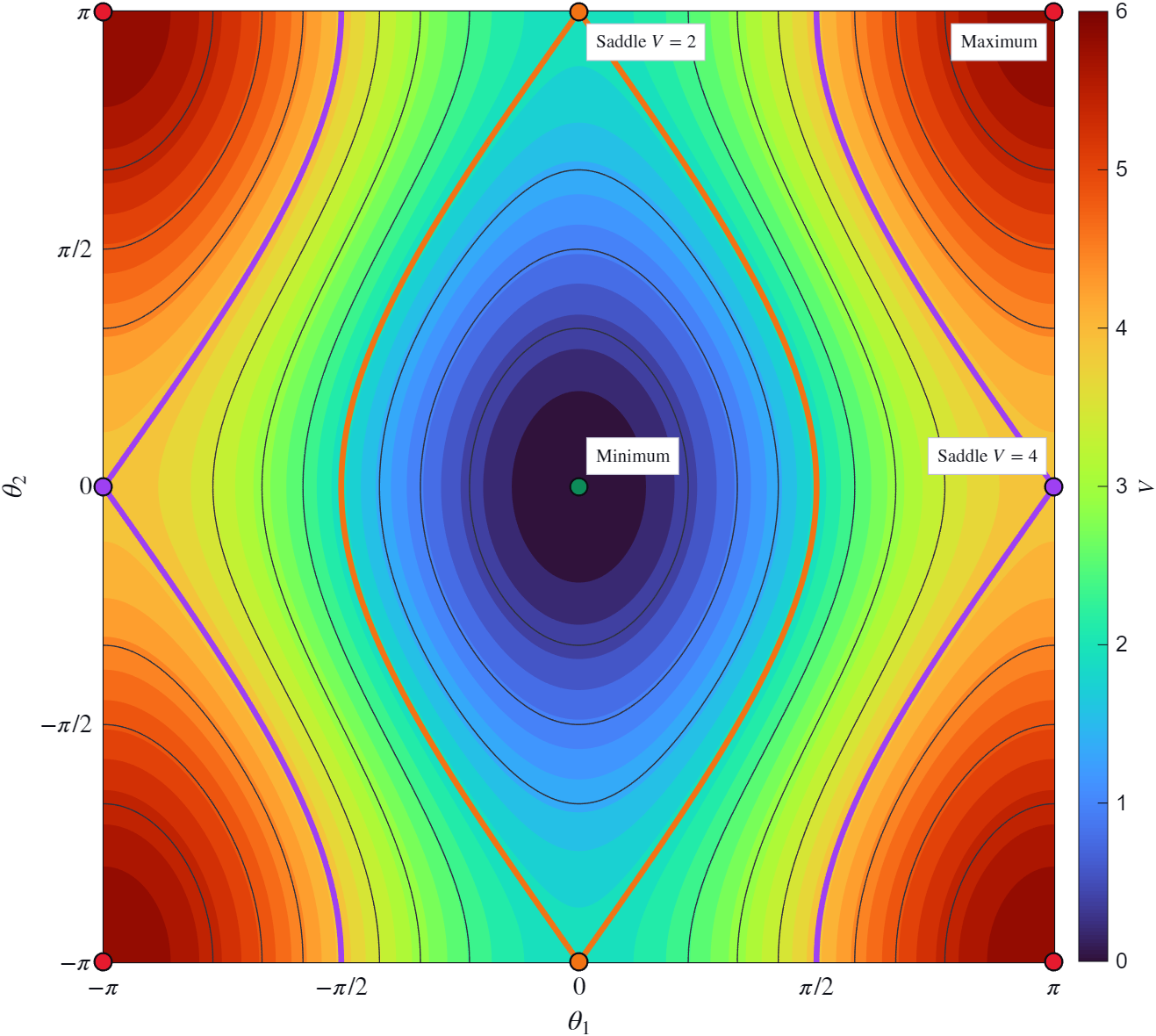}
   \caption{Contour map of the egalitarian double-pendulum potential \eqref{egalV}, with the equilibrium points labeled.}
    \label{fig:potential}
\end{figure}

\paragraph*{Equilibrium points.} The system is in equilibrium precisely when both arms are vertical, with each arm pointing either upward or downward. Thus there are four equilibrium points in total. They are listed in Table~\ref{tab:potential_critical_points}, together with some of their properties. Figure~\ref{fig:potential} shows a contour plot of the potential of the egalitarian double pendulum, with the equilibrium points labeled. 

\begin{figure}[H]
    \centering
    \includegraphics[width=1\linewidth]{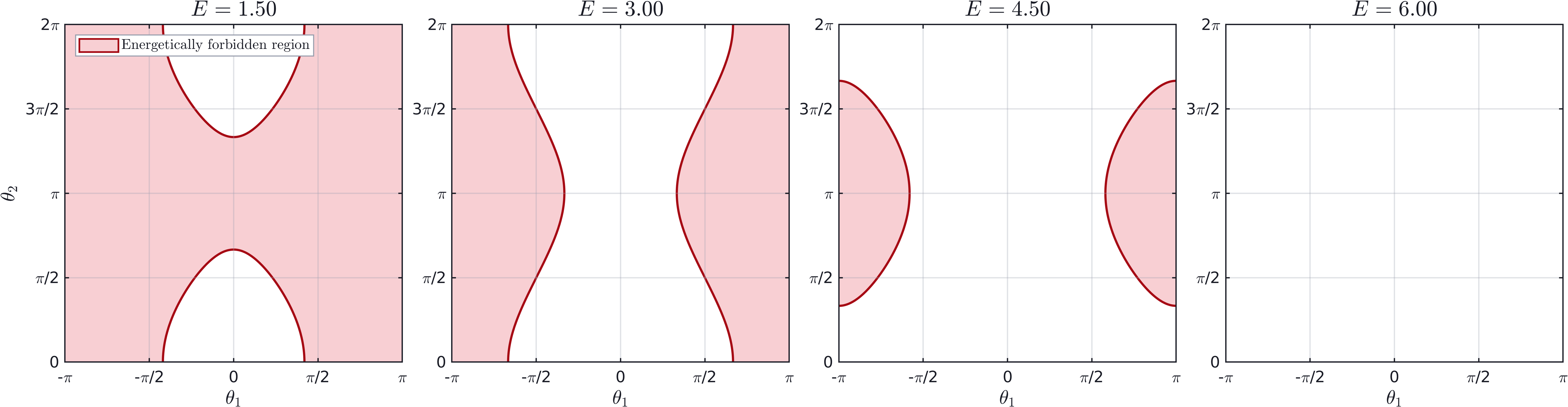}
    \caption{Sequence of possible topologies of the energetically allowed region \eqref{energ_allowed} in configuration space for the egalitarian double pendulum. The panels show representative energies in the intervals separated by the critical values of the potential. The energetically forbidden regions are shaded.}
    \label{fig:allowed_configuration_space}
\end{figure}

Figure~\ref{fig:allowed_configuration_space} shows the sequence of possible topologies of the energetically allowed regions in configuration space, \begin{equation}
     V(\theta_1,\theta_2) \leq E ~.
     \label{energ_allowed}
\end{equation}
For $0<E<2$, the motion is confined to a neighborhood of the minimum of the potential. When the first saddle energy, $E=2$, is crossed, a passage opens through the saddle $(0,\pi)$, allowing motion across the $\theta_2 = \pi$ barrier. Thus, flips of the second arm become energetically accessible. At the second saddle energy, $E=4$, an additional passage opens through $(\pi,0)$, removing the corresponding barrier in the $\theta_1$ direction. Finally, at $E=6$, the energy reaches the maximum of the potential, so that every configuration $(\theta_1,\theta_2)$ becomes energetically accessible. 

We note that $\#\{\text{minima}\} - \#\{\text{saddles}\} + \#\{\text{maxima}\} = 0$, as required by the torus topology of the configuration space. In addition, while the minimum and maximum are global, the ordering of the potential values at the two saddles depends on the system parameters. For $(m_1+m_2) L_1 > m_2 L_2$ (which includes the egalitarian case), the down--up saddle lies below the up--down saddle.

A normal-mode analysis around each equilibrium point is provided in App.~\ref{app:normal_mode}.

\begin{table}[H]
    \centering
    \begin{tabular}{cccc}
        \toprule
        \quad$(\theta_1,\theta_2)$ \quad
        & \quad Name \quad 
        & \quad Type \quad & \quad Egalitarian $V$-value \\
        \midrule
        $(0,0)$       & down--down  & Minimum   & $0$\\
        $(0,\pi)$     & down--up & Saddle    & $2$ \\
        $(\pi,0)$     & up--down & Saddle    & $4$\\
        $(\pi,\pi)$   & up--up   & Maximum   & $6$ \\
        \bottomrule
    \end{tabular}
    \caption{Critical points of the double-pendulum potential, listed with their names, stability types, and potential values in the egalitarian case.}
    \label{tab:potential_critical_points}
\end{table}

\paragraph*{Symmetries and conserved quantities.} As an autonomous Hamiltonian system, the double pendulum conserves its total energy,
\begin{equation}
    H\left( \theta^i, l_i \right) = E ~.
    \label{def_energy}
\end{equation}
In fact, it has been proven that no additional conserved quantities exist \cite{Burov_1986,Dullin_1994}. Thus, the number of conserved quantities (1) is smaller than the number of degrees of freedom (2), and the system is therefore non-integrable.

As usual, conservation of energy is associated with time-translation symmetry. In addition, the system respects the time-reversal symmetry 
\begin{equation}
    t \to -t ~.
    \label{def:time_reversal}
\end{equation}

Finally, the system is invariant under a reflection symmetry $\mathcal{P}$, which acts on configuration space as follows \begin{equation}
    \left( \theta_1,\, \theta_2 \right) \to \left( -\theta_1,\, -\theta_2 \right) ~.
    \label{def:reflection}
\end{equation} 
This is a $\mathbb{Z}_2$ symmetry, namely, $\mathcal{P}^2=1$. 

\paragraph*{Dynamics and energy regimes.} The  qualitative behavior of the double pendulum changes with the total energy, whose possible values satisfy $E \ge 0$. For generic energies the system is non-integrable, but it approaches integrable limits at both low and high energies. 

At low energies, $E \ll 1$, the motion consists of small oscillations around the stable equilibrium. The system can then be linearized, reducing the dynamics to that of two coupled harmonic oscillators, or equivalently to two normal modes. In this approximation, the dynamics is integrable; see App.~\ref{app:normal_mode} for details.

At high energies, $E \gg V \sim 1$, the potential energy becomes negligible compared with the total energy. The system then approaches the free no-gravity double pendulum, for which rotational symmetry implies conservation of total angular momentum. Thus the high-energy limit is again integrable.

\begin{figure}[H]
    \centering
    \includegraphics[width=0.5\linewidth]{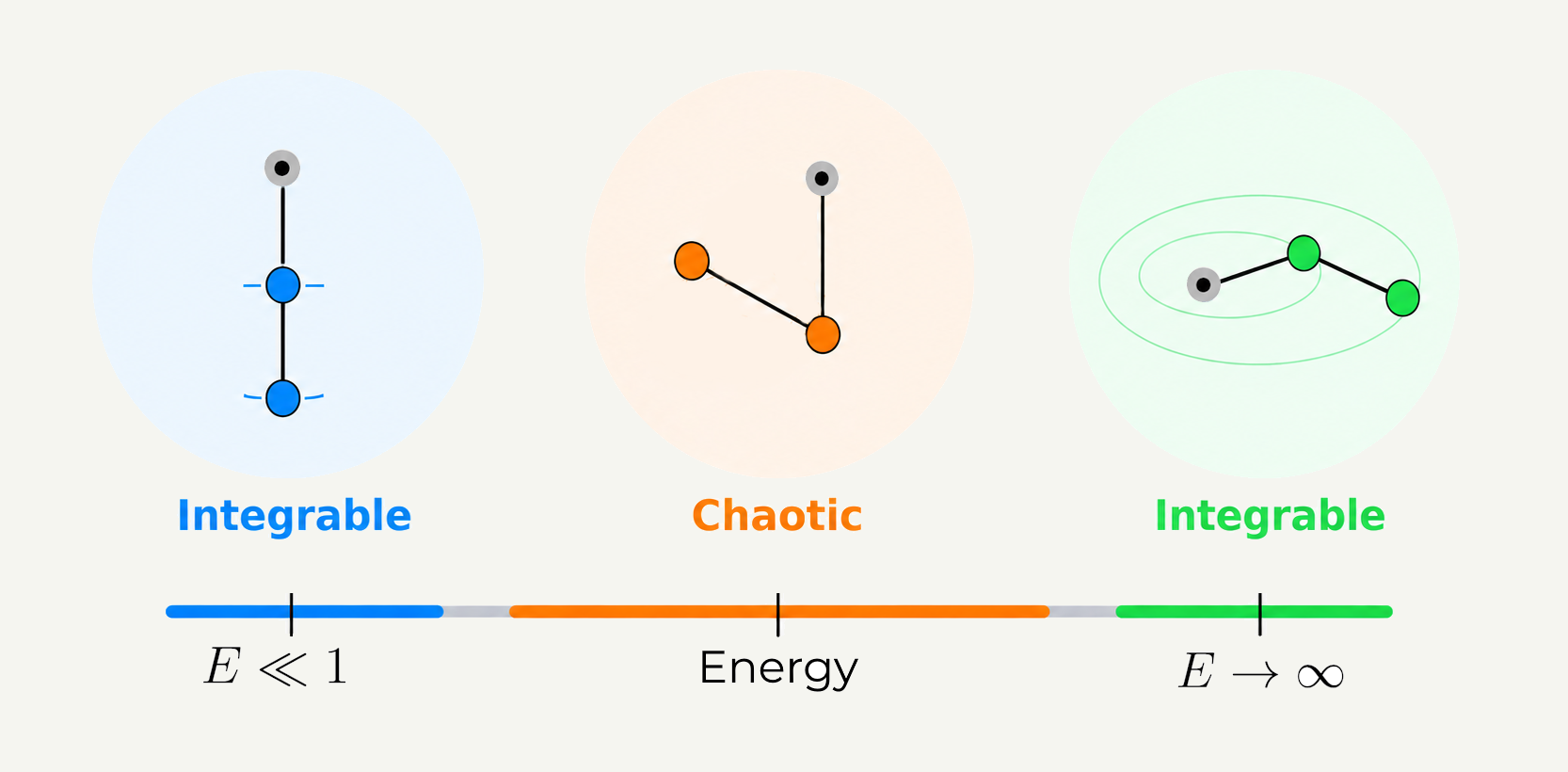}
    \caption{Changes in dynamical character as a function of the total energy $E$. At low energies, the system reduces to small oscillations, while at high energies, gravity becomes negligible and the total angular momentum is approximately conserved.}
    \label{fig:energy regimes}
\end{figure}

These changes in dynamical character along the energy axis are illustrated in Fig.~\ref{fig:energy regimes}.

In this work, we consider the energy range \begin{equation}
    2 \le E \le 6 ~.
\label{def:energy_range}
\end{equation} 
At this point, it is useful to consider Fig.~\ref{fig:allowed_configuration_space}. At $E=2$, the down-up saddle is reached; below this energy, flips are energetically inaccessible. This makes $E=2$ a natural lower bound for the study of arm-2 flips. The value $E=6$ is the maximum of the potential. Above this energy, the saddle-orbit construction used below no longer provides the same natural dividing surface. We therefore take $E=6$ as the upper bound. Nevertheless, we expect that the present approach could be extended to higher energies through a suitable generalization of the saddle orbit. 

We refer to this interval as the chaotic energy range, since it lies away from the two integrable limits. This terminology is meant only as a convenient shorthand: it should not be taken to imply that the dynamics is fully chaotic throughout the range, or that it is regular outside it. Later, in Sec.~\ref{sec:results}, we shall present an empirical subdivision of this range.

\section{Flip criteria and the saddle orbit}
\label{sec:criteria}

The intuitive definition of a flip event as an arm passing over the top refers to a finite time interval. However, both for counting flips in simulated trajectories and for predicting their statistics through phase-space flux, it is useful to define a corresponding \emph{flip moment}, at which a specified threshold is crossed. We therefore define a flip by means of a \emph{dividing surface} in phase space, \begin{equation}
    g (\theta^i,\, l_i) = 0 ~,
    \label{def:g}
\end{equation}
where $g = g (\theta^i,\, l_i)$ is a function on phase space. A trajectory is then said to flip whenever it crosses this surface. Different choices of $g$ therefore lead to different flip criteria, affecting both the statistical prediction and the identification of flip events in simulations.

In the present paper, we consider only dividing surfaces obtained from curves in configuration space. Specifically, we take 
\begin{equation}
 g_a(\theta_1,\theta_2) =\theta_2 - \theta_2^{(a)}(\theta_1)
\label{def:dividing_curve}
\end{equation} 
where the \emph{dividing curve} $\theta_2=\theta_2^{(a)}(\theta_1)$ lies in configuration space, and the label $a$ specifies the choice of curve and the corresponding flip criterion. 


We begin this section with the upright-arm (UA) criterion, the simplest and most immediate choice. We then show that this criterion exhibits the unwanted phenomenon of recrossing, which obscures the intended interpretation of a flip as a passage over the top. This motivates the saddle-orbit (SO) criterion, based on a distinguished periodic orbit associated with the saddle, which provides a natural definition of a flip. Finally, we introduce the approximate-saddle-orbit (ASO) criterion, a simplified and useful approximation to the saddle-orbit criterion.

\subsection{Upright arm: a first attempt}


Focusing on flips of arm 2, the upright-arm (UA) criterion is defined by the dividing curve \begin{equation}
   \theta_2^{\mathrm{UA}} = \pi 
    \label{crit:up}
\end{equation} 
together with \eqref{def:dividing_curve}. This criterion simply records a flip whenever arm 2 points exactly upward. It is perhaps the simplest choice, and the first one that comes to mind.

\subsection{Saddle orbit: the optimal criterion}
\label{subsec:saddle orbit}
One can imagine variants of the upright criterion \eqref{crit:up}. When arm 1 points down ($\theta_1=0$), reflection symmetry~\eqref{def:reflection} singles out the upward direction for arm 2 ($\theta_2=\pi$) as a natural threshold for a flip. For $\theta_1 \neq 0$, however, no analogous symmetry constraint applies. Indeed, as we shall see shortly, a natural flip criterion is given by a curve $\theta_2^{\mathrm{SO}}=\theta_2^{\mathrm{SO}}(\theta_1)$.

\begin{figure}[H]
    \centering
    \includegraphics[width=0.5\linewidth]{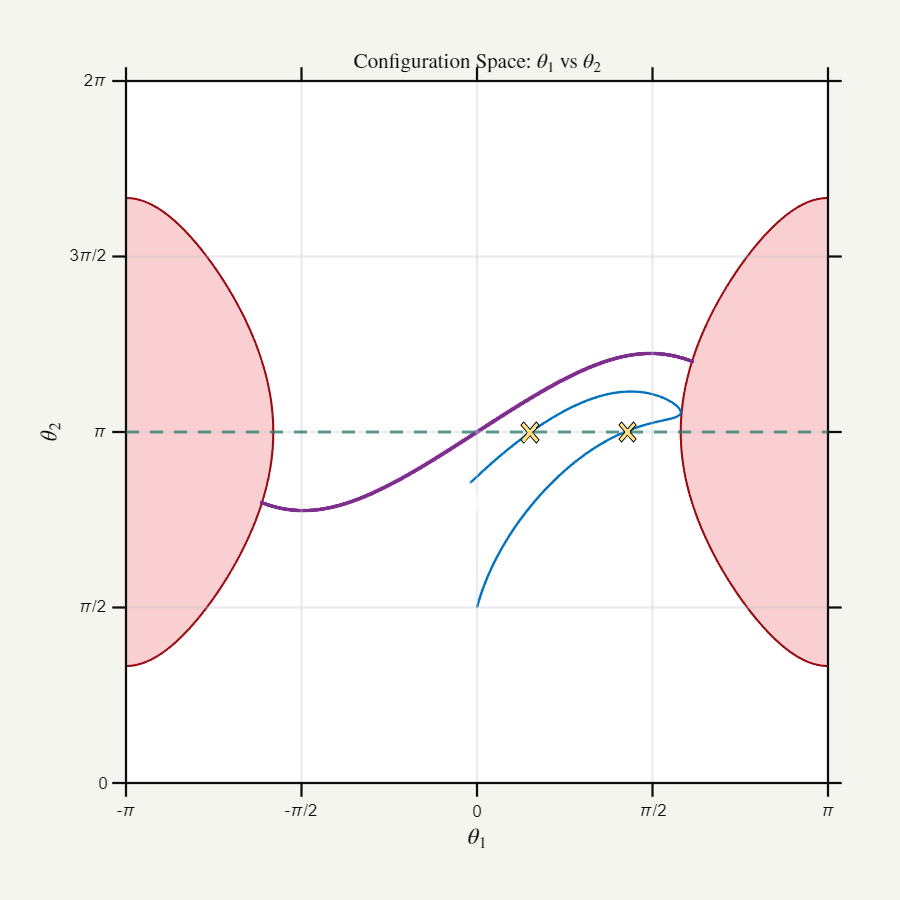}
    \caption{A sample trajectory exhibiting recrossing with respect to the upright-arm criterion. The red region is energetically forbidden in configuration space, while the blue curve shows the trajectory. The dashed green and solid purple lines mark the upright-arm and saddle-orbit criterion curves, respectively. Crossing events are marked by yellow crosses.}
    \label{fig:recrossing}
\end{figure}

\paragraph*{Recrossings and motivation.} The upright flip criterion introduced above exhibits recrossing, by which we mean two successive crossings in opposite directions separated by only a short time interval. Here, ``short'' initially means short compared with a typical period of the system. Once we introduce the gap in the distribution of crossing times, this notion will acquire a more precise meaning: a recrossing is a return occurring on a timescale shorter than the gap time. Such events were termed  ``sneaky returns'' in \cite{MacKay1994}. Figure~\ref{fig:recrossing} shows an example.

\paragraph*{Requirements for an improved criterion.} Recrossings lead to a double-counting of flips and motivate the search for an improved dividing curve. Such a criterion should satisfy the following requirements: \begin{itemize}

\item \emph{Complete separation.}
The dividing curve should bridge the energetically allowed region by connecting appropriate components of its boundary. Consequently, every flipping trajectory must cross the corresponding dividing surface in phase space. 

\item \emph{An invariant dividing curve.}
The dividing curve should itself be an orbit. A dividing curve that is not invariant under the dynamics necessarily admits arbitrarily rapid recrossings, as shown in \cite{MacKay1994}.

Heuristically, this result can be understood as follows. Let $s$ parametrize the dividing curve, and let $g$ be a local coordinate transverse to it \eqref{def:dividing_curve}. Near the dividing curve, the transverse motion of a trajectory may be written, to linear order, as
\begin{equation}
    \ddot{g}=a(s)+b(s)\, g + c(s)\,\dot{g}.
\end{equation}
If the dividing curve is not itself an orbit, then there exists a point $s=s_0$ at which $a(s_0)\neq 0$. Consider a trajectory that crosses the curve at this point at time $t=t_0$, so that $g(t_0)=0$, and choose its transverse velocity $\dot{g}(t_0)$ to be small and of sign opposite to that of $a(s_0)$. Over a sufficiently short time interval, the transverse motion is approximately uniformly accelerated:
\begin{equation}
    g(t_0+\tau) \simeq
\dot{g}(t_0)\, \tau+\frac{1}{2}a(s_0)\, \tau^2.
\end{equation}
In addition to the crossing at $\tau=0$, the trajectory crosses the curve again at
$\tau_{\mathrm{rec}}
\simeq 
-2\, \dot{g}(t_0)/a(s_0).$
The recrossing time can therefore be made arbitrarily short by choosing $\dot{g}(t_0)$ sufficiently small. This argument shows that a dividing curve that is not an orbit necessarily admits arbitrarily rapid recrossings, and hence motivates choosing an orbit as the dividing curve.

\item \emph{Minimal flux.} 
Among dividing surfaces satisfying the preceding conditions, one should prefer a surface with minimal directional phase-flux, also known as a bottleneck. Since recrossings introduce additional counted crossings, they tend to increase the measured flux through a dividing surface.

\end{itemize}

These considerations lead naturally to a dividing curve associated with a periodic orbit near a saddle of the potential. We refer to this orbit as the \emph{saddle orbit}.

\paragraph*{Saddle orbit definition.} The saddle points of the potential were summarized in Table~\ref{tab:potential_critical_points}. We first consider the down--up saddle. Linearization about this equilibrium  reveals two modes: an oscillatory mode in the ``uphill'' direction, and a runaway ``downhill'' mode.
The Lyapunov-center construction for a saddle--center equilibrium yields a family of periodic orbits emanating from the equilibrium as the energy is increased above the saddle energy $E_s=2$; see \cite[Sec.~42]{Lyapunov1992} and \cite{Moser1968LecturesHamiltonian}. These periodic orbits depend continuously on $E$, at least within a finite interval over which the family can be continued regularly. Projecting the saddle orbits onto configuration space gives a family of curves \begin{equation}
    \theta_2^{\mathrm{SO}}=\theta_2^{\mathrm{SO}}(\theta_1 ; E) ~.
\label{crit:saddle}
\end{equation}

We refer to the members of this family as \emph{saddle Lyapunov orbits}, or, more briefly, as \emph{saddle orbits}. As $E$ approaches $E_s$ from above, the periodic orbit contracts to the saddle equilibrium. An analogous definition applies to the up--down saddle and the periodic-orbit family emanating from it.

\begin{figure}
    \centering
    \includegraphics[width=0.35\linewidth]{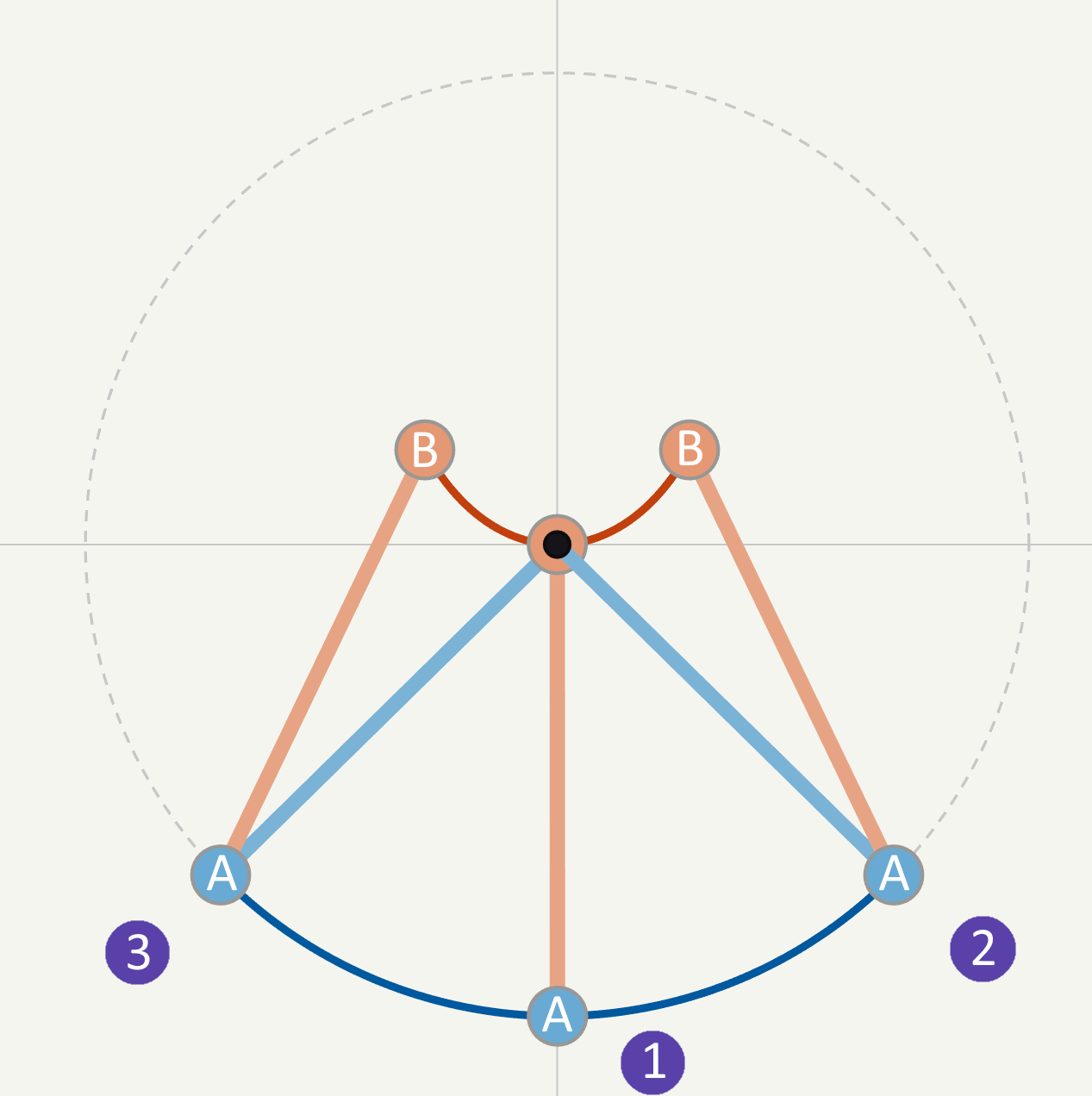}
    \caption{Key frames for the saddle orbit at $E=2.5$ corresponding to the down-up saddle. The frames are numbered from 1 to 3, and the pendulum masses are labeled A and B, corresponding to the first and second masses, respectively. Detailed visualization and videos of the saddle orbit are available in \cite{interactive_viewer}.}
    \label{fig:saddle orbit keyframes E=2.5}
\end{figure}

We illustrate the saddle orbit from two complementary perspectives. Figure~\ref{fig:saddle orbit keyframes E=2.5} shows representative key frames of the physical pendulum motion at a certain energy, while Fig.~\ref{fig:visualization_configuration_space_all_criteria} shows the configuration-space curve for a different value of \(E\). The numerical method that was used to determine these orbits is discussed below.

The role of periodic orbits as the dynamical basis for dividing surfaces in systems with two degrees of freedom was developed by Pechukas, Pollak, and collaborators in the context of transition state theory (TST) in chemistry \cite{Pechukas_McLafferty_1973, Pechukas_1976, Pollak_Pechukas_1978, Pollak_Child_Pechukas_1980}: the 1973 work is a precursor establishing the no-recrossing criterion, the 1976 chapter gives the first published formulation of the periodic-orbit construction, the 1978 paper develops the theory further, and the 1980 paper introduces the term \emph{periodic orbit dividing surface} (PODS).

\paragraph*{Saddle orbit properties.} 
The saddle orbits considered here are time-reversal-symmetric, out-and-back orbits. Their configuration-space projections run between two turning points and are retraced in the opposite direction during the second half of the period. This distinguishes them from rotational periodic orbits.

At each turning point, the momenta vanish; consequently, the turning points lie on the boundary of the energetically allowed region. The configuration-space projection of the saddle orbit consequently connects two points on this boundary and may serve as a dividing curve. It thus satisfies the complete-separation requirement. Because the dividing curve is generated by an orbit, it also satisfies the invariance requirement.

At fixed energy, a mechanical trajectory is, up to reparametrization, a geodesic of the Jacobi--Maupertuis orbit metric and is a stationary point of the abbreviated action. Recall that for a natural Lagrangian ${\cal L}=\half m_{ij}(q) \, \dot{q}^i \,\dot{q}^j - V(q)$, the corresponding Jacobi--Maupertuis orbit metric at energy $E$ is $ds_{JM}^2=\left(E-V(q)\right) m_{ij}(q) \, dq^i \, dq^j$. For a two-degree-of-freedom Hamiltonian system, the directional symplectic flux through a dividing surface bounded by a periodic orbit is equal to this action, see Sec.~\ref{sec:prediction_phase-flux}.

The saddle orbit therefore makes the flux stationary with respect to suitable variations of its boundary. 

\begin{figure}[H]
    \centering
    \includegraphics[width=0.5\linewidth]{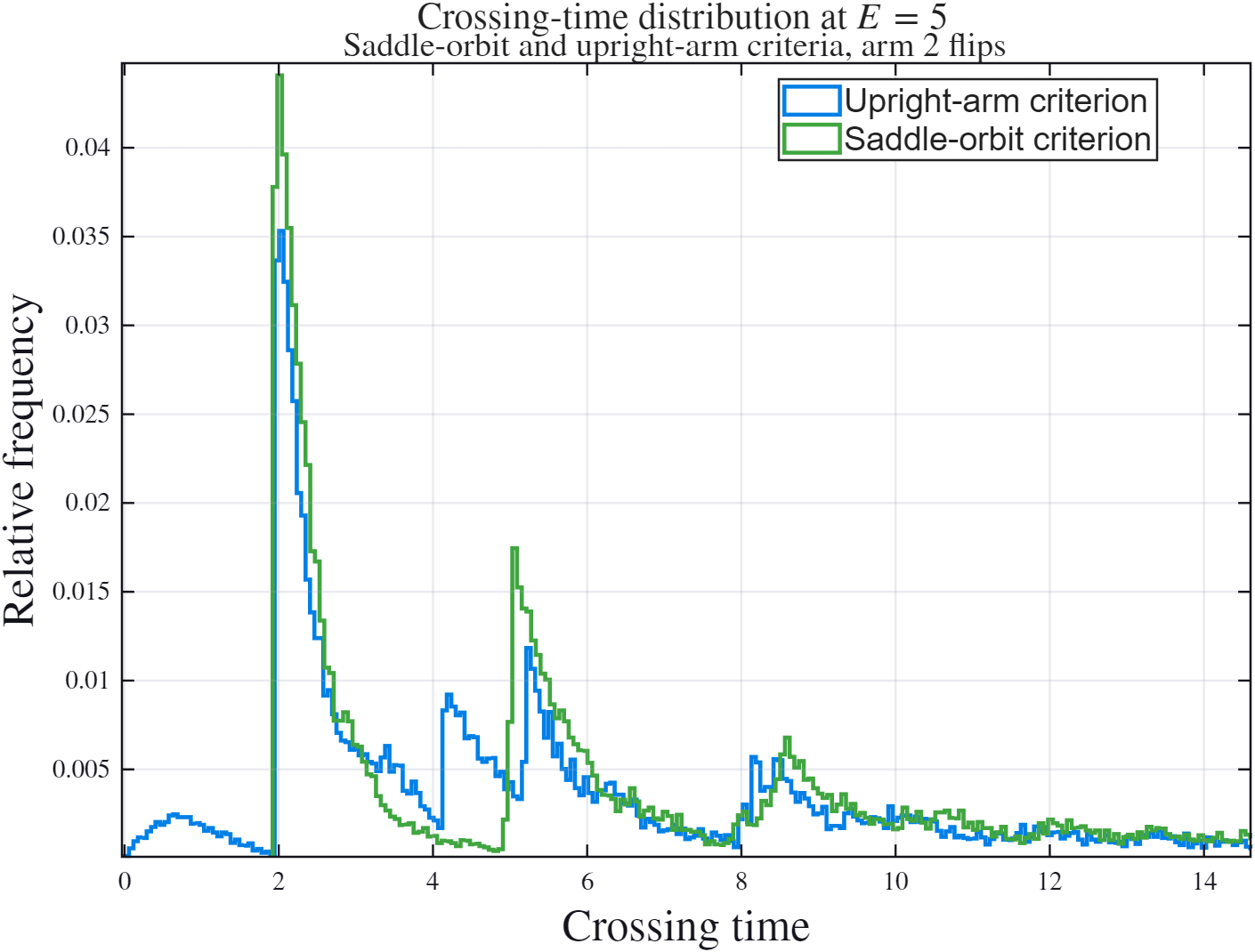}
    \caption{Distribution of crossing times between successive flips for the upright-arm and saddle-orbit criteria at $E = 5$. The distributions are constructed from 30,000 detected flips. An early-time gap is visible for the saddle-orbit criterion, whereas no comparable gap is observed for the upright-arm criterion.}
    \label{fig:crossing_times}
\end{figure}

In the numerical results presented below, the saddle-orbit criterion eliminates the rapid returns observed for the upright criterion: the distribution of crossing times develops a nonzero gap; see Fig.~\ref{fig:crossing_times}. Thus, within the operational definition introduced above, no recrossings occur on timescales shorter than the gap time. These results make the saddle orbit a natural candidate for an optimal dividing curve in configuration space, and the associated surface a natural dividing surface in phase space.

We add a few comments regarding the saddle orbits. Heuristically, a saddle orbit may be viewed as approximately following the potential ridge. This analogy is not exact, however, because the saddle orbit depends on the total energy $E$, whereas the potential ridge does not.

The saddle orbits satisfy a simple geometric bound. By reflection symmetry, it is sufficient to consider $\theta_1>0$.  In this case, the configuration-space curve lies in the range $\pi < \theta_2(\theta_1) <\pi +\theta_1$, where the upper bound, $\theta_2=\pi + \theta_1$, corresponds to folded configurations. 

The saddle orbit also bears similarities to the notions of a \emph{nonlinear normal mode} \cite{Rosenberg1962,Rosenberg1966} and a \emph{brake orbit} \cite{Seifert1948}.
 
\paragraph*{Saddle-orbit observables.} Several quantitative observables are associated with the saddle orbit $\gamma$: 

\begin{itemize}
    \item The period is defined by \begin{equation}
        T(E) := \oint_{\gamma} dt ~.
        \label{def:saddle_period}
    \end{equation}
    \item The directional flux. The abbreviated action of the saddle orbit is defined by \begin{equation}
        \varphi(E) := \oint_{\gamma} p\, dq ~,
        \label{def:saddle_action}
    \end{equation} and will later be identified with the directional flux.
    
The directional flux associated with a periodic orbit and its period are related by the identity
\begin{equation}
    \frac{d\varphi}{dE}=T.
    \label{eq:period_time_flux_relation}
\end{equation}
\textbf{Proof of the identity.}
Consider an autonomous Hamiltonian system with $n$ degrees of freedom.
Hamilton's principal function is defined as
$
    S_{\mathrm{eff}}(\vec q_i,\vec q_f,T)
    =
    \int_0^T
    \left(
        \vec p \cdot \dot{\vec q} - H
    \right) dt,
$
evaluated along a classical trajectory connecting $\vec q_i$ and
$\vec q_f$ in time $T$.
As follows from Hamilton's variational principle
\cite[Sec.~43]{LandauLifshitzMechanics},
its differential is
$
    dS_{\mathrm{eff}}
    =
    \vec p_f \cdot d\vec q_f
    -
    \vec p_i \cdot d\vec q_i
    -
    E\,dT.
$
Now, performing a Legendre transformation with respect to $T$,
we define
$
    \tilde S_{\mathrm{eff}} = S_{\mathrm{eff}} + E\,T.
$
Thus, we obtain
\begin{gather*}
    d\tilde S_{\mathrm{eff}}
    =
    \vec p_f \cdot d\vec q_f
    -
    \vec p_i \cdot d\vec q_i
    +
    T\,dE.
\end{gather*}
The transformed function corresponds to the abbreviated action \cite[Sec.~44]{LandauLifshitzMechanics}.

For a smooth family of periodic orbits $\gamma_E$, each orbit
satisfies
$
    \vec q_f = \vec q_i,
    \, \,
    \vec p_f = \vec p_i.
$
Consequently, the endpoint contributions cancel, yielding
\begin{gather*}
    d\tilde S_{\mathrm{eff}} = T\,dE.
\end{gather*}
Moreover, since $H=E$ along each periodic orbit,
\begin{gather*}
    \tilde S_{\mathrm{eff}}
    =
    S_{\mathrm{eff}} + E\,T
    =
    \int_0^T \vec p \cdot \dot{\vec q}\,dt
    =
    \oint_{\gamma_E} \vec p \cdot d\vec q
    =
    \varphi(E).
\end{gather*}
Therefore,
\begin{gather*}
    \boxed{
        \frac{d\varphi(E)}{dE}=T(E).
    }
\end{gather*}

For a more direct computational proof, see App.~\ref{app:action period relation}. 

    \item Stability multiplier $\mu$. This observable will not be addressed in this paper, and is mentioned for completeness. Given a periodic orbit, such as the saddle orbit, Floquet analysis defines a symplectic monodromy matrix for phase space perturbations. Due to conservation of energy, two of the eigenvalues of this matrix (Floquet multipliers) are equal to unity, while the remaining two form a reciprocal pair,  closed under complex conjugation. For a hyperbolic periodic orbit, these multipliers are real, and we may choose $|\mu|>1$. For an elliptic periodic orbit, the nontrivial multipliers lie on the unit circle and may be written as $\mu=e^{\pm i\, \nu}$.
\end{itemize}

\paragraph*{Numerical determination of the saddle orbit.} 
The saddle orbits must generally be determined numerically. Two approaches are available. The first is a shooting method: one guesses suitable initial conditions, integrates the equations of motion, and adjusts the initial conditions until the periodicity conditions are satisfied. We implement this approach, as described in Sec.~\ref{sec:Implement_saddle_orbit}.

A second approach is variational: one searches for stationary curves of the abbreviated action, or equivalently for geodesics of the Jacobi--Maupertuis metric, subject to the appropriate boundary and symmetry conditions.

\begin{figure}[H]
    \centering
    \includegraphics[width=1\linewidth]{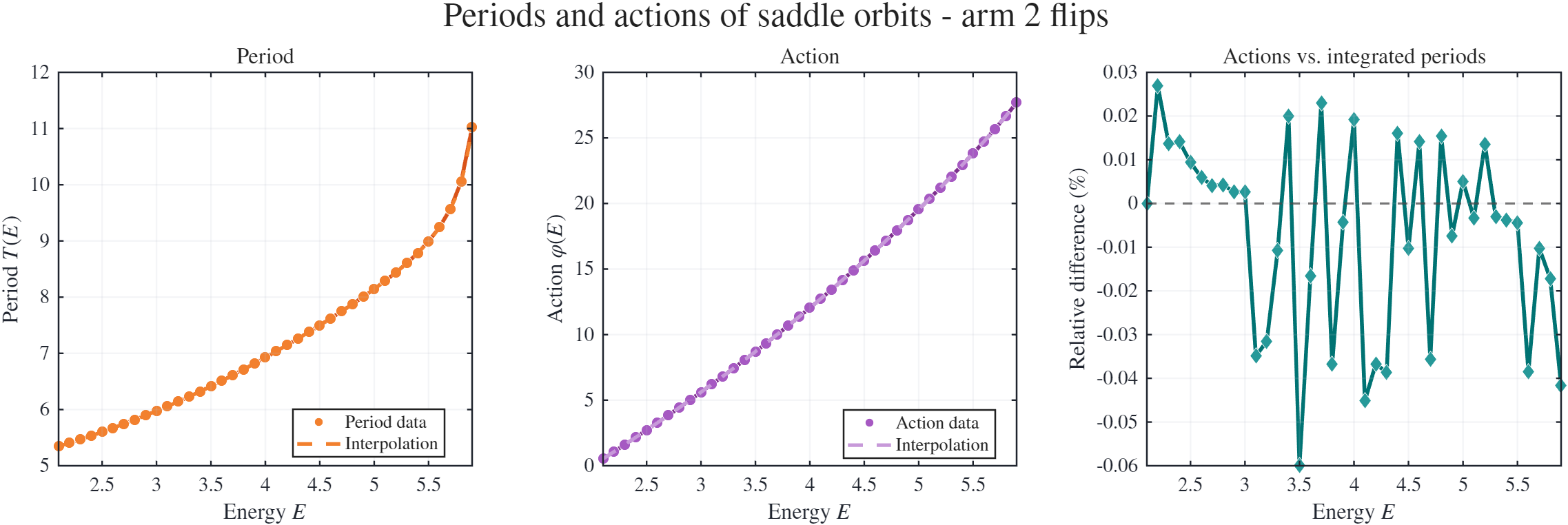}
     \caption{Period (left) and abbreviated action 
     (center) of the arm 2 saddle orbits as functions of energy. The action reconstructed by integrating the measured period with respect to energy is compared with that obtained directly from the saddle-orbit action. The relative residuals between the two are shown on the right and are found to be very small.}
    \label{fig:Time and flux saddle orbit}
\end{figure}

Here, we present properties of the saddle orbits for arm 2 over the chaotic energy range $2 \le E \le 6$.  For each saddle orbit, we evaluate its period and action. The results are shown in Fig.~\ref{fig:Time and flux saddle orbit}. In the next section, we shall see that the bidirectional flux $f_{\mathrm{SO}}(E)$ is twice the action and thus provides an ingredient required for the statistical prediction.

The period and action data provide an internal consistency check through relation \eqref{eq:period_time_flux_relation}. As shown in the figure, this relation is satisfied to very high accuracy.

The corresponding results for the arm 1 saddle orbits will be presented in Sec.~\ref{subsec:results_arm1}. Online videos of both families of saddle orbits, presented as double-pendulum motion and in configuration space, are available in \cite{interactive_viewer}.

We can now assess the occurrence of rapid recrossings (sneaky returns). We define the crossing time (also called the residence time) as the time between successive flips. Figure~\ref{fig:crossing_times} shows the distribution of crossing times for a collection of successive flips, using both the upright-arm (UA) criterion and the saddle-orbit (SO) criterion. As expected, the UA criterion admits arbitrarily short crossing times, corresponding to rapid recrossings. By contrast, the SO criterion eliminates such events and exhibits a finite gap at early times. This behavior is expected, since the SO criterion is based on an invariant dividing surface.

\subsection{Approximate saddle orbit} 

We define the approximate saddle orbit (ASO) criterion by the curve \begin{equation}
    \theta_2^{\mathrm{ASO}}  = \pi + A^{\mathrm{ASO}} \sin \theta_1 ~, \qquad  A^{\mathrm{ASO}}=2-\sqrt{2}
    \label{crit:approx}
\end{equation}

This approximation is constructed to satisfy three simple requirements: (i) $\theta_2$ is periodic in $\theta_1$, as required by the topology of configuration space; (ii) the curve passes through the saddle point; and (iii) the curve is tangent to the oscillatory normal mode as $E \to 2^+$. Indeed, the corresponding normal mode eigenvector of the egalitarian double pendulum at the down--up saddle is given by $(\delta\theta_1, \delta\theta_2)=(1,2-\sqrt{2})$, as derived in \eqref{AASO}, which explains the value of $A^{\mathrm{ASO}}$. 

\paragraph*{Fitted sinusoidal approximation.} 
For comparison, we also record here a more elaborate sinusoidal approximation to the saddle orbit, although it will not be used in the present paper.
\begin{equation}
   \theta_2^{\mathrm{FSA}}(\theta_1; E) = \pi + A(E)\, \sin \! \left( k(E) \,\theta_1 \right) ~.
\label{eq:general_fit_saddle_orbit_approximation}   
\end{equation}
This form involves two numerically fitted functions: an amplitude $A(E)$ and a wavenumber  $k(E)$. 

\begin{figure}[H]
    \centering
\includegraphics[width=.85\linewidth]{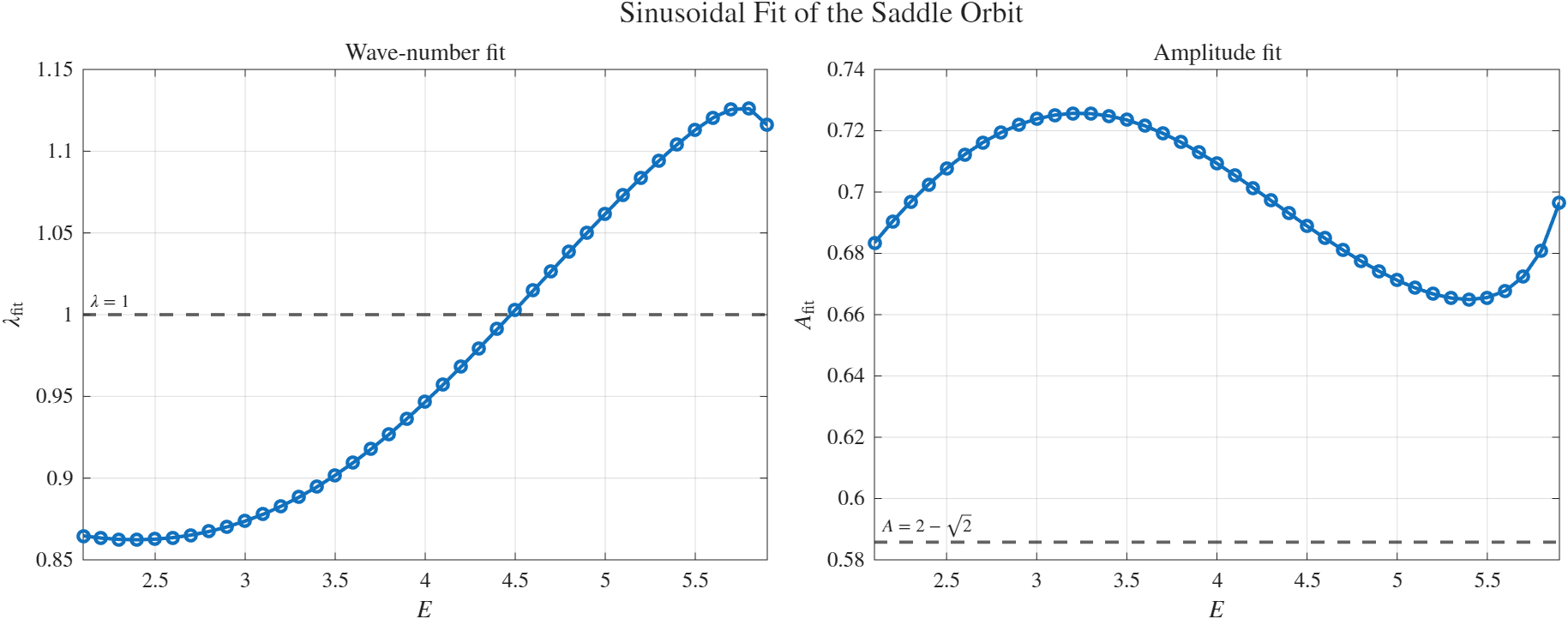}
\caption{Best-fit wavenumber $k(E)$ and amplitude $A(E)$, obtained by fitting the numerical saddle orbit to \eqref{eq:general_fit_saddle_orbit_approximation}.
The reference values $k=1$ and $A=2-\sqrt{2}$ correspond to the saddle-orbit approximation.}
\label{fig:saddle_orbit_approximation_fit}
\end{figure}

The saddle-orbit approximation \eqref{crit:approx} corresponds to the special choice $A(E)=2-\sqrt{2}, \quad k(E) =1$.  Thus, the energy dependence of the fitted parameters provides a measure of how the geometry of the numerical saddle orbit differs from the ASO curve. The results of the numerical fit for $A(E)$ and $k(E)$ are shown in Fig.~\ref{fig:saddle_orbit_approximation_fit}.

Near the saddle energy $E=2$, the numerical saddle orbit approaches the normal mode. In particular, for small $\theta_1$, we expect $\lim_{E \to 2^+} A(E) \, k(E) = A^{\mathrm{ASO}}$. As can be inferred from Fig.~\ref{fig:saddle_orbit_approximation_fit}, the numerical fit is consistent with this limiting behavior.

\subsection{Brief summary}

\begin{figure}[H]
    \centering
    \includegraphics[width=0.5\linewidth]
    {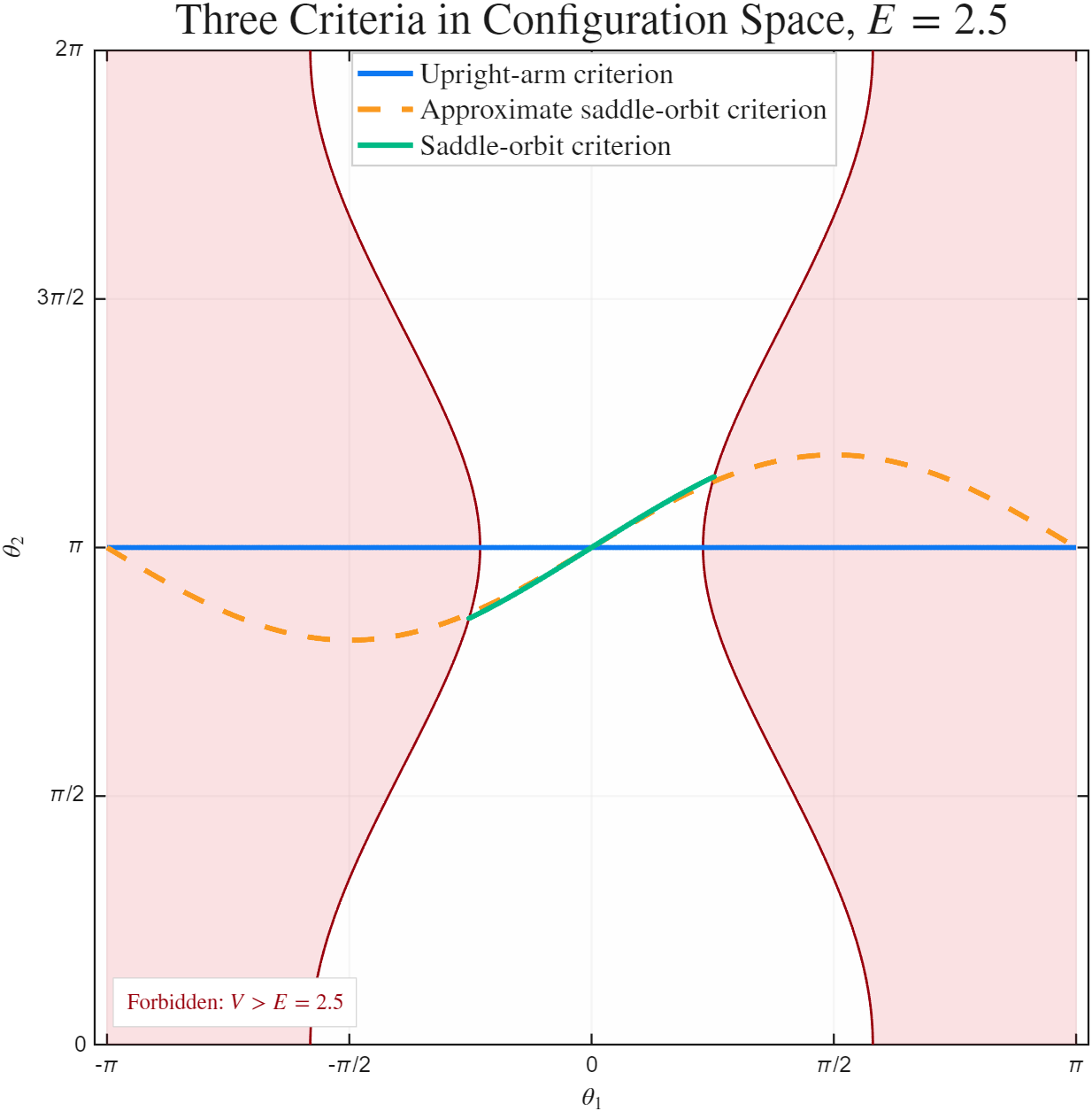}
    \caption{The three dividing curves at $E=2.5$. Each defines a corresponding flip criterion.}
    \label{fig:visualization_configuration_space_all_criteria}
\end{figure}

In this section, we have defined three flip criteria, each specified by a dividing curve through the construction \eqref{def:dividing_curve}:
 \begin{itemize}
    \item the upright-arm (UA) criterion \eqref{crit:up},
    \item the saddle-orbit (SO) criterion \eqref{crit:saddle},
    \item the approximate-saddle-orbit (ASO) criterion \eqref{crit:approx}. 
\end{itemize} 
The three configuration-space curves are illustrated in
Fig.~\ref{fig:visualization_configuration_space_all_criteria} for a representative energy.

\section{Statistical prediction and saddle flux}
\label{sec:prediction}
In this section, we introduce the flux-based method for statistical prediction and apply it to the double pendulum. The central idea is to replace the long-time statistics of a chaotic trajectory by a microcanonical ensemble and to express the rate of a specified event as the probability flux through an associated dividing surface in phase space. 

Let us unpack this statement. Consider a chaotic system with prescribed initial conditions. Under an ergodic or mixing assumption, the system progressively loses memory of the details of its initial conditions, apart from the conserved quantities. This motivates replacing a single long chaotic trajectory by an ensemble of systems sharing the same values of the conserved quantities. Next, consider an event of interest represented by a dividing surface in phase space. The statistical prediction for the event rate is then given by the flux of phase-space probability through this dividing surface.

This idea has a long history, but, to our knowledge, it is applied here for the first time to the double pendulum; a closely related flux-based approach was previously applied to the egalitarian three-body system in \cite{flux-based}. In chemistry, phase-space flux has been used since the 1930s in transition state theory (TST), a statistical theory of reaction rates \cite{Eyring_1935_Activated_Complex, Evans_Polanyi_1935_Transition_State, Wigner_1938_Transition_State_Method}; see also the reviews \cite{TST_rev_1996, Waalkens_Schubert_Wiggins_2008} and the related RRKM theory of unimolecular decomposition \cite{Rice_Ramsperger_1927, Kassel_1928, Marcus_1952}. Early applications to non-chemical mechanical systems include \cite{Jaffe_Farrelly_Uzer_1999} in atomic physics and \cite{Fukushige_Heggie_2000, Asteroid_2002} in celestial mechanics and astrophysics. In 2016, one of us developed the idea independently in the context of the double pendulum, as recounted in App.~\ref{app:story}.
 
The ensemble may be visualized as a gas filling a container that represents phase-space  (it is also interesting to project this picture into configuration space). In this analogy, the dividing surface acts as a membrane, and the flux of gas through it corresponds to the phase-volume flux through the membrane. 

Let us first formulate this idea for a general mechanical system, and then specialize to the double pendulum. Consider a general mechanical system with $n$ degrees of freedom, described by generalized coordinates and momenta $q^i,\, p_i,\, i=1,\dots,n$ and by the Hamiltonian \begin{equation}
    H=H(q,p) ~.
\end{equation} 
Suppose that this system has $k$ conserved quantities $I_a=I_a(q,p),\, a=1,\dots,k$, whose prescribed values are denoted by $J_a$ and collectively by $J=(J_1, J_2, \dots,J_k)$. The corresponding microcanonical phase-space volume element $dV_{J}$ is \begin{equation}
        dV_{J} := d^n q\, d^n p\, \Pi_{a=1}^k \delta\left( I_a(q,p)-J_a\right),
\end{equation}
and the total microcanonical volume is \begin{equation}
     \Omega(J) := \int dV_{J} ~.
 \end{equation}
The normalized probability measure is therefore   \begin{equation}
    dP_{J} := \frac{dV_{J}}{\Omega(J)}~.
\end{equation}

Consider now a dividing surface \begin{equation}
    g(p,q)=0 ~.
\end{equation}
In this paper, it suffices to consider surfaces defined entirely in configuration space, namely $g=g(q)$; however, for now we retain the more general formulation. This definition generalizes \eqref{def:g} to a general Hamiltonian system.

The statistical prediction for the rate of crossing a dividing surface labeled by $a$, $g \equiv g_a$, is given by the probability flux through it \begin{equation}
       R_a(J)= \int dP_{J}\, \delta(g)\, \left| \dot{g}  \right| = \frac{f_a(J)}{\Omega(J)} ~,
   \end{equation}
where the unnormalized phase-volume flux is  \begin{equation}
    f_a(J) := \int dV_{J} \delta(g)\, \left| \dot{g}  \right|
    \label{def:flux_general}
\end{equation}

We note that $\left| \dot{g}  \right|$ measures the bidirectional flux. In order to define the directional flux it should be replaced by $\dot{g}_+$ where the $+$ stands for the ramp function, which is defined below in \eqref{def:ramp}. For all dividing surfaces considered in the present paper, the flux is the same in both directions.

Specializing to the double pendulum, the generalized coordinates and momenta are $\theta^i,\, l_i, \, i=1,2$, the Hamiltonian is given by \eqref{eq:egalitarian double pendulum hamiltonian}, and energy is the only conserved quantity.  
 The statistically predicted rate of crossing the dividing surface $g=0$ is therefore \begin{equation}
    R_a(E) = \frac{f_a(E)}{\Omega(E)}
    \label{eq:flip_rate_formula}
\end{equation}
where the total phase-volume and flux are \begin{align}
    \Omega(E) &:= \int d^2\theta\, d^2l \, \delta\left( H(\theta,l)-E\right) \label{eq:phase_space_volume_double_pendulum} \\
    f_a(E) &:= \int d^2\theta\, d^2l \, \delta\left( H(\theta,l)-E\right)\, \delta(g)\, \left| \dot{g} \right| 
    \label{def:flux_pendulum}
\end{align}  
The different flip criteria discussed in Sec.~\ref{sec:criteria} translate into different choices of the defining functions $g(q)$.

\subsection{Phase-volume}
\label{sec:prediction_phase-volume}

Using \eqref{eq:phase_space_volume_double_pendulum}, and introducing the
available kinetic energy as a function of the configuration space \begin{equation}
    K(\theta_1, \theta_2 ; E) = E-V(\theta_1, \theta_2 ) = E - 3 +2 \cos \theta_1 +\cos \theta_2 ~,
    \label{def:available_kinetic_E}
\end{equation}
the phase-space volume can be written as
\begin{gather*}
    \Omega(E)
    =
    \int d\theta_1\,d\theta_2
    \int dl_1\,dl_2\,
    \delta\!\left(
    \frac{1}{2}l^T M^{-1}l-K(\theta_1, \theta_2)
    \right) ~.
\end{gather*}

The integration over the momenta can be performed analytically.
Define a matrix $L$ to be the square root of $M^{-1}$, namely $M^{-1}=L^T\, L$ and hence $\left| \det L \right| =(\det M)^{-1/2}$. Change integration variables into circularized momenta through 
\begin{gather}
y := L\, l 
\label{def:ciruclarized_momentum}
\end{gather}  
where $y$ lies on a circle with radius-squared $2\, K(\theta_i, E)$. Now, the integral over the momenta $l_i$ becomes \begin{equation}
    \rho(\theta_1,\theta_2;E) := \int d^2l~ \delta\!\left( \frac{1}{2} l^T\, M^{-1} \, l-K \right) =   \sqrt{\det M} \int d^2y ~ \delta\!\left( \frac{1}{2}|y|^2-K \right) = 2\pi \sqrt{\det M(\theta_1,\theta_2)} \; \Theta(K(\theta_1,\theta_2)).
    \label{eq:configuration_space_density}
\end{equation}
Here, $\Theta$ denotes the Heaviside step function. In the last equality we have used $\int d^2y ~ \delta\!\left( \frac{1}{2}|y|^2-K \right) = 2\pi\,\Theta(K)$.
For the egalitarian double pendulum, \eqref{detM} becomes $\det M=1+ \sin^2 \Delta \theta$ and hence
\begin{gather}
    \boxed{
    \Omega(E)
    =
    2\pi
    \int d\theta_1\,d\theta_2\,
    \sqrt{2-\cos^2(\theta_1-\theta_2)}
    \,
    \Theta\!\left(K(\theta_1,\theta_2)\right)
    }
    \label{eq:phase_space_volume_egalitatian_double_pendu}
\end{gather}
where $K$ was defined in \eqref{def:available_kinetic_E}.

For general energies, the configuration-space integral can be evaluated
numerically, as we do later. We comment that it can also be reduced analytically to a one-dimensional integral involving the incomplete elliptic integral of the second kind.
Using the convention
\begin{gather}
    \mathcal E(\phi\mid m)
    :=
    \int_0^\phi
    \sqrt{1-m\sin^2u}\,du,
    \label{def:incomplete_elliptic_integral_second_kind}
\end{gather}
one obtains
\begin{gather}
    \Omega(E)
    =
    2\pi
    \int_{-\pi}^{\pi}
    \left[
    \mathcal E\!\left(
    \theta_1+ \theta_{2,\mathrm{max}} \mid-1
    \right)
    +
    \mathcal E\!\left(
    \theta_{2,\mathrm{max}} -\theta_1 \mid-1
    \right)
    \right]
    d\theta_1,
    \label{eq:phase_space_volume_incomplete_elliptic}
\end{gather}
where
\begin{gather*}
    \theta_{2,\mathrm{max}}(E,\theta_1)
    :=
    \begin{cases}
        0,
        &
        3-E-2\cos\theta_1\geq1,
        \\[4pt]
        \cos^{-1}\!\left(
        3-E-2\cos\theta_1
        \right),
        &
        -1<3-E-2\cos\theta_1<1,
        \\[4pt]
        \pi,
        &
        3-E-2\cos\theta_1\leq-1.
    \end{cases}
\end{gather*}
where the derivation involves a change in the integration variable from $\theta_2$ to the relative angle $\Delta \theta$ defined in \eqref{def:Delta_theta}.

For $E\geq6$, the entire configuration space is energetically accessible,
so that the Heaviside function is equal to unity almost everywhere.
The integral then becomes independent of energy. Using the $\Delta \theta$ variable previously defined gives
\begin{gather*}
    \Omega(E \ge 6)
    =
    4\pi^2
    \int_0^{2\pi}
    \sqrt{2-\cos^2\Delta \theta}\,d\Delta \theta
    =
    4\pi^2
    \int_0^{2\pi}
    \sqrt{1+\sin^2\Delta \theta}\,d\Delta \theta
    =
    16\pi^2\mathcal E(-1)
    ~,
\end{gather*}
 where in the last equality we have used $\mathcal E( \pi/2 \mid -1)=\mathcal E(-1)$, and the expression \eqref{eq:phase_space_volume_incomplete_elliptic} is seen to tend to the current one as $E \to 6$.

\subsection{Phase-flux}
\label{sec:prediction_phase-flux}

In this subsection, we determine the phase-flux (\ref{def:flux_general},\ref{def:flux_pendulum}) for the various dividing curves introduced in Sec.~\ref{sec:criteria}. We begin by showing that quite generally the surface integral reduces to a line integral with a geometric interpretation. 

\paragraph*{Flux reduction.}
The phase-flux integral through a dividing surface $S_E$ \eqref{def:flux_general} naturally divides into two equal parts according to the sign of $\dot{g}$. Let us decompose the dividing surface as follows
\begin{gather*}
    S_E=S_+\cup S_-
\end{gather*}
where the two branches $S_+$ and $S_-$ correspond to crossings where $\dot{g}$ is positive or negative, respectively. 

The positive directional phase-flux is defined as the flux in the direction of growing $g$. It is given by
\begin{gather*}
    f^+
    :=
    \int dq\,dp\,
    \delta(H-E)\,
    \delta(g)\,
    \dot{g}_+ \equiv 
    \int_{S_+} dq\,dp\, \delta(H-E)\, \delta(g) \dot{g} ~.
\end{gather*}
Where the ramp function is defined by 
\begin{gather}
    x_+
    :=
    \begin{cases}
        x, & x\geq0,
        \\
        0, & x<0,
    \end{cases}
    \equiv x\,\Theta(x) ~.
    \label{def:ramp}
\end{gather} 

Similarly, the negative directional flux is defined by $f^- := \int dq\,dp\, \delta(H-E)\, \delta(g)\, (-\dot{g})_+$. For the three dividing surfaces considered here, the two directional fluxes are equal by either time-reversal symmetry or reflection symmetry. Hence, $f^-=f^+$, so that \begin{equation}
    f=f^+ + f^- = 2\, f^+
    \label{non-dir_n_dir}
\end{equation}

For a two-degree-of-freedom Hamiltonian system, the positive directional phase-flux is
equivalently the integral of the symplectic two-form over $S_+$  
\begin{gather*}
    f^+
    =
    \int_{S_+} \omega,
\end{gather*}
where
\begin{gather*}
    \omega
    =
    dp_1\wedge dq_1
    +
    dp_2\wedge dq_2
\end{gather*}
is the fundamental symplectic two-form. A derivation of this equivalence is given in App.~\ref{app:flux expression equivalence}. In other words, the (positive) directional phase-flux is the same as the symplectic flux.  

MacKay's flux-over-a-saddle theorem provides a geometrical expression for the symplectic
flux through such a dividing surface, essentially through Stokes' theorem. For any surface $S\subset\Sigma_E$ in a two-degree-of-freedom Hamiltonian system,
\begin{gather*}
    \int_{S_+} \omega
    =
    \int_{\partial S_+} p\cdot dq 
\end{gather*}

We denote \begin{equation}
    \gamma := \partial S_+ ~.
\end{equation}
It is the boundary of $S_+$, that is, a phase space curve over which both $g=0$ and $\dot{g}=0$. It is a sort of equator which divides $S_E$ into two equal parts. We denote the (abbreviated) action of $\gamma$ by \begin{equation}
    \varphi(\gamma) :=\int_{\gamma} p\cdot dq ~.
    \label{def:varphi}
\end{equation} 
We consider three families of curves $\gamma_i(E), ~i=\mathrm{UA},\mathrm{ASO},\mathrm{SO}$. In general, the curve $\gamma$ is not necessarily an orbit (a solution of the equations of motion). Among our dividing curves, only the saddle orbit is an orbit, while the upright-arm curve and the approximate-saddle-orbit curve are not. 

Collecting (\ref{non-dir_n_dir}--\ref{def:varphi}), we obtain 
\begin{gather}
    \boxed{
    f(S,E)  =
    2\, \varphi(\gamma) } ~.
    \label{phase_flux_n_action}
\end{gather}
Altogether, we have shown that the phase-flux surface integral can be analytically reduced  to a line integral, which has the meaning of the abbreviated action along the curve $\gamma$. 

Next, we proceed to use this equation to specify this phase-flux for each of our flip criteria. 

\paragraph*{Upright arm criterion.}
For the upright arm criterion, based on \eqref{crit:up} the dividing surface is defined by
\begin{equation}
    g_{\mathrm{UA}}(\theta_2)
    =
    \theta_2-\pi
    \label{def:g1}
\end{equation}
so that
$
    \dot g_{\mathrm{UA}}
    =
    \dot\theta_2.
$
The corresponding phase-flux $f_{\mathrm{UA}}$ is given by \eqref{def:flux_pendulum}.

Using the delta function to perform the $\theta_2$ integration and evaluating
the remaining momentum integral analytically reduces the flux to the
one-dimensional expression $f_{\mathrm{UA}}(E) = 8\int_{-\pi}^{\pi} \sqrt{E-4+2\cos\theta_1}\,
    \Theta\!\left(E-4+2\cos\theta_1 \right) d\theta_1 $. This is a special case of the general reduction \eqref{phase_flux_n_action}.
    
Equivalently, using the ramp function, the flux may be written as
\begin{gather}
    \boxed{f_{\mathrm{UA}}(E)
    =
    8\int_{-\pi}^{\pi}
    \sqrt{
    \left(E-4+2\cos\theta_1\right)_+
    }
    \,d\theta_1} ~. 
    \label{eq:upright_arm_flux_integral}
\end{gather}

A derivation for general masses and arm lengths is given in  App.~\ref{sec:rate_computation_general_masses_and_lengths}.

The upright-arm flux can also be expressed in terms of elliptic integrals.
For $E\leq2$, the dividing surface is energetically inaccessible and therefore $f_{\mathrm{UA}}(E)=0$.

For $2<E<6$, only part of the $\theta_1$ interval is allowed, and the
flux is expressed in terms of the incomplete elliptic integral of the second
kind:
\begin{gather}
    \boxed{
    f_{\mathrm{UA}}(E)
    =
    32\sqrt{E-2}\,
    \mathcal E\!\left(
    \sin^{-1}\sqrt{\frac{E-2}{4}}
    \;\middle|\;
    \frac{4}{E-2}
    \right),
    \qquad
    2<E<6.
    }
    \label{eq:upright_arm_flux_incomplete_elliptic}
\end{gather}

For $E\geq6$, the entire interval is accessible, and the incomplete elliptic
integral becomes complete:
\begin{gather}
    \boxed{
    f_{\mathrm{UA}}(E)
    =
    32\sqrt{E-2}\,
    \mathcal E\!\left(
    \frac{4}{E-2}
    \right),
    \qquad
    E\geq6.
    }
    \label{eq:upright_arm_flux_complete_elliptic}
\end{gather}

\paragraph*{Approximate saddle orbit criterion.}

According to \eqref{crit:approx}, the approximate saddle orbit criterion is defined by the configuration-space
curve
\begin{gather}
    g_{\mathrm{ASO}}(\theta_1,\theta_2)
    =
    \theta_2-\pi-A^{\mathrm{ASO}} \sin\theta_1,
    \label{def:g2}
\end{gather}
The velocity normal to this dividing surface is
$ \dot g_{\mathrm{ASO}} = -A \cos\theta_1\,\dot\theta_1 + \dot\theta_2$.
The corresponding phase-flux $f_{\mathrm{ASO}}$ is given by \eqref{def:flux_pendulum}.
The $\theta_2$ integration and the momentum integrations can again be
performed analytically, leaving a one-dimensional integral, following the general reduction \eqref{phase_flux_n_action}:
\begin{gather}
    \boxed{
    f_{\mathrm{ASO}}(E)
    =
    4
    \int_{-\pi}^{\pi}
    \sqrt{2 K_{ASO}(\theta_1)_+}
    \sqrt{
    a(\theta_1)^2
    +
    2a(\theta_1)\xi(\theta_1)
    +
    2
    }
    \,
    d\theta_1
    } ,
    \label{eq:approximate_saddle_flux}
\end{gather}
where
\begin{gather*}
    K_{\mathrm{ASO}}(\theta_1)
    := K(\theta_1, \theta_2=\theta_2^{\mathrm{ASO}}(\theta_1)),
    \quad
    a(\theta_1)
    =
    A^{\mathrm{ASO}} \cos\theta_1,
    \quad
    \xi(\theta_1)
    =
    \cos\!\left(
    \theta_1-\pi-A\sin\theta_1
    \right) ~,
\end{gather*}
where $\theta_2^{\mathrm{ASO}}, A^{\mathrm{ASO}}$ were defined in \eqref{crit:approx}.

Unlike the upright-arm criterion, the remaining integral does not reduce to a
simple standard elliptic form. We therefore evaluate \eqref{eq:approximate_saddle_flux} numerically.

\paragraph*{The saddle orbit criterion.}
For the saddle orbit criterion, the dividing surface is constructed from the
numerically determined saddle orbit, which we denote by $\gamma_{SO}$. The numerical construction of this periodic orbit is described in Sec.~\ref{sec:Implement_saddle_orbit}.  The corresponding flip criterion is 
\begin{gather}
    g_{\mathrm{SO}} (\theta_1 , \theta_2) = \theta_2 - \theta_2^{\mathrm{SO}}(\theta_1).
    \label{def:g3}
\end{gather} 

After specializing to the double pendulum, where the action 1-form becomes
$ p\cdot dq = l_1\,d\theta_1 + l_2\,d\theta_2$, the general reduction of flux into action \eqref{phase_flux_n_action} implies
\begin{gather}
    \boxed{
    f_{\mathrm{SO}}(E)
    =
    2\displaystyle\oint_{\gamma_{SO}}
    \left(
    l_1\,d\theta_1+l_2\,d\theta_2
    \right)
    }.
    \label{eq:saddle_flux}
\end{gather}
Thus, once the saddle orbit $\gamma_{SO}$ is determined, the corresponding
phase-flux follows directly from its action.

\subsection{Determination of statistical prediction}
For convenience, we close this section by summarizing its results.
The three statistical predictions have the common form
\begin{gather}
    \boxed{
    R_a(E)
    =
    \frac{f_a(E)}{\Omega(E)} = 2\, \frac{\varphi_a(E)}{\Omega(E)},
    \qquad
    a=\mathrm{UA},\mathrm{ASO},\mathrm{SO}
    }
    \label{eq:three_flip_rate_predictions}
\end{gather}
where $\Omega(E)$ is the common phase-space volume at fixed energy and
$f_a(E)$ is the bidirectional phase-flux associated with the
corresponding flip criterion --- see \eqref{eq:flip_rate_formula}.

The phase-volume $\Omega(E)$ is given by \eqref{eq:phase_space_volume_egalitatian_double_pendu}.

For the upright-arm criterion, the phase flux $f_{\mathrm{UA}}(E)$ is given by \eqref{eq:upright_arm_flux_integral}. This integral is available analytically in terms of an incomplete elliptic
integral for $2<E<6$ and a complete elliptic integral for $E\geq6$.

For the approximate saddle orbit criterion, the phase-flux $f_{\mathrm{ASO}}(E)$ is given by \eqref{eq:approximate_saddle_flux}.

Finally, for the saddle orbit criterion, the phase-flux $f_{\mathrm{SO}}(E)$ is given by \eqref{eq:saddle_flux}.

Thus, all three criteria share the same phase-space normalization
$\Omega(E)$, while they differ in the dividing surface used to define a flip
and, consequently, in the associated phase-flux. The upright-arm flux
admits an analytic representation in terms of elliptic integrals, the
approximate saddle-orbit flux reduces to a one-dimensional numerical
integration, and the exact saddle-orbit flux is obtained from the action of the
numerically determined periodic orbit.

\paragraph*{Evaluation of statistical prediction.} 
According to \eqref{eq:three_flip_rate_predictions}, 
the statistical prediction requires the evaluation of both the phase-volume function $\Omega(E)$ and the phase-flux functions $f_a (E)$, $a=\mathrm{UA},\mathrm{ASO},\mathrm{SO}$.

The phase-volume function \eqref{eq:phase_space_volume_egalitatian_double_pendu}
is readily evaluated by numerical integration, as described at the end of Sec.~\ref{sec:Implement_saddle_orbit}.

\begin{figure}[H]
    \centering
    \includegraphics[width=0.7\linewidth]{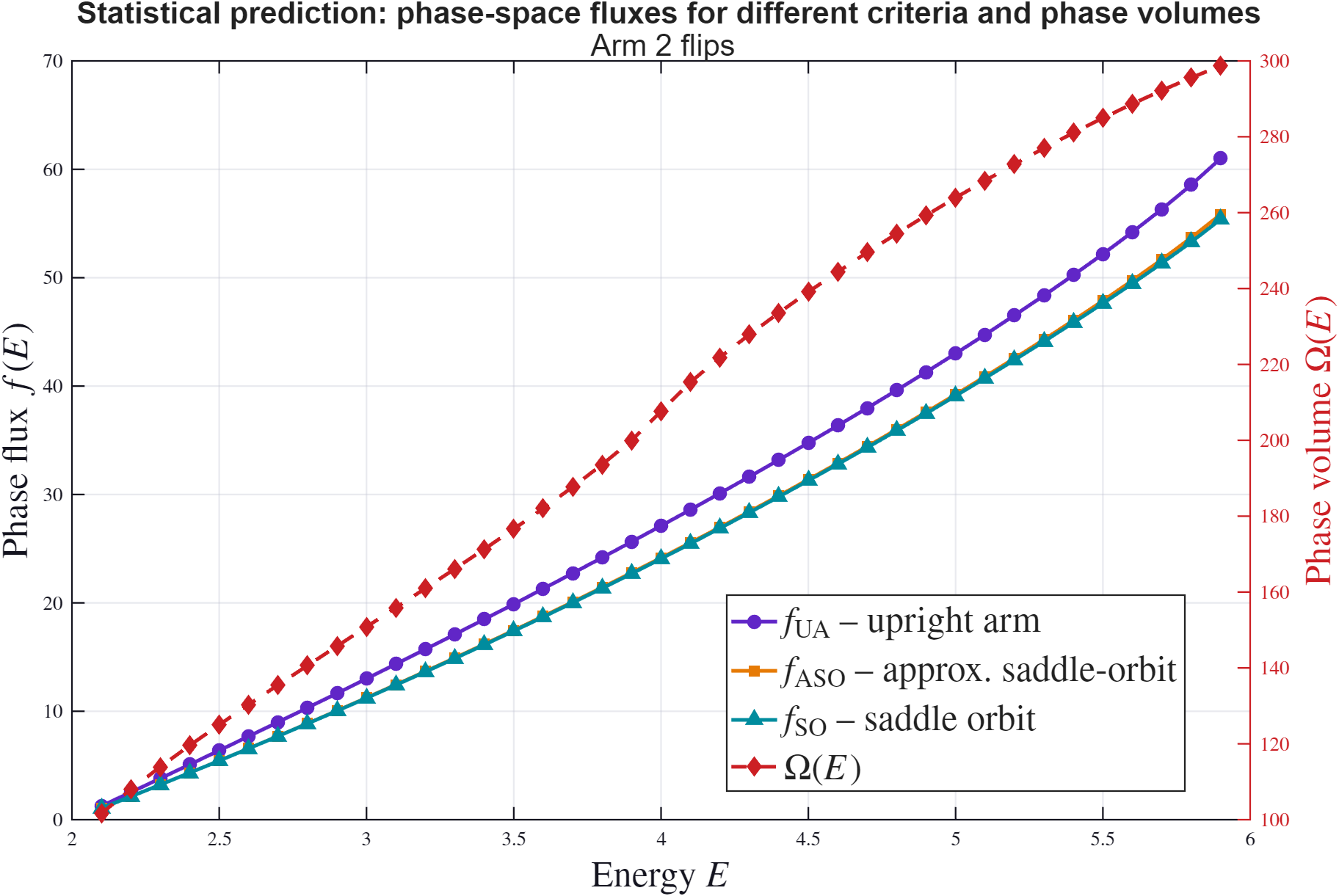}
    \caption{Phase-flux associated with the three flip criteria, together with the microcanonical phase-space volume, as functions of energy. The fluxes corresponding to the saddle-orbit approximation and the saddle-orbit criterion are nearly indistinguishable on the scale of the figure.}
    \label{fig:phase space flux and criteria flux}
\end{figure}

We can now collect all the ingredients required for the statistical prediction; they are displayed in Fig.~\ref{fig:phase space flux and criteria flux}.

\paragraph*{Numerical integration in phase space.}

The phase-space integrals for the phase flux of the upright-arm criterion~\eqref{eq:upright_arm_flux_integral} and the approximate saddle-orbit criterion~\eqref{eq:approximate_saddle_flux}, as well as the phase-volume integral~\eqref{eq:phase_space_volume_egalitatian_double_pendu}, were evaluated numerically using MATLAB's \textit{integral} and \textit{integral2} functions. The \textit{integral} function was used for the one-dimensional integrations, while \textit{integral2} was used for the two-dimensional phase-volume integral. For both integrations, a relative tolerance of $10^{-10}$ was prescribed.

\section{Numerical implementation}
\label{sec:implementation}
The numerical implementation consists of four main components. First, the
equations of motion are integrated to obtain individual trajectories. Second,
flip events are identified along each trajectory according to the three
criteria introduced previously. Third, an ensemble of initial conditions is
sampled from the microcanonical energy surface in order to estimate the
statistical flip rates. Finally, the saddle orbit required for the exact
saddle-orbit criterion is computed independently using the shooting algorithm and
continuation procedure.

\subsection{Simulated rate}

\paragraph*{Solving for the trajectory evolution.} The double-pendulum equations of motion \eqref{Euler-Lagrange} with prescribed initial conditions are integrated numerically in MATLAB using \textit{ode45}, a variable-step Runge-Kutta $(4,5)$ method. 
For each initial condition, the numerical solution provides the time evolution
of the configuration and velocity variables,
\begin{gather*}
    \theta_1(t),\qquad
    \theta_2(t),\qquad
    \dot{\theta}_1(t),\qquad
    \dot{\theta}_2(t).
\end{gather*}

\paragraph*{Counting flips.}
A flip is identified as a crossing of the dividing surface associated with a
specified flip criterion. The same numerically propagated trajectory can
therefore be analyzed using several different definitions of a flip.

In the present work, three criteria are considered.
The upright-arm, saddle-orbit, and approximate saddle-orbit criteria are defined by \eqref{def:g1}, \eqref{def:g3}, and \eqref{def:g2}, respectively.

For each trajectory and each criterion, the total number of detected
crossings during the simulation time is denoted by
$
    N_{\mathrm{flips}}(T_{\mathrm{sim}}).
$
The corresponding time-averaged flip rate of an individual trajectory is
\begin{gather*}
    R_{\mathrm{sim}}
    =
    \frac{N_{\mathrm{flips}}(T_{\mathrm{sim}})}
    {T_{\mathrm{sim}}}.
\end{gather*}

\paragraph*{Ensemble statistics and microcanonical sampling.} The statistical prediction derived in Sec.~\ref{sec:prediction} requires averaging over the microcanonical ensemble at fixed energy. We perform this average in two steps: first, by defining and sampling the corresponding marginal distribution on configuration space, and then by sampling the conditional distribution of momentum directions.

Recall from \eqref{eq:configuration_space_density} and
\eqref{eq:phase_space_volume_egalitatian_double_pendu} that
\begin{gather*}
    \Omega(E)
    =
    \int d\theta_1\,d\theta_2\,
    \rho(\theta_1,\theta_2;E),
\end{gather*}
where $\rho$ is defined by \eqref{eq:configuration_space_density}.
It is convenient to write
$
    \rho(\theta_1,\theta_2;E)
    = 2\pi \, 
    w(\theta_1,\theta_2)
    \,
    \Theta\!\left(
    K(\theta_1,\theta_2;E)
    \right),
$
with
$
    w(\theta_1,\theta_2)
    =
    \sqrt{
    2-\cos^2(\theta_1-\theta_2)
    },
$
where $K$ is defined by \eqref{def:available_kinetic_E}.

The microcanonical initial conditions are generated as follows:

\begin{enumerate}

    \item Sample a configuration
    $
        \vec{\theta}
        =
        (\theta_1,\theta_2)
    $
    from the energetically accessible region
    $
        K(\vec{\theta};E)>0
    $
    with probability density proportional to
    $
        w(\vec{\theta})
        =
        \sqrt{
        2-\cos^2(\theta_1-\theta_2)
        }.
    $
    \item For the sampled configuration, calculate the available kinetic
    energy,
    $
        K(\vec{\theta})
        =
        E-V(\vec{\theta}).
    $
    \item Sample a uniformly distributed angle
    $
        \alpha
        \sim
        \operatorname{Uniform}(0,2\pi)
    $
    and define the unit vector
    $
        u
        =
        \begin{pmatrix}
            \cos\alpha\\
            \sin\alpha
        \end{pmatrix}.
    $
    \item Determine the canonical momenta corresponding to the sampled circularized
    direction and the prescribed kinetic energy, see Sec.~\ref{sec:prediction_phase-volume}. Let
    $
        M^{-1}(\vec{\theta})
        =
        L^TL.
    $
    Then
    $
        K
        =
        \frac{1}{2}l^TM^{-1}l
        =
        \frac{1}{2}|Ll|^2.
    $
    Hence,
    $
        |Ll|
        =
        \sqrt{2K},
    $
    and the momenta are chosen as
    $
        l
        =
        L^{-1}
        \sqrt{2K}\,u.
    $
\end{enumerate}

This procedure ensures that the initial conditions are sampled according to the natural
microcanonical measure on the energy surface.

For each energy, an ensemble of
$
    N_{\mathrm{IC}}=1{,}000
$
independently sampled initial conditions is evolved up to
$T_{\mathrm{sim}}=10{,}000$.
The flip rate is then computed separately for each trajectory and each flip criterion.
For the subsequent analysis of the chaotic subensemble in Sec.~\ref{sec:results}, we use an additional set of simulations of $N_{\mathrm{IC},\chi}  = 100 $ independently sampled initial conditions, each evolved up to a longer simulation time of $T_{\chi}  = 10^{5}$.

The simulation estimate of the ensemble-averaged flip rate is then
\begin{gather}
    R_{\mathrm{sim}}(E)
    =
    \left\langle
    \frac{N_{\mathrm{flips}}(T_{\mathrm{sim}})}
    {T_{\mathrm{sim}}}
    \right\rangle,
    \label{eq:simulation_flip_rate_estimate}
\end{gather}
where the average is taken over the sampled initial conditions on the chosen ensemble.
In addition to the ensemble mean, the trajectory-to-trajectory variability is
recorded in order to quantify the spread of the measured long-time flip rates
within the ensemble.

\paragraph*{Computational resources.}
All numerical simulations were performed on a desktop computer equipped with an Intel Core Ultra 7 265 processor (20 CPU cores, 2.4 GHz). Parallel computing was employed to distribute the independent trajectory integrations across multiple CPU cores.

\subsection{Saddle orbit and shooting}
\label{sec:Implement_saddle_orbit}

The saddle-orbit criterion requires the determination of the saddle orbit at each energy. 
This constitutes a separate numerical problem from the propagation of the ensemble trajectories described above.
For each energy
$
    E>E_c,
$
we compute the periodic orbit associated with the saddle point, which we refer
to as the saddle orbit. For the down-up saddle at
$
    E_c=2,
$
the corresponding equilibrium configuration is
$
    q_*
    =
    (\theta_1,\theta_2)
    =
    (0,\pi).
$

For the shooting calculation, the system is written in canonical variables $(\theta_1,\theta_2,l_1,l_2),$ with the Hamiltonian \eqref{eq:egalitarian double pendulum hamiltonian}.

The evolution used in the shooting procedure is governed by Hamilton's
equations,
\begin{gather*}
    \dot q
    =
    \frac{\partial H}{\partial l}
    =
    M^{-1}(q)l,
    \qquad
    \dot l
    =
    -\frac{\partial H}{\partial q}.
\end{gather*}

The shooting problem is reduced by fixing the initial configuration at the saddle,
$
    q_0=q_*.
$ This choice exploits the reflection symmetry \eqref{def:reflection} of the system and restricts the shooting calculation to the symmetric periodic-orbit family passing through the saddle configuration.

At fixed energy, the magnitude of the initial canonical momentum is determined by
the energy constraint once its direction is specified. We therefore parametrize the initial momentum as
\begin{gather*}
    l_0
    =
    \chi\,u(\beta),
    \qquad
    u(\beta)
    =
    \begin{pmatrix}
        \cos\beta\\
        \sin\beta
    \end{pmatrix},
\end{gather*}
where $\beta$ denotes the direction of the initial canonical momentum in the $(l_1,l_2)$ plane, and $\chi(E,\beta)$ is its Euclidean magnitude.  

Substitution into the energy constraint gives
\begin{gather*}
    E
    =
    V(q_*)
    +
    \frac{1}{2}
    \chi^2
    u(\beta)^TM^{-1}(q_*)u(\beta),
\end{gather*}
and therefore
\begin{gather*}
    \chi(E,\beta)
    =
    \sqrt{
    \frac{
    2K(q_*;E)
    }{
    u(\beta)^TM^{-1}(q_*)u(\beta)
    }
    },
\end{gather*}
where $K(q_*;E) = E-V(q_*) = E-E_c.$

Thus, at a prescribed energy, the initial momentum is completely determined by the single parameter $\beta$.

The shooting variables are
$
    \eta
    =
    \begin{pmatrix}
        \beta \\ T
    \end{pmatrix}
$
where $T$ is the unknown period of the orbit.
For every trial pair $(\beta,T)$, the initial condition
$
    q_0=q_*,
    \, \,
    l_0=l_0(E,\beta)
$
is constructed, and Hamilton's equations are integrated from $t=0$ to
$t=T$. Periodicity requires closure in phase space,
\begin{gather*}
    q(T)=q(0),
    \quad
    l(T)=l(0).
\end{gather*}
The shooting residual is defined as a four-dimensional vector
\begin{gather}
    r(\beta,T)
    =
    \begin{pmatrix}
        q(T)-q(0)\\
        l(T)-l(0)
    \end{pmatrix}.
\end{gather}
Its first two components measure the mismatch in the generalized coordinates, while the remaining components measure the mismatch in the corresponding canonical momenta.

The saddle orbit is then obtained by minimizing the squared norm of this residual with respect to $\beta$ and $T$,
$
    \min_{\beta,T}
    \|r(\beta,T)\|^2.
$

For the first energy above the saddle, the initial estimate of the period is obtained from the linearized center-mode frequency $\omega$. The corresponding linearized oscillation has period
$
    T_{\mathrm{lin}}
    =
    \frac{2\pi}{\omega}.
$
The periodic-orbit family is subsequently continued in energy. Denoting the solution at energy $E$ by 
$\eta(E) = (\beta(E),T(E))^T$, the initial prediction at $E_k$, given solutions at $E_{k-2}$ and $E_{k-1}$, is obtained by secant extrapolation:
\begin{gather*}
    \eta(E_k)
    \simeq
    \eta(E_{k-1})
    +
    (E_k-E_{k-1})
    \frac{
        \eta(E_{k-1})-\eta(E_{k-2})
    }{
        E_{k-1}-E_{k-2}
    }.
\end{gather*}
The predicted solution is then corrected by the shooting solver. If the algorithm fails to converge at a new energy, an intermediate energy is inserted, reducing the continuation step.

The numerical propagation used in the shooting calculation is performed with
the fourth-order, two-stage Gauss--Legendre method (\textit{GL2}), a symplectic implicit integrator. This method preserves the Hamiltonian phase-space structure during the periodic-orbit
calculation and is therefore preferred over the \textit{ode45} propagation employed for the ensemble simulations.

Once the saddle orbit has been obtained, its action is evaluated by \eqref{eq:saddle_flux}.
By the flux-over-a-saddle result \eqref{phase_flux_n_action}, this action determines the one-way flux through the dividing surface bounded by the saddle orbit. The resulting periodic orbit is also used to construct the saddle-orbit flip criterion employed in the trajectory analysis above.

\subsection{Quality indices}

To quantify both the accuracy of the numerical implementation and the quality of the statistical prediction, we introduce two quality indices. The first measures the agreement between the numerical implementation and the corresponding microcanonical ensemble average, while the second takes into account the ensemble variability in quantifying the agreement. 

\paragraph*{Implementation quality index.}
The implementation quality index (IQI) measures the agreement of the full microcanonical ensemble average with the numerically evaluated statistical prediction. For each energy $E$, we define the relative deviation of the simulated mean from the theoretical ensemble average as
\begin{equation}
\mathcal{R}(E)
=
\left|
\frac{R_{\mathrm{sim}}(E)}
{R_{\mathrm{pred}}(E)}
-1
\right|.
 \label{def:RE}
\end{equation}
The implementation quality index is then defined as the root-mean-square (RMS) of this relative deviation over the set of sampled energies,
\begin{equation}
IQI
=
\sqrt{
\frac{1}{N_E}
\sum_{i=1}^{N_E}
\mathcal{R}(E_i)^2
},
\label{eq:implementation_quality_index}
\end{equation}
where $N_E$ denotes the number of energy values considered. A smaller value of IQI indicates a more accurate numerical implementation. 

The theory behind the flux-based method predicts that the IQI should vanish, apart from numerical errors. It therefore quantifies the cumulative numerical error across the entire implementation chain: the numerical integrations required to evaluate the prediction, the numerical solution of the equations of motion, and the estimation of flip rate statistics over the chosen ensemble. The low value we shall find will thus provide an empirical validation of the numerical implementation and support the underlying mathematical reasoning.

\paragraph*{Prediction quality index.}
The prediction quality index (PQI) is designed to quantify the agreement between the statistical prediction and individual orbits. Unlike the IQI, it is not determined by the present flux-based theory; rather, it is a property of the chaotic dynamics and can only be evaluated through simulations.

For each energy $E$, we define the relative deviation of the simulated mean $
\mathcal{R}(E)$ through \eqref{def:RE}, and the relative statistical uncertainty,
$
\mathcal{S}(E)
=
\frac{\sigma_{\mathrm{R,sim}}(E)}
{R_{\mathrm{pred}}(E)},
$
where $\sigma_{\mathrm{R,sim}}(E)$ is the standard deviation of the trajectory-averaged flip rates within the ensemble. We combine these two contributions into a single quantity,
\begin{equation}
\mathcal{Q}^2(E)
= \left\langle \left( \frac{R_{\mathrm{sim}}}{R_{\mathrm{pred}}}-1 \right)^2 \right\rangle 
=  \mathcal{R}(E)^2
+
\mathcal{S}(E)^2
.
\end{equation}
The prediction quality index is defined as
\begin{equation}
PQI
= \sqrt{
\frac{1}{N_E}
\sum_{i=1}^{N_E}
\mathcal{Q}^2(E_i)}.
\label{eq:prediction_quality_index}
\end{equation}

Thus, the PQI incorporates both the discrepancy between the statistical prediction and the simulated mean and the statistical variability within the ensemble. A smaller PQI therefore indicates a more effective statistical prediction.

\section{Results}
\label{sec:results}

In this section, we compare the statistical predictions for the mean flip rates of the egalitarian double pendulum derived in Sec.~\ref{sec:prediction} with the mean flip rates obtained from numerical simulations of the equations of motion. The two are expected to agree, up to numerical errors, after averaging  the simulation initial conditions over the constant-energy surface in phase space, namely, over the microcanonical ensemble. 

In analyzing the dependence on initial conditions, we find a sharp increase in the trajectory-to-trajectory variability over a certain energy range. This observation motivates a physically meaningful restriction of the ensemble of initial conditions, under which the validity of the prediction extends beyond ensemble averages to individual trajectories within the resulting subensemble.

\subsection{Microcanonical-ensemble average}

\begin{figure}[H]
    \centering
    \includegraphics[width=0.9\linewidth]{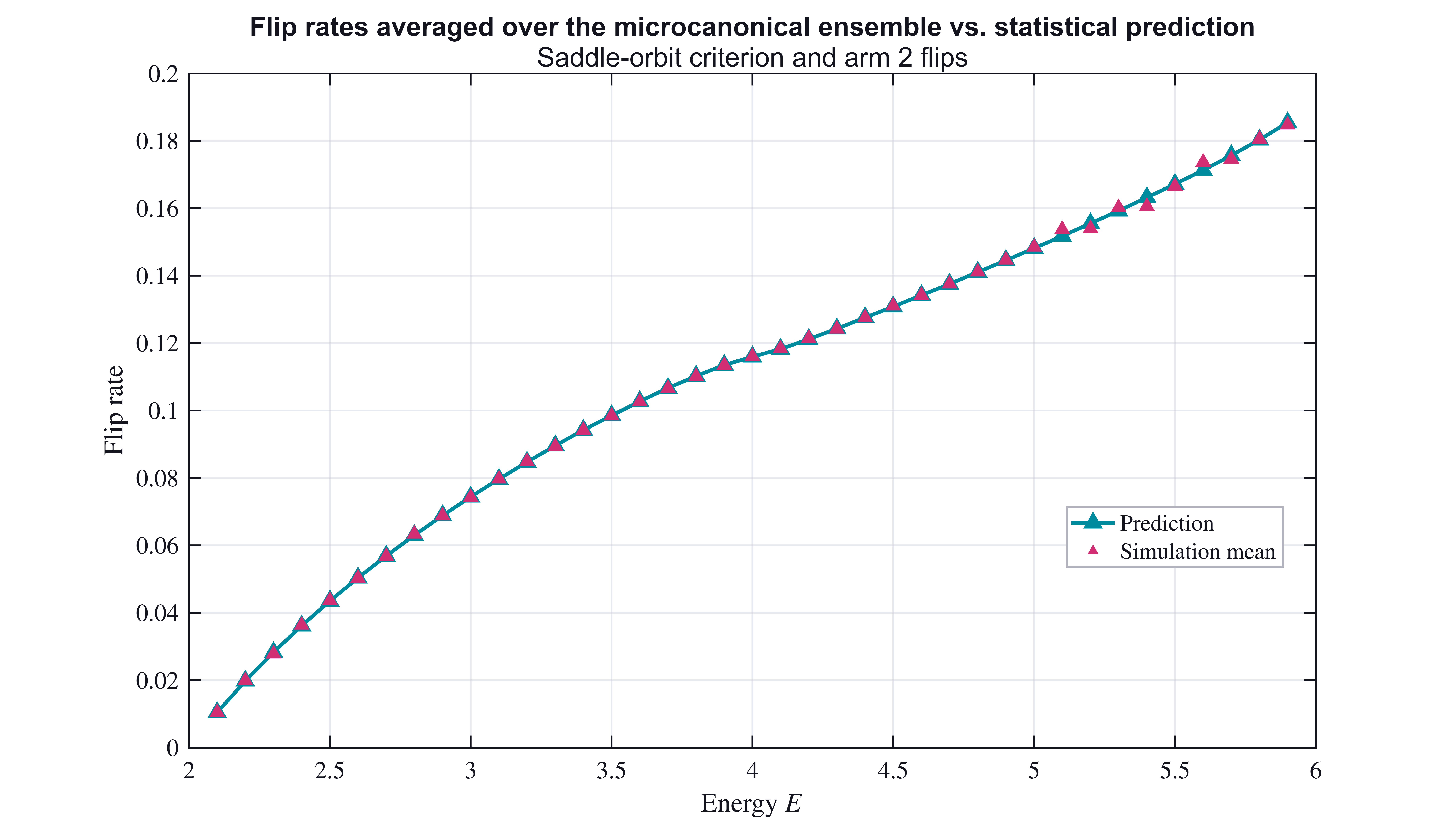}
    \caption{Flip rate of arm 2 according to the saddle-orbit criterion, averaged over the microcanonical ensemble: flux-over-a-saddle prediction and numerical simulation. The numerical mean is obtained from $1{,}000$ independently sampled initial conditions, each propagated for $10{,}000$ simulation time units.}
\label{fig:saddle_orbit_criterion_second_pendulum_all_trajectories_mean_only}
\end{figure}

We simulated trajectories, measured their flip rates, and averaged over initial conditions as explained in the previous sections. Flips were counted for both arm 2 and arm 1, using the various flip criteria introduced in Sec.~\ref{sec:criteria}. We begin with arm-2 flips, which are more frequent, and with the saddle-orbit criterion, which provides the most natural definition of a flip. Results for arm 1 and alternative flip criteria are presented later in this section. Figure~\ref{fig:saddle_orbit_criterion_second_pendulum_all_trajectories_mean_only} shows the resulting microcanonical-ensemble-averaged flip rates together with the statistical prediction.   

\paragraph*{Energy ranges.} To characterize the quality of agreement between prediction and simulation, we define the following energy subranges. 

We refer to the energy range \eqref{def:energy_range} as the chaotic energy regime. For later use, we further divide it into three subranges: \begin{itemize}
    \item $2 \le E  < E_1:=2.8$ -- low chaotic regime;
    \item $E_1 \le E  < E_2:= 5$ -- central chaotic regime;
    \item $E_2 \le E  < 6$ -- high chaotic regime.
\end{itemize}
This subdivision is motivated as follows. As $E \to 2$, the flip rate tends to zero, leading to correspondingly larger \emph{relative} discrepancies between prediction and simulation. For $E \ge E_1$, this effect becomes less pronounced, motivating our choice of $E_1$ as the boundary between the low and central regimes. The central chaotic regime is characterized by particularly pronounced chaotic behavior and, as we shall see, by the best agreement with the flux-based prediction. At $E=E_2$, a  regular component becomes discernible and needs to be taken into account, motivating our choice of this value as the boundary between the central and high regimes. Clearly, the precise choices of the boundaries $E_1$ and $E_2$ retain some degree of arbitrariness. 

When considering flips of arm 1, we use a different energy range,
\begin{equation}
4 \le E \le 6 ~,
\label{def:energy_range_arm1}
\end{equation}
whose lower bound, $E=4$, is the energy of the up-down saddle. We further divide this range into two subranges at $E=4.6$, motivated by a similar increase in the relative discrepancies between prediction and simulation at lower energies.

\paragraph*{Implementation quality index.} We quantify the agreement between the numerical microcanonical-ensemble flip rate and the statistical prediction using the implementation quality index (IQI), defined in \eqref{eq:implementation_quality_index}. Essentially, the IQI is the relative root-mean-square (RMS) deviation over energy. Table~\ref{tab:implementation_quality} gives its values over the full range $2<E<6$ \eqref{def:energy_range}, as well as separately within the low-, central-, and high-energy chaotic regimes.

\begin{table}[H]
\centering
\begin{tabular}{c|c}
Energy regime & Implementation quality index (IQI)[\%]  \\
\hline
Full chaotic range ($2<E<6$) & 0.56 \\
Low chaotic ($2<E<2.8$) & 0.72 \\
Central chaotic ($2.8\leq E<5$) & 0.17 \\

High chaotic ($5 \leq E<6$) & 0.91 \\
\end{tabular}
\caption{Implementation quality index (IQI) of the numerical microcanonical-ensemble flip rate of arm 2, evaluated over the full energy range and separately within the three chaotic energy regimes.}
\label{tab:implementation_quality}
\end{table}

\paragraph*{Comparison.} The IQI remains below $1\%$ throughout the full energy range and reaches a minimum of $0.17\%$ in the central chaotic regime. This close agreement supports the statistical prediction and provides a stringent validation of the numerical implementation in all its components.

\paragraph*{Convergence.} The data for Fig.~\ref{fig:saddle_orbit_criterion_second_pendulum_all_trajectories_mean_only} were obtained from $N_{\mathrm{IC}}=1{,}000$ independently sampled initial conditions, each propagated to $T_{\mathrm{sim}}=10{,}000$ simulation time units. We investigated convergence with respect to both parameters. In a later subsection, we present evidence that the associated uncertainty is $\lesssim 1\%$, which is sufficient for our purposes. In the central chaotic regime, we estimate it to be even smaller, $\lesssim 0.2\%$. Further improvements in statistical precision can be achieved by increasing $N_{\mathrm{IC}}$ or $T_{\mathrm{sim}}$.

\subsection{Ensemble variability}

\begin{figure}[H]
    \centering
    \includegraphics[width=0.80\linewidth]{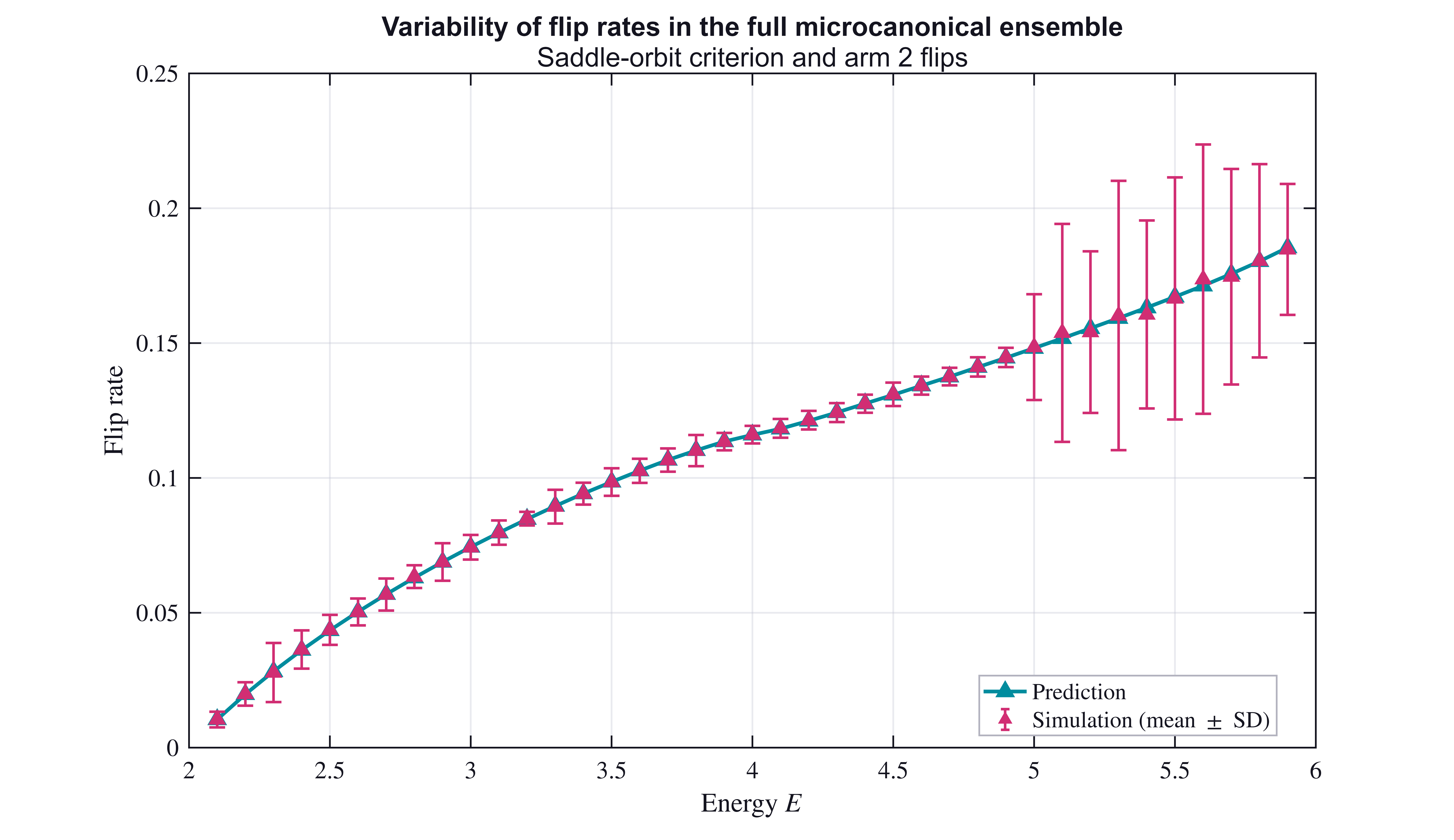}
    \caption{Mean flip rates with ensemble variability. Microcanonical-ensemble mean flip rate of arm 2 according to the saddle-orbit criterion, compared with the flux-over-a-saddle prediction. The simulation points show the mean over $1{,}000$ independently sampled initial conditions, while the vertical bars indicate one standard deviation of the trajectory-to-trajectory variability. Each trajectory is propagated for $10{,}000$ simulation time units.}
\label{fig:saddle_orbit_criterion_second_pendulum_all_trajectories_with_scattering}
\end{figure}

Although the ensemble mean follows the statistical prediction closely, the trajectory-to-trajectory variability increases substantially at the upper end of the investigated energy range, as shown in Fig.~\ref{fig:saddle_orbit_criterion_second_pendulum_all_trajectories_with_scattering}.

\begin{figure}[H]
    \centering
    \includegraphics[width=0.7\linewidth]{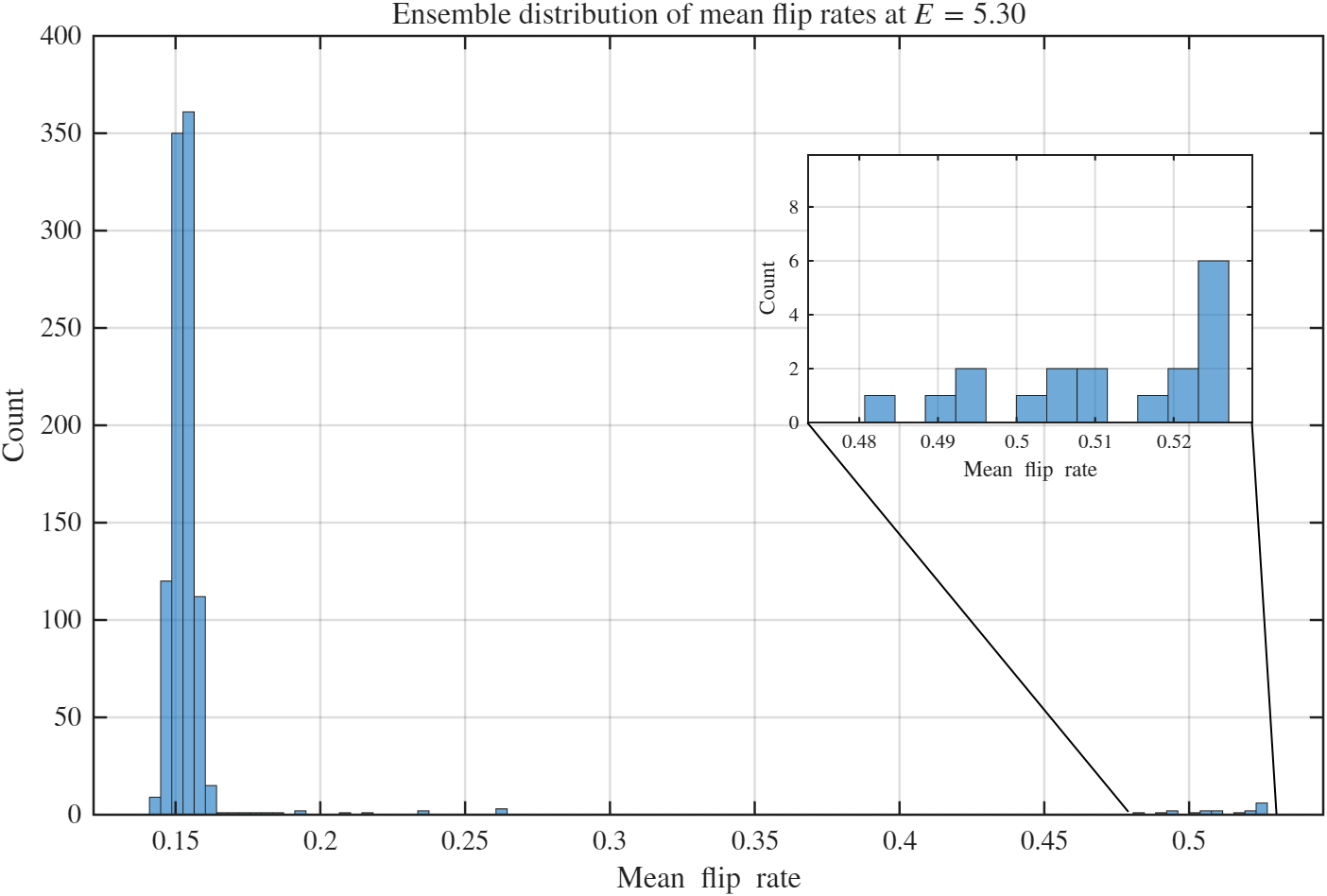}
    \caption{Distribution of trajectory-averaged arm-2 flip rates at $E = 5.3$, according to the saddle-orbit criterion. The dominant population forms a narrow peak near the ensemble mean, while a small, well-separated high-rate population extends to values of order three times the mean. The inset magnifies the high-rate population.}
    \label{fig:flip_rate_histogram}
\end{figure}

The distribution of trajectory-averaged flip rates reveals the origin of this increased variability. At $E=5.3$, for example, the histogram in Fig.~\ref{fig:flip_rate_histogram} exhibits a small, well-separated high-rate population, with flip rates reaching roughly three times the ensemble mean. This behavior is consistent with the mixed phase-space structure observed in this energy range, where regular islands coexist with the chaotic sea. Trajectories confined to different dynamical regions can consequently exhibit substantially different long-time flip rates.

\paragraph*{Chaotic subensemble.}
Restricting the analysis to the chaotic component substantially reduces the trajectory-to-trajectory variability. For sufficiently long integration times, the measured flip rates of most such trajectories cluster within a narrow range.

This observation motivates separating the dominant chaotic population from the regular component. The dominant narrow population in histograms such as Fig.~\ref{fig:flip_rate_histogram} corresponds to the chaotic component, while the well-separated high-rate population corresponds to the regular component. Using the histogram itself to define the separation, however, would require choosing a numerical cutoff in the flip rate. We therefore adopt an operational definition based instead on the comparison of flip counts obtained using the different flip criteria. We identify as regular those trajectories for which all the flip criteria yield exactly the same total number of flips over the same integration interval, and define the \emph{chaotic subensemble} as the complement of this set. Because this construction does not privilege any one flip criterion, the resulting classification is independent of the particular criterion subsequently used to evaluate the flip rate.

To validate this operational classification, we independently examined Poincaré sections at representative energies. 
The trajectories excluded from the chaotic subensemble were found to trace closed curves in the Poincaré sections, consistent with regular motion. This observation supports the effectiveness of the proposed criterion in identifying the regular component of the ensemble over the energy range considered.   
\subsection{Convergence}

In this subsection, we discuss the convergence of the numerical computations and estimate the associated numerical uncertainties.  

\paragraph*{Convergence of individual-trajectory flip rates.}
We first assess the finite-time uncertainty in the flip rates of individual trajectories. The production simulations were performed up to $T_{\mathrm{sim}}=10{,}000$ simulation time units.

\begin{figure}[H]
    \centering
\includegraphics[width=0.60\linewidth]{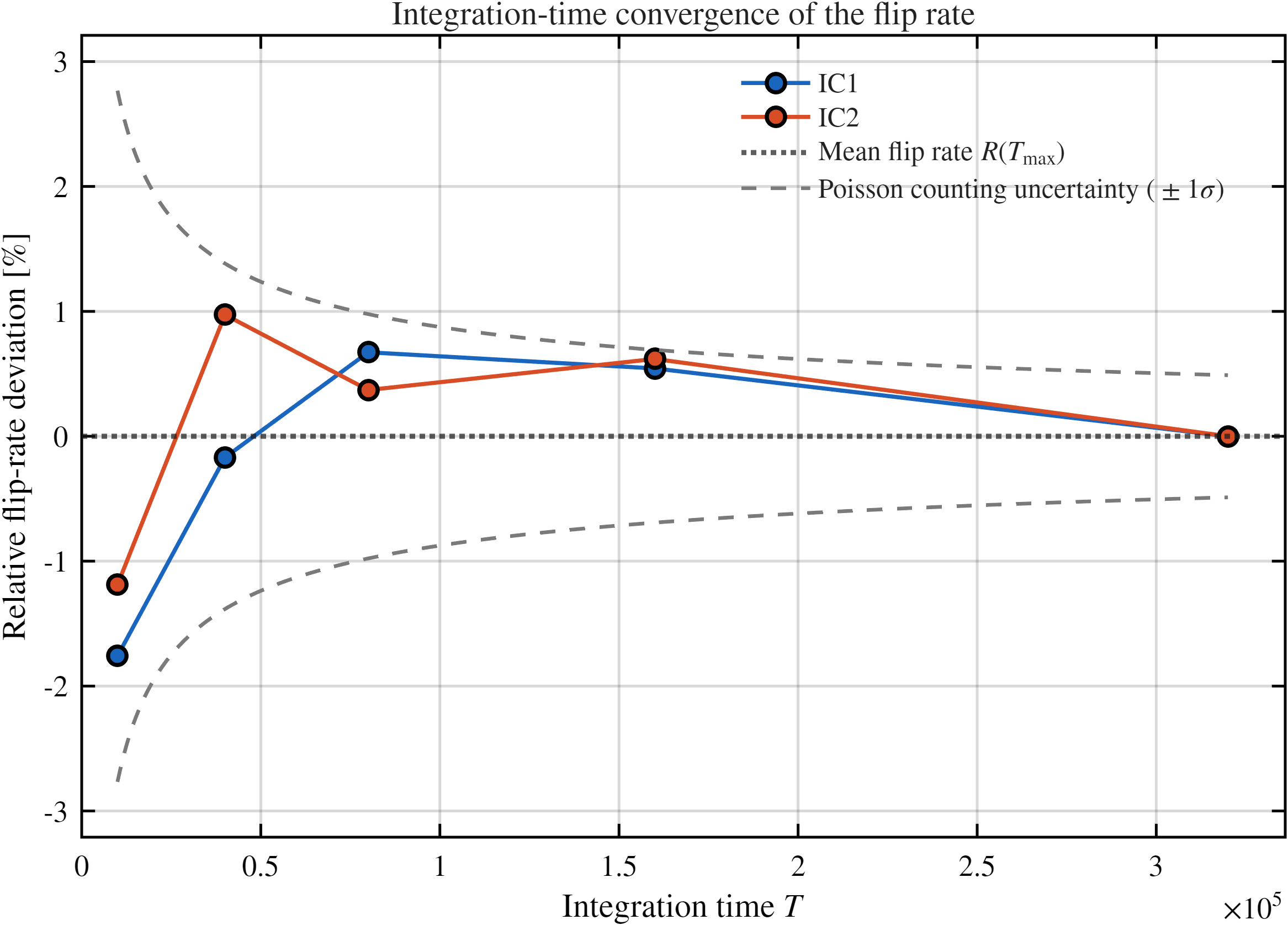}
    \caption{Relative deviations of the running flip rates of two representative chaotic trajectories at $E=4.5$ from their respective long-time reference values $R(T_{\max})$. The dashed curves indicate the Poisson counting scale $\pm\epsilon_{\mathrm{Poisson}}$, \eqref{eq:poisson_error}.}
\label{fig:convergence_test_integration_time}
\end{figure}

To assess the residual finite-time variability of chaotic trajectories, we select two representative chaotic initial conditions at $E=4.5$ and integrate their trajectories over a substantially longer interval, $T_{\max} > T_{\mathrm{sim}}$.
As a benchmark, we consider the relative counting uncertainty associated with a Poisson process,
\begin{gather}
    \epsilon_{\mathrm{Poisson}} = \frac{1}{\sqrt{N_{\mathrm{flips}}}}~,
    \label{eq:poisson_error}
\end{gather}
where $N_{\mathrm{flips}}$ is the number of flips recorded during the observation interval. This provides a characteristic scale for finite-time relative fluctuations. For each trajectory, we compute the relative deviation of the running flip rate $R(T)$ from its long-time reference value $R(T_{\max})$, 
\begin{gather}
    \epsilon_R(T) := \frac{R(T) - R(T_{\max})}{R(T_{\max})} ~.
\end{gather}
Figure~\ref{fig:convergence_test_integration_time} shows these deviations, expressed as percentages, together with the corresponding Poisson counting scale. Over the displayed interval, the deviations remain within the Poisson envelope. We therefore take the characteristic convergence error for an individual chaotic trajectory to scale as \begin{equation}
    \epsilon_{R,\chi}(T_{\text{sim}}) \sim \frac{1}{\sqrt{N_{\mathrm{flips}}}} = \frac{1}{\sqrt{R\, T_{\text{sim}}}}~.
    \label{chaotic_traj_error}
\end{equation}

We also studied the convergence of the flip rates of individual regular trajectories with respect to $T_{\text{sim}}$ and found that they converge at least as fast as those of chaotic trajectories, namely, \begin{equation}
    \epsilon_{R,\text{reg}} (T_{\text{sim}}) \le \epsilon_{R,\chi} (T_{\text{sim}})~.
\end{equation}

\paragraph*{Convergence of microcanonical-ensemble-averaged mean flip rates.}
We now return to the convergence and estimated uncertainty of the microcanonical-ensemble-averaged mean flip rates shown in Fig.~\ref{fig:saddle_orbit_criterion_second_pendulum_all_trajectories_mean_only}.

At each energy, the microcanonical-ensemble average is obtained from $N_{\mathrm{IC}}$ independently sampled initial conditions. Its statistical uncertainty is therefore governed by the trajectory-to-trajectory standard deviation $\sigma_R(E)$. The corresponding relative standard error of the ensemble average is
\begin{equation}
    \epsilon_{\overline{R}}(E)
    \simeq
    \frac{\sigma_R(E)}
    {\overline{R}(E)\sqrt{N_{\mathrm{IC}}}}.
\end{equation}
Figure~\ref{fig:saddle_orbit_criterion_second_pendulum_all_trajectories_with_scattering} shows that the relative spread depends strongly on energy. In the high chaotic regime, $E\geq5$, we find conservatively
\begin{equation}
    \frac{\sigma_R(E)}{\overline{R}(E)}
    \lesssim0.33.
\end{equation}
For the production value $N_{\mathrm{IC}}=1{,}000$, this gives
\begin{equation}
    \epsilon_{\overline{R}}(E)
    \lesssim
    \frac{0.33}{\sqrt{1{,}000}}
    \simeq1.0\%.
\end{equation}
Thus, in the high-energy regime the statistical uncertainty is of order $1\%$ at its largest.

In the central chaotic regime, $2.8\leq E<5$, the trajectory-to-trajectory spread is much smaller. From Fig.~\ref{fig:saddle_orbit_criterion_second_pendulum_all_trajectories_with_scattering}, we estimate
\begin{equation}
    \epsilon_{\overline{R}}(E)\lesssim0.2\%.
\end{equation}
At still lower energies, $2<E<2.8$, the relative uncertainty increases again as the mean flip rate tends to zero.

For comparison, we estimate separately the characteristic finite-time counting scale. Using the Poisson counting estimate established above, and noting that the total number of flips is $N_{\mathrm{flips,tot}}=N_{\mathrm{flips}}N_{\mathrm{IC}}$, we obtain
\begin{gather}
    \epsilon_{\overline{R},\mathrm{count}}
    \lesssim
    \frac{1}{\sqrt{N_{\mathrm{flips,tot}}}}
    =
    \frac{1}{\sqrt{N_{\mathrm{flips}}N_{\mathrm{IC}}}}
    =
    \frac{1}{\sqrt{\overline{R}\,T_{\mathrm{sim}}N_{\mathrm{IC}}}},
    \label{eq:full_ensemble_mean_flip_rate_variability}
\end{gather}
where $N_{\mathrm{flips}}=\overline{R}\,T_{\mathrm{sim}}$ is the mean number of flips per trajectory. For $E\geq E_1=2.8$, we have $\overline{R}(E)\geq0.06$, and with $T_{\mathrm{sim}}=10{,}000$ and $N_{\mathrm{IC}}=1{,}000$ this gives
\begin{equation}
    \epsilon_{\overline{R},\mathrm{count}}
    \lesssim
    \left(0.06\cdot10{,}000\cdot1{,}000\right)^{-1/2}
    \simeq0.13\%.
\end{equation}

The finite-sampling uncertainty therefore clearly dominates in the high-energy regime, where it reaches values of order $1\%$. In the central chaotic regime, the Poisson counting scale is comparable to the overall statistical uncertainty and may account for a substantial part of it.

Finally, since the measured $\sigma_R$ already includes finite-time fluctuations, the two estimates should not be regarded as independent contributions and combined in quadrature. Rather, $\sigma_R/(\overline{R}\sqrt{N_{\mathrm{IC}}})$ provides the relevant overall statistical uncertainty of the ensemble average, while the Poisson estimate identifies the characteristic contribution expected from finite-time counting fluctuations.

\paragraph*{Convergence of chaotic-ensemble statistics.}
We next examine the dependence of the chaotic-subensemble statistics on the integration time $T$ and the number of sampled initial conditions $N_{\mathrm{IC}}$.

Two distinct effects are relevant. At finite $T$, the observed trajectory-to-trajectory standard deviation may contain finite-time fluctuations. If dynamically distinct chaotic components with different asymptotic flip rates are present, it may also contain a persistent contribution that survives at long integration times. By contrast, the finite value of $N_{\mathrm{IC}}$ affects the precision with which the ensemble mean and standard deviation are estimated, rather than the underlying trajectory-to-trajectory spread itself.

To distinguish these effects, we consider the representative energies $E=2.5,4.5,5.4$ and $5.9$ and compare the four simulation configurations
\begin{align}
N_{\mathrm{IC}} &= 1{,}000 \text{ or } 2{,}000, \nonumber \\
T &= 10{,}000 \text{ or } 40{,}000.
\label{eq:total_convergence_test_simulation_parameters}
\end{align}
The resulting trajectory-to-trajectory standard deviations are given in Table~\ref{tab:total_convergence_test_std}.

\begin{table}[H]
\centering
\renewcommand{\arraystretch}{1.15}
\setlength{\tabcolsep}{12pt}
\begin{tabular}{ccccc}
    \toprule
    & \multicolumn{4}{c}{
        Standard deviation $\sigma_{\chi}$
    } \\
    \cmidrule(lr){2-5}
    & \multicolumn{2}{c}{$T=10{,}000$}
    & \multicolumn{2}{c}{$T=40{,}000$} \\
    \cmidrule(lr){2-3}
    \cmidrule(lr){4-5}
    Energy $E$
    & $N_{\mathrm{IC}}=1{,}000$
    & $N_{\mathrm{IC}}=2{,}000$
    & $N_{\mathrm{IC}}=1{,}000$
    & $N_{\mathrm{IC}}=2{,}000$ \\
    \midrule
    2.5 & 0.0046 & 0.0045 & 0.0023 & 0.0023 \\
    4.5 & 0.0033 & 0.0033 & 0.0017 & 0.0017 \\
    5.4 & 0.0085 & 0.0085 & 0.0041 & 0.0047 \\
    5.9 & 0.0041 & 0.0041 & 0.0020 & 0.0020 \\
    \bottomrule
\end{tabular}
\caption{Trajectory-to-trajectory standard deviation $\sigma_{\chi}$ of arm-2 flip rates within the chaotic subensemble at representative energies, for the integration times $T$ and ensemble sizes $N_{\mathrm{IC}}$ specified in \eqref{eq:total_convergence_test_simulation_parameters}.}
\label{tab:total_convergence_test_std}
\end{table}

Two trends are apparent. At fixed integration time, doubling the ensemble size produces little change in the measured standard deviation, indicating that $N_{\mathrm{IC}}=1{,}000$ is sufficient to obtain a stable estimate of the trajectory-to-trajectory spread. By contrast, increasing the integration time by a factor of four reduces the standard deviation by approximately a factor of two.

At each fixed energy, this behavior is consistent with
\begin{gather*}
\sigma_{\chi}(T)\propto T^{-1/2},
\end{gather*}
or, in terms of the relative spread,
\begin{gather*}
\frac{\sigma_{\chi}(T)}{\overline{R}_{\chi}(T)}
\propto
\frac{1}{\sqrt{\overline{N}_{\mathrm{flips},\chi}(T)}}
\propto T^{-1/2},
\end{gather*}
where $\overline{R}{\chi}$ and $\overline{N}{\mathrm{flips},\chi}$ denote the mean flip rate and mean flip count within the chaotic subensemble. The observed scaling, together with the absence of a detectable $T$-independent contribution, indicates that finite-time fluctuations dominate the measured spread over the integration times examined.
\paragraph*{Choice of chaotic-subensemble simulation parameters $T_{\chi},\, N_{\mathrm{IC},\chi}$.}
Eq.~\eqref{chaotic_traj_error} and the parameter-sensitivity study above show that increasing the integration time directly reduces the finite-time spread of the chaotic-subensemble flip rates. Increasing the number of initial conditions, on the other hand, improves the sampling of the subensemble but does not reduce the underlying trajectory-to-trajectory spread.

For the subsequent chaotic-subensemble analysis, we therefore favor longer integrations while retaining a sufficiently large sample of initial conditions. Relative to the production simulations, we increase the integration time tenfold and decrease the number of initial conditions tenfold,
\begin{gather*}
T_{\chi}=100{,}000,
\qquad
N_{\mathrm{IC},\chi}=100.
\end{gather*}
This choice keeps the total integration effort $N_{\mathrm{IC},\chi}T_{\chi}$ unchanged while substantially reducing finite-time fluctuations. We do not reduce $N_{\mathrm{IC},\chi}$ further because the ensemble size serves a second purpose: it provides a check that the regular component of phase space has been successfully excluded.

\subsection{Statistics over the chaotic subensemble}

\begin{figure}[H]
    \centering
    \includegraphics[width=1\linewidth]{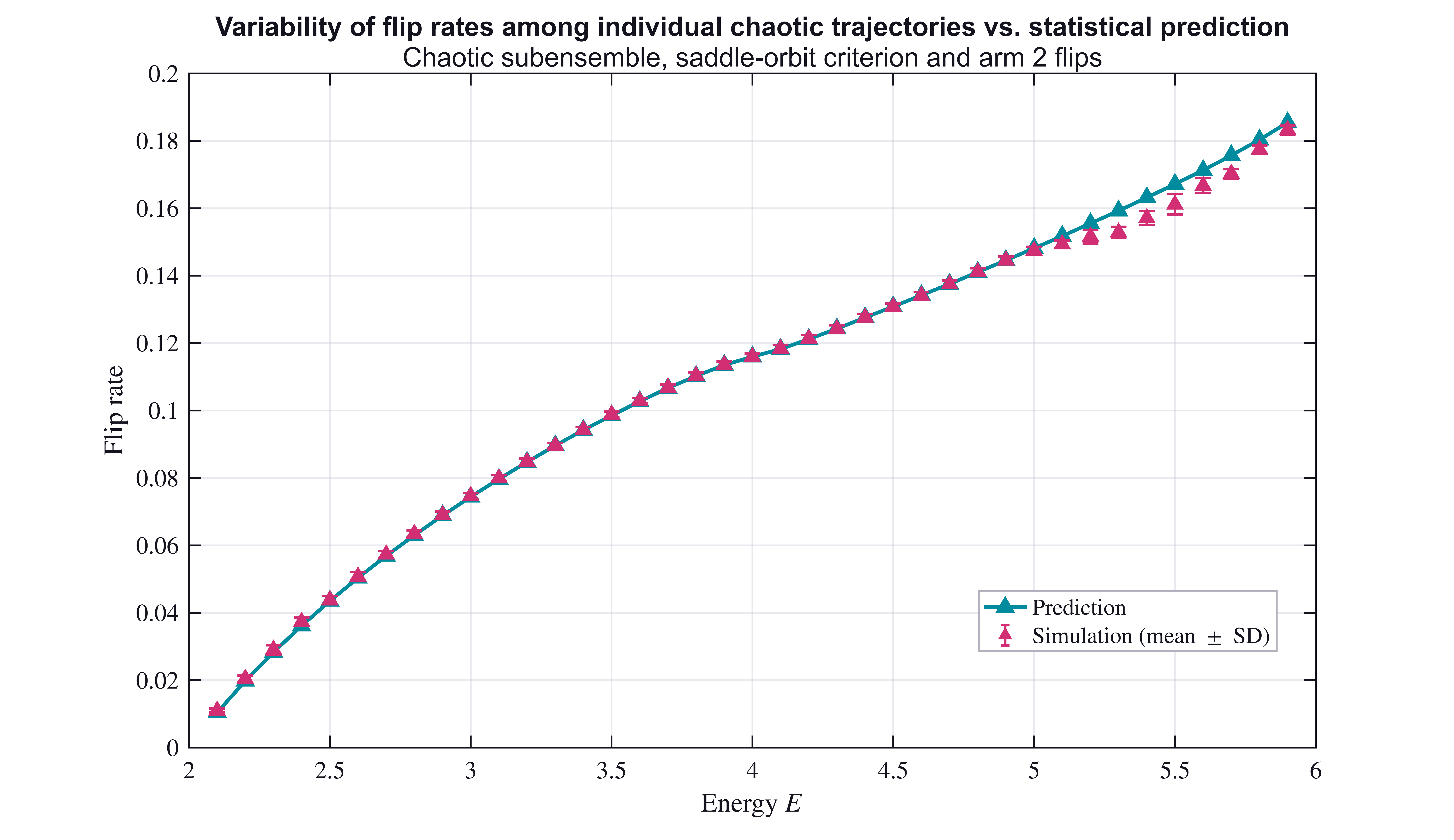}
    \caption{Mean flip rate of arm 2 within the chaotic subensemble, according to the saddle-orbit criterion, compared with the flux-over-a-saddle prediction. The vertical bars indicate the trajectory-to-trajectory standard deviation within the chaotic subensemble. The results are obtained from $N_{\mathrm{IC},\chi}=100$ independently sampled initial conditions, each propagated for $T_{\chi}=100{,}000$ simulation time units.}
\label{fig:saddle_orbit_criterion_second_pendulum_chaotic_trajectories}
\end{figure}

Having defined the chaotic subensemble and the corresponding simulation parameters $T_{\chi}$ and $N_{\mathrm{IC},\chi}$, we now examine whether the statistical prediction can be extended beyond ensemble averages to long-time individual chaotic trajectories. Since the trajectory-to-trajectory spread within the chaotic subensemble becomes small for sufficiently long integration times, agreement between the chaotic-subensemble mean and the statistical prediction would imply approximate agreement also for individual trajectories. Figure~\ref{fig:saddle_orbit_criterion_second_pendulum_chaotic_trajectories} presents the corresponding results for arm-2 flips according to the saddle-orbit criterion.

The flux-based prediction corresponds to an average over the full microcanonical energy surface, whereas the chaotic-subensemble mean is conditioned on a particular dynamical component. The two averages are therefore not expected to coincide exactly. In the present simulations, the chaotic-subensemble mean lies slightly below the full microcanonical prediction. This is consistent with the distribution of trajectory-averaged flip rates shown in Fig.~\ref{fig:flip_rate_histogram}, where the excluded regular component constitutes a well-separated high-rate population.

\paragraph*{Prediction quality index.}
We quantify the overall agreement between the statistical prediction and the numerical chaotic-subensemble results using the prediction quality index (PQI), defined in \eqref{eq:prediction_quality_index}. The resulting values are summarized in Table~\ref{tab:prediction_quality_saddle_orbit_arm2}. As for the implementation quality index, the PQI is evaluated over the full chaotic energy range $2<E<6$ \eqref{def:energy_range}, as well as separately within the low-, central-, and high-energy chaotic regimes. The largest deviation occurs in the low-energy regime, whereas the central chaotic regime exhibits the best agreement, with a PQI of $1.02\%$.

\begin{table}[H]
\centering
\begin{tabular}{c|c}
Energy regime & Prediction quality index (PQI)[\%]  \\
\hline
Full chaotic range ($2<E<6$) & 2.66 \\
Low chaotic ($2<E<2.8$) & 4.91 \\

Central chaotic ($2.8\leq E<5$) & 1.02 \\

High chaotic ($5 \leq E<6$) & 2.91 \\

\end{tabular}
\caption{Prediction quality index (PQI) for the arm-2 flip rate within the chaotic subensemble, according to the saddle-orbit criterion, evaluated over the full chaotic energy range and separately within the three chaotic energy regimes.}
\label{tab:prediction_quality_saddle_orbit_arm2}
\end{table}

\paragraph*{Chaotic quality index.}
The PQI characterizes the overall agreement of the chaotic-subensemble results with the statistical prediction. To isolate the systematic shift of the chaotic-subensemble mean from the full microcanonical prediction, we define a second measure. Denoting the mean flip rate measured over the chaotic subensemble by $R_{\chi,\mathrm{sim}}$, we define its relative deviation from the statistical prediction at each energy by
\begin{equation}
\mathcal R_{\chi}(E)
:=
\left|
\frac{R_{\chi,\mathrm{sim}}(E)}
{R_{\mathrm{pred}}(E)}
-1
\right|.
\end{equation}
The \emph{chaotic quality index} is then defined as the root-mean-square of this deviation over energy,
\begin{gather}
\chi\mathrm{QI}
:=
\sqrt{
\frac{1}{N_E}
\sum_{i=1}^{N_E}
\mathcal R_{\chi}(E_i)^2
}.
\label{eq:chaotic_quality_index}
\end{gather}
Thus, a smaller $\chi\mathrm{QI}$ indicates a smaller systematic difference between the chaotic-subensemble mean and the full microcanonical prediction. The resulting values are given in Table~\ref{tab:chaotic_quality_saddle_orbit_arm2}.

\begin{table}[H]
\centering
\begin{tabular}{c|c}
Energy regime & Chaotic quality index $\chi \mathrm{QI}$[\%]  \\
\hline
Full chaotic range ($2<E<6$) & 1.84 \\
Low chaotic ($2<E<2.8$) & 2.91 \\

Central chaotic ($2.8\leq E<5$) & 0.20 \\

High chaotic ($5 \leq E<6$) & 2.69 \\

\end{tabular}
\caption{Chaotic quality index $\chi\mathrm{QI}$, measuring the relative deviation of the chaotic-subensemble mean arm-2 flip rate from the full microcanonical prediction according to the saddle-orbit criterion, evaluated over the full chaotic energy range and separately within the three chaotic energy regimes.}
\label{tab:chaotic_quality_saddle_orbit_arm2}
\end{table}

Comparing Tables~\ref{tab:prediction_quality_saddle_orbit_arm2} and \ref{tab:chaotic_quality_saddle_orbit_arm2} shows that the offset between the chaotic-subensemble mean and the full microcanonical prediction contributes substantially to the PQI. This is particularly evident in the high-energy regime, where the two indices are close, $2.91\%$ and $2.69\%$, respectively. In the central chaotic regime, by contrast, the chaotic-subensemble mean differs from the full microcanonical prediction by only $0.20\%$, while the overall PQI remains $1.02\%$.

Together, Fig.~\ref{fig:saddle_orbit_criterion_second_pendulum_chaotic_trajectories} and Tables~\ref{tab:prediction_quality_saddle_orbit_arm2} and \ref{tab:chaotic_quality_saddle_orbit_arm2} constitute one of the central quantitative results of this work.

\subsection{Arm-1 flips} 
\label{subsec:results_arm1}

As discussed in Sec.~\ref{subsec:saddle orbit}, the saddle associated with flips of arm 1 also possesses a saddle orbit. The corresponding dividing surface therefore provides a saddle-orbit criterion for identifying arm-1 flips. We can consequently perform the same comparison between the flux-over-a-saddle prediction and the numerically measured flip rate.

\begin{figure}[H]
    \centering
    \includegraphics[width=0.85\linewidth]{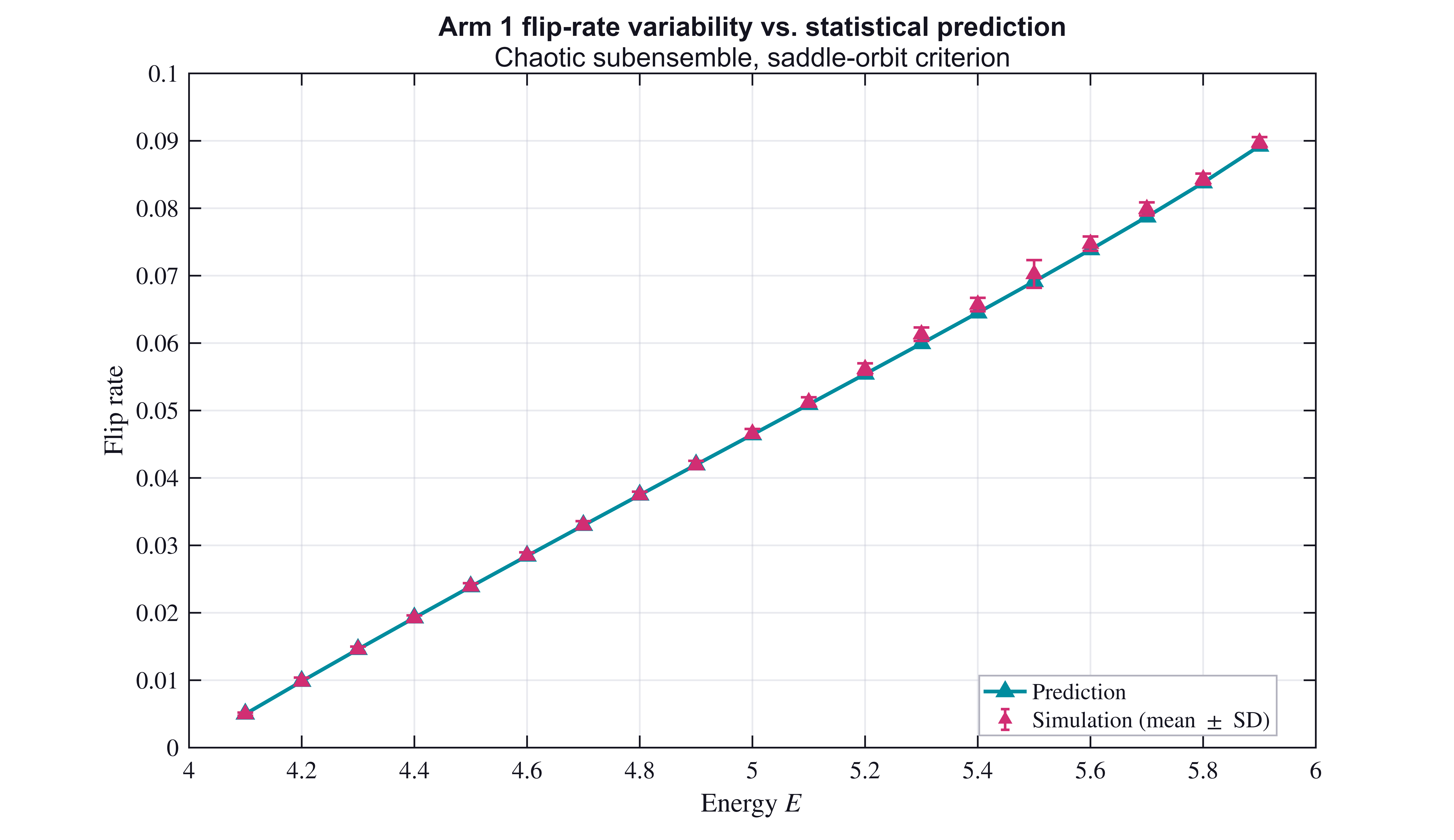}
    \caption{Mean flip rate of arm 1 within the chaotic subensemble, according to the saddle-orbit criterion, compared with the flux-over-a-saddle prediction. The vertical bars indicate the trajectory-to-trajectory standard deviation within the chaotic subensemble. The results are obtained from $N_{\mathrm{IC},\chi}=100$ independently sampled initial conditions, each propagated for $T_{\chi}=100{,}000$ simulation time units.}
    \label{fig:saddle_orbit_pendulum1_theory_vs_simulation}
\end{figure}

Figure~\ref{fig:saddle_orbit_pendulum1_theory_vs_simulation} shows close agreement between the flux-over-a-saddle prediction and the flip rate measured within the chaotic subensemble. Recall that for $5 \le E <6$, removing the regular component from the ensemble is important, as illustrated by the flip-rate histogram in Fig.~\ref{fig:flip_rate_histogram}. The regular component exhibits no arm-1 flips; consequently, excluding it slightly increases the simulated flip rate relative to the  statistical prediction for the full microcanonical ensemble, as seen in the figure.

To assess the accuracy of the numerical implementation independently of the chaotic-subensemble restriction, we evaluate the implementation quality index over the full arm-1 chaotic energy range $4<E<6$ \eqref{def:energy_range_arm1}, using the full microcanonical ensemble. We obtain an implementation quality index of $0.99\%$, indicating that the numerical procedure reproduces the corresponding microcanonical-ensemble average with high accuracy.

To assess the quality of the statistical prediction for the chaotic subensemble, we evaluate the prediction quality index over the same energy range \eqref{def:energy_range_arm1}. Since the agreement changes near the threshold energy at which arm 1 becomes capable of flipping, we divide the arm-1 chaotic range into low- and high-energy arm-1 chaotic regimes at $E=4.6$. The resulting prediction quality indices are summarized in Table~\ref{tab:prediction_quality_saddle_orbit_arm1}. The agreement improves in the higher-energy regime, where the PQI decreases from $3.79\%$ to $1.97\%$.

\begin{table}[H]
\centering
\begin{tabular}{c|c}
Energy regime & Prediction quality index (PQI) [\%] \\
\hline
Full arm-1 chaotic range ($4<E<6$) & 2.57 \\
Low arm-1 chaotic ($4<E<4.6$) & 3.79 \\
High arm-1 chaotic ($4.6\leq E<6$) & 1.97 \\
\end{tabular}
\caption{Prediction quality index (PQI) for the arm-1 flip rate within the chaotic subensemble, according to the saddle-orbit criterion, evaluated over the full energy range and separately within the low- and high-energy chaotic regimes.}
\label{tab:prediction_quality_saddle_orbit_arm1}
\end{table}

\subsection{Alternative flip criteria}

We next examine the dependence of the results on the choice of flip criterion. In addition to the exact saddle-orbit criterion, we consider the upright-arm criterion and the saddle-orbit approximation introduced in Sec.~\ref{sec:criteria}. The fluxes associated with these two alternative criteria are evaluated by numerical integration of \eqref{eq:upright_arm_flux_integral} and \eqref{eq:approximate_saddle_flux}, respectively.

\begin{figure}[]
    \centering
    \includegraphics[width=0.85\linewidth]{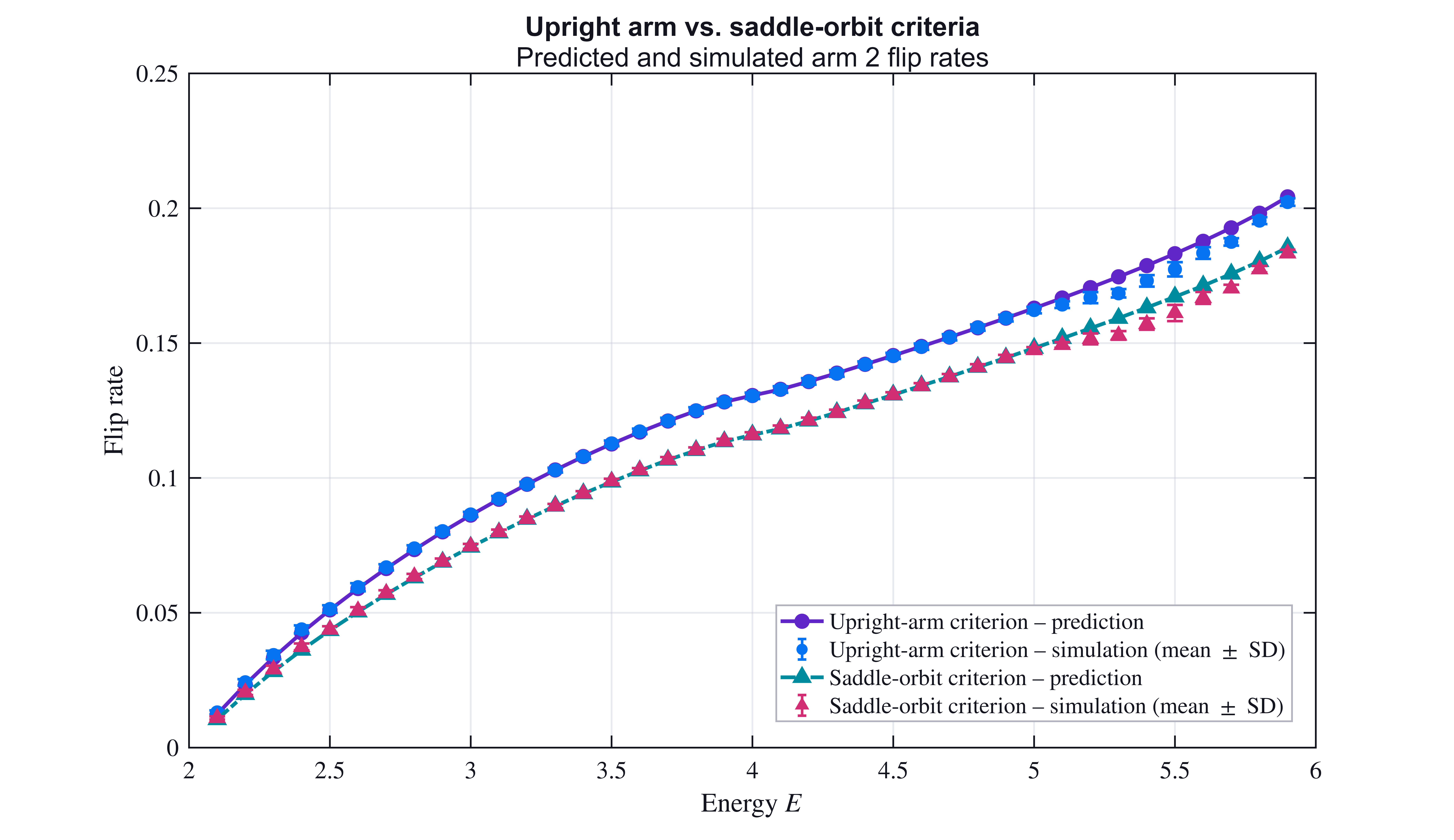}
    \caption{Theoretical and simulated arm-2 flip rates according to the upright-arm and saddle-orbit criteria, with the numerical ensemble restricted to the chaotic subensemble. The vertical bars indicate the trajectory-to-trajectory standard deviation within the chaotic subensemble. The results are obtained from $N_{\mathrm{IC},\chi}=100$ independently sampled initial conditions, each propagated for $T_{\chi}=100{,}000$ simulation time units.}
\label{fig:saddle_orbit_and_upright_criterion_theory_vs_simulation}
\end{figure}

We first compare the statistical prediction with the simulated flip rates according to the upright-arm criterion. Figure~\ref{fig:saddle_orbit_and_upright_criterion_theory_vs_simulation} presents this comparison alongside the corresponding results for the exact saddle-orbit criterion, shown previously in Fig.~\ref{fig:saddle_orbit_criterion_second_pendulum_chaotic_trajectories}. The figure shows close agreement between the statistical prediction and the simulated rates also for the upright-arm criterion.

The saddle-orbit dividing surface yields a \emph{lower} crossing rate by eliminating the additional short-time crossings associated with local recrossings of the simpler upright-arm surface. It therefore provides the physically relevant no-local-recrossing definition of passage through the saddle region.

To quantify the effect of these local recrossings, we compute the relative difference between the theoretical flip rates obtained using the upright-arm and saddle-orbit criteria and take its root-mean-square value over the investigated energy range. The resulting RMS relative difference is $13.46\%$. This provides a quantitative measure of the effect of the choice of dividing surface on the flip rate and, in particular, of the additional short-time recrossings counted by the upright-arm criterion.

\begin{figure}[H]
    \centering
    \includegraphics[width=0.75\linewidth]{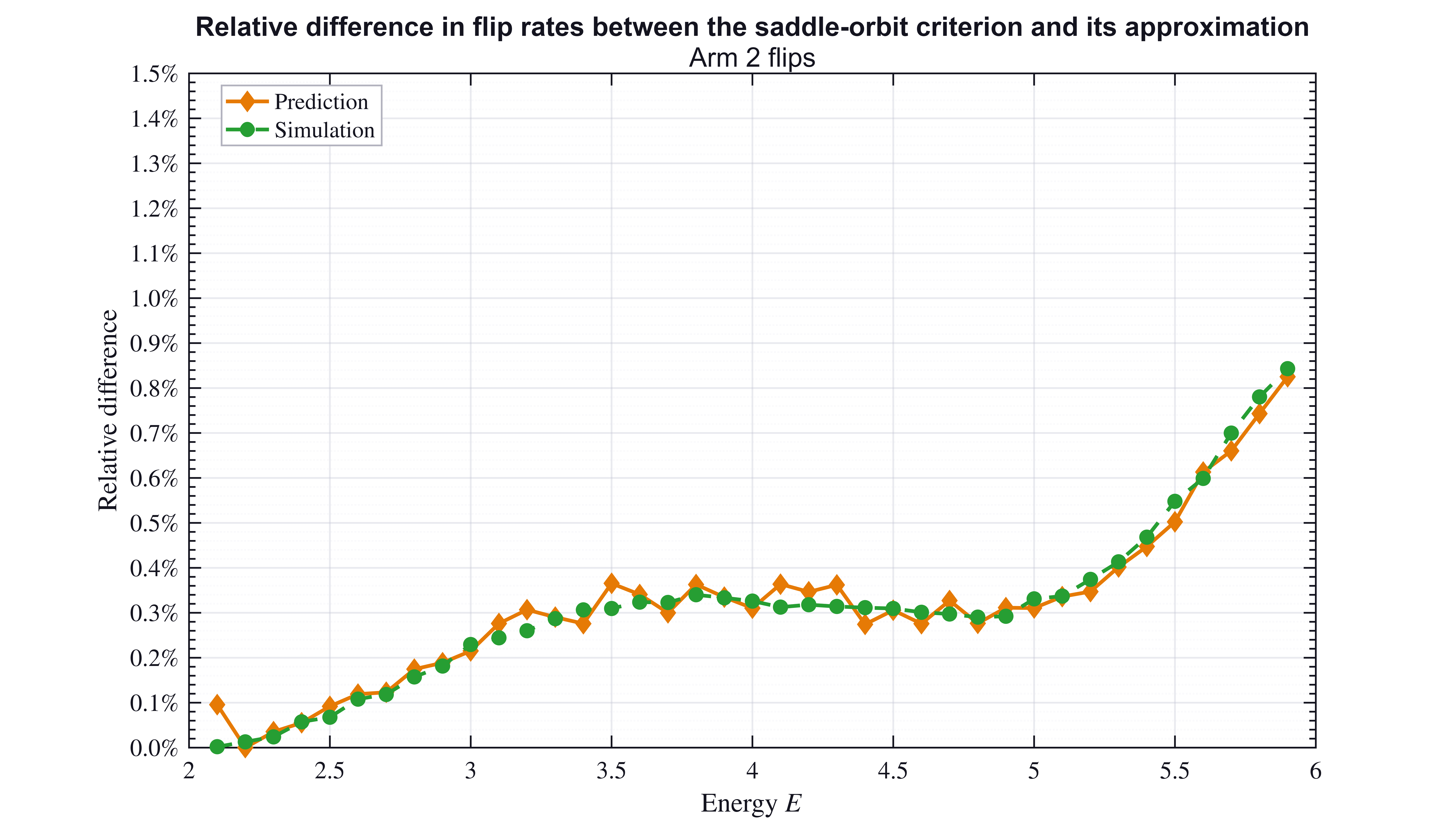}
    \caption{Relative difference between the saddle-orbit approximation and the exact saddle-orbit criterion as a function of energy. Results are shown separately for the statistical predictions and the corresponding simulation means, with the exact saddle-orbit criterion used as the reference.}
\label{fig:relative_error_saddle_orbit_approximation}
\end{figure}

We next assess the accuracy of the saddle-orbit approximation. The corresponding flip rates turn out to be almost indistinguishable from those obtained using the exact saddle-orbit criterion. Figure~\ref{fig:relative_error_saddle_orbit_approximation} therefore shows their relative difference directly. Over the investigated energy range, the relative difference between the two statistical predictions remains below $1\%$. The corresponding difference between the simulation means is likewise below $1\%$, approaching this value only at the largest deviations.

Computing the root-mean-square relative difference between the two theoretical flip rates over the investigated energy range, we obtain $0.36\%$. Thus, although the approximate dividing curve is considerably simpler than the numerically determined saddle-orbit dividing surface, it reproduces the theoretical flip rate of the exact criterion with high accuracy. The close agreement of the corresponding simulation means further shows that the approximation introduces only a sub-percent change in the measured flip rate.

Finally, we use the prediction quality index to compare the agreement of each alternative statistical prediction with its corresponding numerical simulations. The resulting PQI values for the saddle-orbit approximation and the upright-arm criterion are presented in Tables~\ref{tab:prediction_quality_saddle_orbit_approximation_arm2} and \ref{tab:prediction_quality_upright_arm_arm2}, respectively. Over the full chaotic energy range, the saddle-orbit approximation yields $\mathrm{PQI}=2.65\%$, while the upright-arm criterion gives $\mathrm{PQI}=2.58\%$. These values are also essentially identical to the value $2.66\%$ obtained for the exact saddle-orbit criterion in Table~\ref{tab:prediction_quality_saddle_orbit_arm2}. Thus, all three criteria exhibit very similar overall agreement with the simulations. In each case, the largest deviations occur in the low-energy chaotic regime, while the agreement improves substantially in the central and high-energy regimes.

\begin{table}[H]
\centering
\begin{tabular}{c|c}
Energy regime & Prediction quality index (PQI) [\%] \\
\hline
Full chaotic range ($2<E<6$) & 2.65 \\
Low chaotic ($2<E<2.8$) & 4.90 \\
Central chaotic ($2.8\leq E<5$) & 1.01 \\
High chaotic ($5\leq E<6$) & 2.89 \\
\end{tabular}
\caption{Prediction quality index (PQI) for the saddle-orbit-approximation criterion: arm-2 flip rate within the chaotic subensemble, evaluated over the full chaotic energy range and separately within the three chaotic energy regimes.}
\label{tab:prediction_quality_saddle_orbit_approximation_arm2}
\end{table}

\begin{table}[H]
\centering
\begin{tabular}{c|c}
Energy regime & Prediction quality index (PQI) [\%] \\
\hline
Full chaotic range ($2<E<6$) & 2.58 \\
Low chaotic ($2<E<2.8$) & 4.97 \\
Central chaotic ($2.8\leq E<5$) & 0.99 \\
High chaotic ($5\leq E<6$) & 2.53 \\
\end{tabular}
\caption{Prediction quality index (PQI) for the upright-arm criterion: arm-2 flip rate within the chaotic subensemble, evaluated over the full chaotic energy range and separately within the three chaotic energy regimes.}
\label{tab:prediction_quality_upright_arm_arm2}
\end{table}

\section{Summary and discussion}
\label{sec:summary}

\subsection*{Summary of results}

We conclude by summarizing the main results of this paper.

\paragraph*{Flip criterion and saddle-orbit flux.}  
Following \cite{MacKay1990,MacKay1994}, we adopted crossing of the saddle orbit to be a natural criterion for a flip. This criterion has two important advantages: the saddle orbit is periodic, and it minimizes the phase-flux, thereby avoiding what MacKay termed “sneaky returns,” namely spurious rapid recrossings. 
In addition, as shown in Fig.~\ref{fig:crossing_times}, this criterion displays a gap in the residence-time distribution, which is an even stronger indication that it gives a clean separation between successive flips. We believe that this is the unique flip criterion with this property. 

We determined the saddle orbit of the egalitarian double pendulum for both the down-up and up-down saddles. The bidirectional phase-flux is given by twice the integrated symplectic form, which in the present case reduces to the saddle orbit action.  We evaluated the resulting action and period functions, $\varphi(E)$ and $T(E)$, as shown in Fig.~\ref{fig:Time and flux saddle orbit}.

We also introduced two alternative flip criteria: the upright criterion and an analytic approximation to the saddle curve. For each of these criteria, we evaluated the corresponding flux function, see Fig.~\ref{fig:phase space flux and criteria flux}. These criteria are useful approximations: they require less computation while illustrating the minimal-flux property of the saddle orbit.  

\paragraph*{Statistical prediction.}
The flux-based statistical theory predicts the mean flip rate at fixed energy $E$ as the ratio of the relevant phase-flux to the accessible phase volume \eqref{eq:flip_rate_formula}. We computed the phase-volume function $\Omega(E)$, as shown in Fig.~\ref{fig:phase space flux and criteria flux}. Together with the flux functions described above, this provides a complete statistical prediction for the mean flip rate.

\paragraph*{Rate statistics.}
We implemented numerical simulations of the double pendulum and collected flip statistics from long-time trajectories, with initial conditions sampled from the natural uniform distribution on phase space (with prescribed energy $E$). We evaluated both the mean and the standard deviation of the ensemble distribution, see Figs.~\ref{fig:saddle_orbit_criterion_second_pendulum_all_trajectories_mean_only} and \ref{fig:saddle_orbit_criterion_second_pendulum_all_trajectories_with_scattering}. We also presented the full flip-rate distribution in a certain case of interest, see Fig.~\ref{fig:flip_rate_histogram}.

\paragraph*{Comparison between prediction and simulations.}
Figure~\ref{fig:saddle_orbit_criterion_second_pendulum_all_trajectories_mean_only} presents the comparison between the statistical prediction and the ensemble-averaged simulation statistics of the flip rate for the down-up configuration saddle (the saddle at $E=2$).
The agreement is at the high precision level of order $\sim 0.2 - 1\%$, see Table~\ref{tab:implementation_quality}. 

Figure~\ref{fig:saddle_orbit_criterion_second_pendulum_chaotic_trajectories} further compares the prediction with the mean and scatter obtained after selecting chaotic time evolutions.  The selected chaotic data agree particularly well with the prediction in the central chaotic regime, where the prediction quality index is approximately 1\%; over the full energy range, it is approximately 2.6\%, see Table~\ref{tab:prediction_quality_saddle_orbit_arm2}. While agreement with the full ensemble average is expected from the statistical theory, up to numerical error, the selected data show that the prediction is also an accurate description of individual chaotic trajectories, without requiring averaging over initial conditions. This provides quantitative evidence for the validity of the ergodic approximation, particularly in the central chaotic regime.

This comparison between the flux-based prediction and the chaotic flip-rate statistics is the central quantitative result of the paper.

\paragraph*{Arm-1 flips.} A similar comparison of statistical prediction and simulation statistics was carried out for arm-1 flips, see Fig.~\ref{fig:saddle_orbit_pendulum1_theory_vs_simulation}. The prediction quality index is approximately 2.6\% over the full energy range.

\paragraph*{Dependence on the choice of flip criterion.} In addition to the natural saddle-orbit criterion, we considered two alternative criteria: the upright criterion and the one based on the approximate saddle orbit. For both, we find excellent agreement between prediction and simulation data, see Figs.~\ref{fig:saddle_orbit_and_upright_criterion_theory_vs_simulation} and \ref{fig:relative_error_saddle_orbit_approximation}. This demonstrates that the success of the flux-based method is robust with respect to the choice of dividing surface.

Moreover, the predictions obtained with the alternative criteria are rather close to those based on the saddle orbit: the approximate-saddle-orbit criterion differs by only ~0.4\%, while the upright-arm criterion differs by up to ~13\%, see Figs.~\ref{fig:saddle_orbit_and_upright_criterion_theory_vs_simulation} and \ref{fig:relative_error_saddle_orbit_approximation}. Thus, these simpler criteria, which require less computational effort, may provide useful approximations.

\subsection*{Discussion}

\paragraph*{Chaotic clock perspective.}
The ordinary simple pendulum famously serves as the basic physical mechanism of a clock: its periodic motion makes  the number of oscillations proportional to the elapsed time. Moreover, the near-independence of the small-oscillation period on the energy makes this measurement of time robust.

At first sight, one would not expect a chaotic and irregular system to be useful for measuring time. However, the existence of a well-defined mean flip rate shows that the double pendulum can effectively measure time intervals that are long enough to contain many flips. In this sense, the mean flip rate of the double pendulum may be regarded as a \emph{chaotic clock}. This perspective can be generalized to other chaotic systems with identifiable recurrent events. Unlike the simple pendulum, however, the flip rate is sensitive to energy dissipation; if dissipation is significant, it must be accounted for through the energy dependence of the predicted flip rate.

\paragraph*{Significance and context.} 
The main significance of this work lies in providing a statistical prediction for the mean flip rate of the double pendulum, a key observable of a prototypical chaotic system. The prediction is formulated in terms of phase-flux through a dividing surface naturally defined by a distinguished periodic orbit of the system---the saddle orbit---and is validated by its excellent agreement with numerical simulations. The flux-based method has appeared before, see the third paragraph of Sec.~\ref{sec:prediction}, and so did the flux-over-a-saddle realization, see the third paragraph below \eqref{crit:saddle} . To our knowledge, they are applied here for the first time to the double pendulum. A closely related flux-based approach, inspired in part by an earlier effort in the direction of this work \cite{Kol_Marmor_2016}, was applied to the egalitarian three-body system in \cite{flux-based}. The present results suggest that this framework may be useful more broadly for predicting significant observables in chaotic systems that admit suitable dividing surfaces.

We would like to end this discussion by mentioning a few open questions. 

\paragraph*{Beyond the ergodic approximation.}
The agreement between the present statistical prediction and the chaotic mean flip rate is limited by the fact that phase space also contains regular regions. Regular trajectories exhibit non-random flip sequences, and are not expected to be well described by a statistical theory. It is thus natural to exclude them from the ensemble of considered time evolutions. However, such a selection also biases the statistical prediction, producing an effect that lies beyond the simple ergodic approximation.

It would be interesting to extend the theory so as to correct for such effects. This would require knowledge of both the phase-space fraction occupied by the various regular components and the mean flip rate associated with each of them. 

\paragraph*{The surprising accuracy of the approximate saddle orbit.} 
We found that the approximate saddle orbit (Fig.~\ref{fig:phase space flux and criteria flux}) remains surprisingly close to the exact saddle orbit, even at energies significantly above the saddle energy. This leads to accurate analytic approximations for the saddle orbits and their associated fluxes. It would be interesting to understand the reason for this unexpected accuracy.

\begin{acknowledgments}
    It is a pleasure to thank R. MacKay and A. Ori for discussions. This research was funded by the Israel Science Foundation (grant no. 2058/25) and by a grant from Israel’s Council of Higher Education.
\end{acknowledgments}

\appendix
\section{The story behind this work}
\label{app:story}
The work on this project developed in a somewhat unusual way.  In this appendix, we give a short account of how the problem first arose over a decade ago, and how it later became connected with the three-body problem.

In 2015-2016, one of us (BK) taught Analytical Mechanics. He showed the class a demonstration of the double pendulum in order to emphasize that, while the course focuses mainly on integrable systems, most systems with two or more degrees of freedom are non-integrable, as the double pendulum clearly illustrates. When he watched the demonstration, the flips of the pendulum appeared as dramatic, random-like events. At first sight, predicting them seemed beyond hope. 

Later, BK found himself thinking that even when no clear path is in sight, a scientist may still be able to identify patterns and make progress. He thought about the fact that flips persisted, despite their irregularity. This suggested that an average flip rate could be defined. He then wondered whether a theoretical prediction for this rate might be possible. 

After some thought, he found a statistical prediction in terms of phase-flux, and recorded the idea in his research notebook in January 2016 \cite{Kol2016Notes}.

In the summer of that year, BK suggested this problem to Andrew Marmor, a visiting undergraduate summer student. The simple upright criterion was used to define flips. Andrew evaluated the numerical phase-space integrals required for the statistical predictions and compared them with simulation statistics. He found rather good agreement (in the 3\%–-9\% range) and summarized the results in a short document \cite{Kol_Marmor_2016}, all within only a few weeks. At the time, the document was not submitted for publication; as an outsider to the field, BK mistakenly assumed that such a result was probably already known in the literature.

In December 2019, BK learned about the work of Stone and Leigh \cite{Stone_Leigh_2019}, who had presented a closed-form statistical prediction for the egalitarian (non-hierarchical) three-body problem. He became drawn into this problem and worked on it intensively during COVID-19 lockdowns. Familiarity with the earlier flux-based analysis of the double pendulum helped guide the formulation of the flux-based statistical theory for the three-body system \cite{flux-based}, see also the related papers \cite{DynRed, testing, measurement, sigmaE, sigmaEL, story}. This theory replaced the foundations of previous statistical theories and led to a leap in agreement with simulations. It rests on the same basic premise: that an event rate can be predicted through phase-flux. 

In 2021, BK interacted with R. MacKay and learned about references \cite{MacKay1990,MacKay1994}.

Finally, in November 2025, BK described several possible research problems for PH and mentioned the double pendulum as background. PH expressed interest in reproducing and revisiting the 2016 results, and this marked the beginning of the present work, returning to the original problem roughly ten years after the first ideas were recorded, and extending them in a meaningful way.

\section{Normal mode analysis at extrema of the potential}
\label{app:normal_mode}

For the general double pendulum, we derive the normal modes at the four extrema of the potential.
As seen in Fig.~\ref{fig:potential}, there are four extremum points for the double pendulum potential. 
For $(\theta_1^*,\theta_2^*) \in \{(0,0),(0,\pi),(\pi,0),(\pi,\pi)\}$ define $\epsilon_i = \cos(\theta_i^*)$. Thus,
\begin{table}[H]
    \centering
    \begin{tabular}{c|c|c|c}
        $(\theta_1^*,\theta_2^*)$ & $\epsilon_1$  & $\epsilon_2$  & $\epsilon \equiv \epsilon_1 \epsilon_2$
        \\
        \hline
         $(0,0)$& $+1$  & $+1$  & $+1$ \\
         $(0,\pi)$& $+1$  & $-1$  & $-1$ \\
         $(\pi,0)$& $-1$  & $+1$  & $-1$ \\
         $(\pi,\pi)$& $-1$  & $-1$  & $+1$ \\
    \end{tabular}
    \label{tab:epsilon values for extremum points of the potential}
\end{table}
We use $\mu := \frac{m_2}{m_1}, \quad \eta := \frac{L_2}{L_1}$ and $c := \cos(\theta_1 - \theta_2)$. The mass matrix is:
\begin{gather*}
    M(\theta_1 , \theta_2) = m_1 L_1^2 \begin{pmatrix}
        1+ \mu & \mu \eta c \\ 
        \mu \eta c & \mu \eta^2
    \end{pmatrix},
\end{gather*}
From the fact that $c^* = \cos(\theta_1^* - \theta_2^*) = \epsilon_1 \epsilon_2 \equiv \epsilon$, the equilibrium mass matrix is 
\begin{gather*}
    M_* = m_1 L_1^2 \begin{pmatrix}
        1+\mu & \mu\eta \epsilon \\ 
         \mu\eta \epsilon & \mu\eta^2
    \end{pmatrix}
\end{gather*}
Now, we want to derive the Hessian of the potential.
The potential of the general double pendulum is:
\begin{gather*}
  V(\theta_1,\theta_2)= -(m_1 + m_2) L_1 g \cos(\theta_1) - m_2 L_2 g \cos(\theta_2)  + (m_1 +m_2)L_1 g + m_2L_2g    
\end{gather*}
Therefore, 
\begin{gather*}
    \frac{\partial V}{\partial \theta_1} = (m_1 + m_2) L_1 g \sin(\theta_1), \qquad \frac{\partial V}{\partial \theta_2} = m_2 L_2 g \sin(\theta_2)
\end{gather*}
Both derivatives vanish at all four extrema, since
$\sin\theta_i^*=0$ for $\theta_i^*\in\{0,\pi\}$.
The Hessian is:
\begin{gather*}
    K(\theta_1,\theta_2) = D^2 V = \begin{pmatrix}
        (m_1+m_2)L_1 g \cos(\theta_1) & 0 \\ 
        0 & m_2 L_2 g \cos(\theta_2)
    \end{pmatrix}
    = m_1 L_1 g \begin{pmatrix}
        (1+ \mu) \cos \theta_1 & 0 \\ 
        0 & \mu \eta \cos \theta_2
    \end{pmatrix}
\end{gather*}
At an arbitrary extremum,
\begin{gather*}
    K_* 
     = m_1L_1 g \begin{pmatrix}
         (1 +\mu) \epsilon_1 & 0 \\ 
         0 & \mu \eta \epsilon_2
     \end{pmatrix}
\end{gather*}
This immediately determines the character of each extremum:
$(0,0)$ is a minimum, $(\pi,\pi)$ is a maximum, and the two mixed
configurations $(0,\pi)$ and $(\pi,0)$ are index-one saddles, since
$K_*$ has one positive and one negative eigenvalue there.
At an arbitrary extremum, 
\begin{gather*}
    E_* = m_1L_1 g \left[(1+\mu)(1-\epsilon_1) + \mu \eta (1-\epsilon_2) \right]
\end{gather*}
Thus, 
\begin{gather*}
    E_{DD} = 0, \quad E_{DU} = 2\mu \eta m_1 L_1 g, \quad E_{UD} = 2(1+\mu)m_1L_1g, \quad E_{UU} = 2 (1+\mu + \mu \eta) m_1 L_1 g
\end{gather*}
The normal-mode equation is: 
\begin{gather*}
    M_* \ddot{\xi} + K_* \xi = 0.
\end{gather*}
We seek a normal-mode solution of the form
$
    \vec{\xi}(t)=\vec{a}e^{i\Omega t},
$
which gives
\begin{gather*}
    (K_*-\Omega^2M_*)\vec{a}=0.
\end{gather*} 
Introduce the dimensionless squared frequency: $x= \frac{L_1}{g} \Omega^2$.
Dividing out the common factor $m_1 L_1g$, the eigenvalue problem becomes
\begin{gather*}
    \begin{pmatrix}
        (1+ \mu)(\epsilon_1-x) & -x\mu\eta \epsilon \\ 
        -x \mu \eta \epsilon & \mu\eta (\epsilon_2 - \eta x)
    \end{pmatrix} 
    \begin{pmatrix}
        a_1 \\ a_2
    \end{pmatrix} 
    =0
    \\
    \implies \det (K_* - \Omega ^2 M_* ) = 0 \implies (1+\mu)(\epsilon_1-x)(\epsilon_2-\eta x)(\mu\eta) - (\mu \eta)^2 x^2 \epsilon^2 = 0 
\end{gather*}
After simplification, using that $\epsilon^2=1$
\begin{gather*}
    \eta x^2 - (1+\mu) (\eta \epsilon_1 + \epsilon_2) x + (1+\mu) \epsilon_1 \epsilon_2 = 0 . 
\end{gather*}
Solving the quadratic equation, 
\begin{gather*}
\boxed {
    x_{\pm} = \frac{(1+\mu) (\eta \epsilon_1 + \epsilon_2) \pm \sqrt{(1+\mu)^2 (\eta \epsilon_1 + \epsilon_2)^2 - 4\eta (1+\mu) \epsilon_1 \epsilon_2} }{2\eta} 
    .}
\end{gather*}
Since
$
    x=\frac{L_1}{g}\Omega^2,
$
a positive root $x>0$ corresponds to an oscillatory mode with frequency
$
    \omega=\sqrt{\frac{g}{L_1}x}
    ,
$
and linearized eigenvalues $\lambda=\pm i\omega$.
A negative root $x<0$ instead corresponds to a hyperbolic mode. In this case,
writing $\Omega^2=-\alpha^2$ gives
$
    \alpha=\sqrt{-\frac{g}{L_1}x}
    ,
$
with linearized eigenvalues $\lambda=\pm\alpha$.

We can also determine the relative motion of the two arms near each equilibrium point.
From the second row of the equation $(K_* - \Omega^2 M_*) \vec a  =0 $, we obtain 
\begin{gather*}
    -x \mu \eta \epsilon_1 \epsilon_2 a_1 + \mu\eta (\epsilon_2 - \eta x) a_2 = 0 \implies \frac{a_2}{a_1} = \frac{x \epsilon_1 \epsilon_2}{\epsilon_2  - \eta x}
\end{gather*}
Thus, a normal-mode eigenvector associated with either root can be chosen as 
\begin{gather*}
    \vec{a}(x) = \begin{pmatrix}
        1 \\ 
        \frac{x \epsilon_1 \epsilon_2}{\epsilon_2 - \eta x}
    \end{pmatrix}
\end{gather*}
For the egalitarian case, at the $(0,\pi)$ saddle, the equation for $x$ is 
$x^2-2 =0$ so $x_{\pm} = \pm \sqrt{2} \implies \omega^2 = \sqrt{2}$. The center mode has $\frac{a_2}{a_1} = \frac{\sqrt{2}}{1+ \sqrt{2}} = 2 - \sqrt{2},$ so 
\begin{gather}
    a_{osc} \propto \begin{pmatrix}
        1 \\ 
        2- \sqrt{2}
    \end{pmatrix}.
    \label{AASO}
\end{gather}
This eigenvector is used in the saddle orbit criterion approximation.
\section{Action--period relation for periodic orbits: further proofs} 
\label{app:action period relation}
In this appendix, we provide two alternative derivations of the action--period identity 
\begin{gather*}
    \frac{d \varphi}{dE} = T(E),
\end{gather*}
where $\varphi(E) = \oint_{\gamma_E} \vec p \cdot d \vec q $ is the canonical action of the periodic orbit $\gamma_E$, and $T(E)$ is its period.

We consider an autonomous Hamiltonian system and assume that $\gamma_E$ belongs to a smooth, energy parametrized family of periodic orbits.

\paragraph*{One-degree-of-freedom systems.}
Consider a one-degree-of-freedom system with Hamiltonian $H(q,p)= E$.
\begin{gather*}
    \varphi(E)=\oint_{\gamma_E}p\,dq.
\end{gather*}

Differentiating with respect to the energy gives
\begin{gather}
    \frac{d\varphi}{dE}
    =
    \oint_{\gamma_E}
    \left.\frac{\partial p}{\partial E}\right|_q\,dq.
    \label{eq:action_derivative_1d_appendix}
\end{gather}

For a librating orbit, this relation follows by writing
the action as an integral over the two momentum branches,
$p_+(q,E)$ and $p_-(q,E)$, between the turning points
$q_-(E)$ and $q_+(E)$:
\begin{gather*}
    \varphi(E)
    =
    \int_{q_-(E)}^{q_+(E)}
    \left[p_+(q,E)-p_-(q,E)\right]\,dq.
\end{gather*}
When differentiating this expression, the boundary terms
vanish because the two momentum branches coincide at
the turning points. Equation
\eqref{eq:action_derivative_1d_appendix}
therefore follows. The corresponding result for a
rotating orbit follows directly by integrating over
one complete angular cycle.

Since $H(q,p(q,E))=E$, differentiation with respect to
$E$ at fixed $q$ yields
$
    \frac{\partial H}{\partial p}
    \left.\frac{\partial p}{\partial E}\right|_q
    =1.
$

Using Hamilton's equation,
$
    \dot q=\frac{\partial H}{\partial p},
$
we obtain
$
    \left.\frac{\partial p}{\partial E}\right|_q
    =
    \frac{1}{\dot q}.
$
Substituting into \eqref{eq:action_derivative_1d_appendix},
we find
\begin{gather}
    \boxed{
    \frac{d\varphi}{dE}
    =
    \oint_{\gamma_E}\frac{dq}{\dot q}
    =
    \int_0^{T(E)}dt
    =
    T(E).
    }
\end{gather}
The identity holds
for regular periodic orbits for which the period is finite.

\paragraph*{Systems with $n$ degrees of freedom.}
Consider an autonomous Hamiltonian system with $n \ge 1$ degrees of freedom and a smooth family of periodic orbits $\gamma_E$.
The canonical action is
$\varphi(E) = \oint_{\gamma_E} \vec p \cdot d \vec q.$
We parameterize each periodic orbit by $s \in [0,1]$ such that $\vec q = \vec q (s,E),\,\, \vec p = \vec p (s,E)$ and $dt = T(E) \, ds.$
The action can then be expressed as
\begin{gather}
    \varphi(E)
    =
    \int_0^1
    \vec p(s,E)\cdot
    \frac{\partial\vec q(s,E)}{\partial s}\,ds.
    \label{eq:action_parametrized_appendix}
\end{gather}
Differentiating with respect to the energy at fixed $s$,
we obtain 
\begin{gather*}
    \frac{d\varphi}{dE}
    =
    \int_0^1 ds\,
    \left[
    \frac{\partial\vec p}{\partial E}
    \cdot
    \frac{\partial\vec q}{\partial s}
    +
    \vec p\cdot
    \frac{\partial^2\vec q}{\partial E\,\partial s}
    \right].
\end{gather*}
Integrating the second term by parts gives
\begin{gather*}
    \frac{d\varphi}{dE}
    =
    \left[
    \vec p\cdot
    \frac{\partial\vec q}{\partial E}
    \right]_0^1
    +
    \int_0^1 ds\,
    \left[
    \frac{\partial\vec p}{\partial E}
    \cdot
    \frac{\partial\vec q}{\partial s}
    -
    \frac{\partial\vec p}{\partial s}
    \cdot
    \frac{\partial\vec q}{\partial E}
    \right].
\end{gather*}
Since the trajectories are periodic,
$
    \vec q(s=1,E)=\vec q(s=0,E),
    \, \,
    \vec p(s=1,E)=\vec p(s=0,E),
$
the boundary term vanishes. Consequently,
\begin{gather*}
    \frac{d\varphi}{dE}
    =
    \int_0^1 ds\,
    \left[
    \frac{\partial\vec p}{\partial E}
    \cdot
    \frac{\partial\vec q}{\partial s}
    -
    \frac{\partial\vec p}{\partial s}
    \cdot
    \frac{\partial\vec q}{\partial E}
    \right].
\end{gather*}
Using the parameterization $dt=T(E)\,ds$, we have
\begin{gather*}
    \frac{\partial\vec q}{\partial s}
    =
    T(E)\dot{\vec q},
    \qquad
    \frac{\partial\vec p}{\partial s}
    =
    T(E)\dot{\vec p}.
\end{gather*}

Therefore,
\begin{gather*}
    \frac{d\varphi}{dE}
    =
    \int_0^1 T(E)\,ds\,
    \left[
    \frac{\partial\vec p}{\partial E}
    \cdot\dot{\vec q}
    -
    \dot{\vec p}\cdot
    \frac{\partial\vec q}{\partial E}
    \right].
\end{gather*}
Applying Hamilton's equations,
\begin{gather*}
    \dot{\vec q}
    =
    \frac{\partial H}{\partial\vec p},
    \qquad
    \dot{\vec p}
    =
    -\frac{\partial H}{\partial\vec q},
\end{gather*}
gives
\begin{gather*}
    \frac{d\varphi}{dE}
    =
    \int_0^1 T(E)\,ds\,
    \left[
    \frac{\partial H}{\partial\vec p}
    \cdot
    \frac{\partial\vec p}{\partial E}
    +
    \frac{\partial H}{\partial\vec q}
    \cdot
    \frac{\partial\vec q}{\partial E}
    \right].
\end{gather*}
Along the family of periodic orbits,
$
    H\bigl(\vec q,\vec p\bigr)=E.
$

Differentiating this relation with respect to $E$
at fixed $s$, and applying the chain rule, yields
$
    \frac{\partial H}{\partial\vec p}
    \cdot
    \frac{\partial\vec p}{\partial E}
    +
    \frac{\partial H}{\partial\vec q}
    \cdot
    \frac{\partial\vec q}{\partial E}
    =1.
$

It follows that
\begin{gather}
    \boxed{
    \frac{d\varphi}{dE}
    =
    \int_0^1 T(E)\,ds
    =
    T(E).
    }
    \label{eq:action_period_identity_ndof_appendix}
\end{gather}

Thus, the action--period identity holds for a smooth
family of periodic orbits in an autonomous Hamiltonian
system with an arbitrary number of degrees of freedom.

\section{Upright arm criterion flips for general masses and arm-lengths}\label{sec:rate_computation_general_masses_and_lengths}
We introduce the dimensionless ratios: 
\begin{gather*}
    \mu = \frac{m_2}{m_1}, \quad \eta = \frac{L_2}{L_1}
\end{gather*}
For the general double pendulum, 
\begin{gather*}
    T = \frac 1 2 \dot{\theta}^T M(\theta) \dot{\theta}, \quad M(\theta) = m_1 L_1^2 \begin{pmatrix}
        1+\mu & \mu \eta c \\ 
        \mu \eta c & \mu \eta^2
    \end{pmatrix}, \quad c= \cos(\theta_1-\theta_2)
\end{gather*}
Its determinant is:
\begin{gather*}
    \det M =  (m_1 L_1^2)^2 \left[ \mu\eta^2 (1+\mu) - (\mu \eta c)^2
    \right] = (m_1 L_1^2)^2 \mu \eta^2 (1+ \mu - \mu c^2) 
\end{gather*}
Its inverse is 
\begin{gather*}
    M^{-1}  =  m_1 L_1^2 \frac{1}{\det M} \begin{pmatrix}
        \mu \eta^2 & -\mu \eta c \\ 
        - \mu \eta c & 1+\mu
    \end{pmatrix} 
     = \frac{1}{m_1 L_1^2} \frac{1}{\mu \eta^2 (1+ \mu - \mu c^2) } \begin{pmatrix}
        \mu \eta^2 & -\mu \eta c \\ 
        - \mu \eta c & 1+\mu
    \end{pmatrix} 
\end{gather*}
The potential for the general case: 
\begin{gather*}
      V(\theta_1,\theta_2)= -(m_1 + m_2) L_1 g \cos(\theta_1) - m_2 L_2 g \cos(\theta_2)  + (m_1 +m_2)L_1 g + m_2L_2g   
      \\
      =m_1 L_1 g \left[(1+\mu)(1-\cos \theta_1) + \mu \eta (1-\cos \theta_2)\right]
\end{gather*}

\paragraph*{Phase-volume \texorpdfstring{$\Omega(E)$}{Omega(E)}.}
First, we consider 
\begin{gather*}
    \Omega(E) = \int d^2\theta \; d^2 l \; \delta(H-E) 
\end{gather*}
For fixed $\theta$, 
\begin{gather*}
    K(\theta_1,\theta_2;E) = E- V(\theta_1, \ \theta_2)
\end{gather*}
Applying the transformation $y=Ll$ where $L^TL = M^{-1}$ gives
\begin{gather*}
    \int d^2l \; \; \delta\left(\frac{1}{2}l^T M^{-1}l - K \right) = 2\pi \sqrt{\det M} \;\Theta(K).
\end{gather*}
Using the determinant 
\begin{gather*}
      \det M = (m_1 L_1^2)^2 \mu \eta^2 (1+ \mu - \mu c^2) \\
      \implies \sqrt{\det M} = m_1 \, L_1^2 \, \eta \, \sqrt{\mu \left(1+\mu -\mu c^2 \right)} = m_1 \, L_1^2 \, \eta \, \sqrt{\mu (1+\mu \sin^2(\theta_1 - \theta_2))} 
\end{gather*}
Thus 
\begin{gather*}
\boxed{
    \Omega(E) = 2\pi m_1 L_1^2 \eta \sqrt{\mu} \int_{-\pi}^{\pi} d\theta_1 \int_{-\pi}^{\pi} d\theta_2 \; \sqrt{1+ \mu \sin^2(\theta_2- \theta_1)} \; \Theta(K)
    }
\end{gather*}
where 
\begin{gather*}
\boxed{
    K = E - m_1 L_1 g \left[(1+\mu)(1-\cos \theta_1) + \mu \eta (1-\cos \theta_2)\right]
    }
\end{gather*}
For $L_1 = L_2 = 1, \quad m_1 =m_2 = 1$ 
\begin{gather*}
    \sqrt{1+ \mu \sin^2 (\theta_1 - \theta_2 )} = \sqrt{2 -\cos^2 (\theta_1 - \theta_2)}
\end{gather*}
and for the available kinetic energy, $K$: 
\begin{gather*}
    K(\theta) = E - g [2(1-\cos(\theta_1)) + (1-\cos(\theta_2))] = E - 3g +2g\cos(\theta_1 ) +g\cos(\theta_2)
\end{gather*}
Thus the original integrand for the egalitarian case and $g=1$
\begin{gather*}
    \Omega(E)^{(eg)} = 2\pi \int_{-\pi}^{\pi} d\theta_1 \int_{-\pi}^{\pi} d\theta_2 \sqrt{2-\cos^2 (\theta_2 - \theta_1)} \; 
    \Theta(E-3 + 2\cos \theta_1 + \cos \theta_2 )
\end{gather*}
is recovered. 

\paragraph*{Phase-flux \texorpdfstring{$f_1(E)$}{f1(E)}.}
The flip flux is 
\begin{gather*}
    f_1 (E)  = \int d\theta_1 \; d\theta_2 \; dp_1 \; dp_2 \; \delta(H-E) |\dot{\theta}_2| \delta(\theta_2 - \pi)
\end{gather*}
After integrating over $\theta_2$, 
\begin{gather*}
    f_1 (E) = \int_{-\pi}^{\pi} d\theta_1 \; I(\theta_1; E) , \qquad I(\theta_1;E) = \int d^2 p \; \delta \left(\frac{1}{2} p^T M^{-1} p - K_{\pi} \right) |(M^{-1}p)_2|
\end{gather*}
The available kinetic energy on the wall is 
\begin{gather*}
    K_{\pi} (\theta_1 ; E) = E -V(\theta_1, \theta_2 = \pi) = E - m_1 g L_1 \left[(1+\mu)(1-\cos \theta_1) + \mu \eta (1-\cos \pi)\right] 
    \\ 
    = E - m_1 g L_1 \left[(1+\mu)(1-\cos \theta_1) + 2\mu \eta\right]
\end{gather*}
The saddle energy associated with the second-pendulum flip is 
\begin{gather*}
    E_c^{DU} = 2\mu \eta m_1 g L_1
\end{gather*}
For the case where $g=1, \quad m_1= m_2 = 1, \quad L_1 = L_2 = 1$ this becomes 
$E_c^{DU} = 2 $.
Now take $M^{-1} = L^T L$ and define $y = Lp$. Then, $\frac 1 2 p^T M^{-1} p = \frac{1}{2}y^T y$ and $d^2p = \frac{d^2y }{|\det L|} = \sqrt{\det M} \; d^2y $

For the velocity, 
\begin{gather*}
    \dot{\theta_2} = e_2^T M^{-1} p.
\end{gather*}
Define 
\begin{gather*}
    w = Le_2 \implies \dot{\theta}_2 = w^T y \implies |w|^2 = e_2^T M^{-1}e_2 = (M^{-1})_{22}
\end{gather*}
Therefore, the momentum integral becomes 
\begin{gather*}
    I = \sqrt{\det M} \int d^2y \; \delta\left(\frac{1}{2}y^2 - K_{\pi}\right) |w \cdot y |
\end{gather*}
Rotating $y$ so $w$ points along one axis yields:
\begin{gather*}
    I  = 4 \sqrt{2K_{\pi}} \sqrt{\det M} \sqrt{(M^{-1})_{22}}\;\;\Theta(K_{\pi})
\end{gather*}
For a symmetric $2 \times2$ mass matrix, 
\begin{gather*}
    (M^{-1})_{22} = \frac{M_{11}}{\det M} \implies \sqrt{M_{11}} = \sqrt{\det M} \sqrt{(M^{-1})_{22}}
\end{gather*}
Here, $M_{11} = m_1 L_1^2 (1+ \mu)$ 
Hence, 
\begin{gather*}
    I(\theta_1; E) = 4 \sqrt{2K_{\pi}} \sqrt{M_{11}} \;\Theta(K_{\pi}) = 4L_1 \sqrt{m_1 (1+\mu)} \sqrt{2K_{\pi}} \; \Theta(K_{\pi})
\end{gather*}
Therefore, 
\begin{gather*}
    \boxed{
    f_1(E) = 4L_1 \sqrt{2m_1 (1+\mu)} \int_{-\pi}^{\pi} d\theta_1 \sqrt{K_{\pi}(\theta_1;E)}_{+}  = 4L_1 \sqrt{2m_1 (1+\mu)} \int_{-\pi}^{\pi} d\theta_1 \sqrt{K_{\pi}(\theta_1;E)} \;\Theta(K_{\pi})
    }
\end{gather*}

\section{Flux expression equivalence} 
\label{app:flux expression equivalence} 

Define the Hamiltonian and the section function 
\begin{gather*}
    H(q,p) = E, \qquad g(q,p) = 0
\end{gather*}
For a two-degree-of-freedom system, phase space is four-dimensional 
\begin{gather*}
    d^2 p \; d^2q
\end{gather*}
The phase-space expression for the signed flux through $g=0$ is:
\begin{gather*}
    f(E,S) = \int d^2q \; d^2p \; \delta(H-E) \; \delta(g) \; \dot{g}
\end{gather*}
We show that this is the same geometrical object as 
\begin{gather*}
    f(E,S) = \int_{S} \omega.
\end{gather*}
Assume that $dg \neq 0$ on the dividing surface. Then locally, we can choose canonical coordinates 
$
    (Q_1,P_1,Q_2,P_2)
$
such that \begin{gather*}
    Q_1 = g.
\end{gather*}
Canonical transformations preserve the phase space measure 
\begin{gather*}
    d^2q \; d^2p = dQ_1 \; dP_1 \;  dQ_2 \; dP_2
\end{gather*}
Moreover, from Hamilton's equation 
\begin{gather*}
    \dot{g} = \dot{Q}_1 = \frac{\partial H}{\partial P_1}.
\end{gather*}
Therefore, 
\begin{gather*}
    f(E,S) = \int dQ_1 \; dP_1 \; dQ_2 \; dP_2 \; \delta(H-E) \; \delta(Q_1) \; \frac{\partial H}{\partial P_1} .
\end{gather*}
Integrating over $Q_1$ yields 
\begin{gather*}
    f(E,S) = \int   dP_1 \; dQ_2 \; dP_2 \; \delta(H(0,P_1,Q_2,P_2)-E)  \; \frac{\partial H}{\partial P_1}
\end{gather*}
Let $P_1^{\star}$ denote the value of $P_1$ on the chosen branch of the dividing surface, satisfying 
\begin{gather*}
    H(0,P_1^{\star},Q_2,P_2) = E
\end{gather*}
for fixed $(Q_2,P_2).$
Using the delta function identity $\delta(F(x)) = \sum_i \frac{\delta(x-x_i)}{|F'(x_i)|}$
\begin{gather*}
    \delta(H-E) = \frac{\delta(P_1 - P_1^{\star})}{|\frac{\partial H}{\partial P_1}|}.
\end{gather*}
Returning to the flux expression:
\begin{gather*}
      f(E,S) = \int   dP_1 \; dQ_2 \; dP_2 \; \frac{\delta(P_1 - P_1^{\star})}{|\frac{\partial H}{\partial P_1}|} \; \frac{\partial H}{\partial P_1} = \int dQ_2 \; dP_2 \; \frac{\partial H/ \partial P_1}{|\partial H/ \partial P_1|}.
\end{gather*}
The remaining factor is the orientation of the crossing, $sgn(\dot g) = \frac{\partial H/ \partial P_1}{|\partial H/ \partial P_1|}$. If we consider only positive crossings, $\dot g >0$, then 
\begin{gather}
    f(E,S_+) = \int_{S_+} dQ_2 \; dP_2
\label{eq:usual flux expression}
\end{gather}
Consider the symplectic form, $\omega = dp_1 \wedge dq_1 + dp_2 \wedge dq_2 = dP_1 \wedge dQ_1 + dP_2 \wedge dQ_2.$ 
On the dividing surface, $Q_1 = 0$, so $dQ_1 = 0.$
Thus, with the chosen orientation 
\begin{gather*}
    \int_{S_+} \omega = \int_{S_+} dP_2 \wedge dQ_2 = \int_{S_+} dQ_2 \; dP_2 ~.
\end{gather*}
From \eqref{eq:usual flux expression}, 
\begin{gather*}
    \boxed{
    \int_{S_+} \omega = \int d^2q \; d^2p \; \delta(H-E) \; \delta(g) \; \dot{g}
    } ~,
\end{gather*}
where the phase-space integral is understood to be restricted to the branch $S_+$ that is equivalent to positive crossings with $\dot{g}>0$.

\bibliography{pend_flux}
\end{document}